\documentclass[a4paper,fleqn,usenatbib]{mnras}

\usepackage[T1]{fontenc}
\usepackage{ae,aecompl}

\usepackage{graphicx}	% Including figure files
\usepackage{amsmath}	% Advanced maths commands
\usepackage{amssymb}	% Extra maths symbols
\usepackage{pdflscape}
\title[Mass loading of magnetized jets by stellar winds]{On the deceleration of Fanaroff-Riley Class I jets: mass loading of magnetized jets by stellar winds.}

\author[Angl\'es, Perucho, Mart\'{\i}, Laing]{
Andreu Angl\'es-Castillo$^{1,2}$, Manel Perucho,$^{1,3}$\thanks{E-mail: manel.perucho@valencia.edu}
Jos\'e Mar\'{\i}a Mart\'{\i}$^{1,3}$, Robert A. Laing$^{4}$
\\
$^{1}$Departament d'Astronomia i Astrof\'{\i}sica, Universitat de Val\`encia, C/ Dr. Moliner, 50, 46100, Burjassot, Val\`encia, Spain.\\
$^{2}$Departament de F\'{\i}sica Te\`orica \& IFIC, Universitat de Val\`encia-CSIC, C/ Dr. Moliner, 50, 46100, Burjassot, Val\`encia, Spain.\\
$^{3}$Observatori Astron\`omic, Universitat de Val\`encia, C/ Catedr\`atic Jos\'e Beltr\'an 2, 46980, Paterna, Val\`encia, Spain.\\
$^{4}$Square Kilometre Array Organisation, Jodrell Bank Observatory, Lower Withington, Macclesfield, Cheshire SK11 9FT, UK
}

\date{Accepted XXX. Received YYY; in original form ZZZ}

\pubyear{2015}

\begin{document}
\label{firstpage}
\pagerange{\pageref{firstpage}--\pageref{lastpage}}
\maketitle

\begin{abstract}
In this paper we present steady-state RMHD simulations that include a mass-load term to study the process of jet deceleration. The mass-load mimics the injection of a proton-electron plasma from stellar winds within the host galaxy into initially pair plasma jets, with mean stellar mass-losses ranging from $10^{-14}$ to $10^{-9}\,{M_\odot\,yr^{-1}}$. The spatial jet evolution covers $\sim 500\,{\rm pc}$ from jet injection in the grid at 10~pc from the jet nozzle. Our simulations use a relativistic gas equation of state and a pressure profile for the ambient medium. We compare these simulations with previous dynamical simulations of relativistic, non-magnetised jets. Our results show that toroidal magnetic fields can prevent fast jet expansion and the subsequent embedding of further stars via magnetic tension. In this sense, magnetic fields avoid a runaway deceleration process. Furthermore, when the mass-load is large enough to increase the jet density and produce fast, differential jet expansion, the conversion of magnetic energy flux into kinetic energy flux (i.e., magnetic acceleration), helps to delay the deceleration process with respect to non-magnetised jets. We conclude that the typical stellar population in elliptical galaxies cannot explain jet deceleration in classical FRI radio galaxies. However, we observe a significant change in the jet composition, thermodynamical parameters and energy dissipation along its evolution, even for moderate values of the mass-load. 
\end{abstract}


%%%%%%%%%%%%%%%%%%%%%%%%%%%%%%%%%%%%%%%%%%%%%%%%%%

%%%%%%%%%%%%%%%%% BODY OF PAPER %%%%%%%%%%%%%%%%%%

\section{Introduction}
%###################

Jets from radio-loud AGN are separated into two main classes, namely FR\,I and FR\,II, following their large-scale morphology and radio luminosity \citep{fr74}. FR\,I jets are decollimated at kiloparsec scales and show diffuse emission at their edges  \citep[e.g., 3C~31,][]{lb02a}. On the contrary, the more powerful FRII jets remain highly collimated and supersonic until the reverse shock \citep[hot-spots, e.g., Cyg~A,][]{cb96}. Despite their different morphology at large-scales both types of sources show relativistic speeds at parsec-scales, with differences associated to the transversal structure of the  axial velocity and with a trend towards smaller velocities in the case of FR\,I jets \citep{gi01,cg08,me11}.    

The deceleration of these jets is deduced by the symmetrization of the jet and counter-jet brightness, which implies non-relativistic flow velocity at kiloparsec scales \citep[][and references therein]{lb14}. The most widely accepted explanation of FRI jet deceleration at kpc scales relates it to entrainment of colder and denser ambient gas \citep{bi84,la93,la96} via a mixing shear layer \citep{dy86,dy93,bi94,wa09}. The growth of Kelvin-Helmholtz instability pinching modes after a strong recollimation shock \citep{PM07} has also been proposed as a destabilizing mechanism. \citet{ro08} discussed the deceleration of FR\,I jets by the growth of helical instabilities in a jet propagating through a homogeneous ambient medium. From an observational point of view, \cite{wo07} studied the possible relation between the X-ray helical structure observed in the powerful FR\,I jet ($\sim 10^{44}$\,erg\,s$^{-1}$) in NGC~315 and the development of instabilities, but no clear conclusions could be derived.

A recent paper \citep{gk18a} proposes the development of the relativistic centrifugal instability \citep{gk18b}, after the generation of a conical recollimation shock at hundreds of parsecs from the jet source. The recollimation shock is produced due to jet overpressure when propagating through a galactic atmosphere with decreasing pressure. The development of the centrifugal instability would imply entrainment by small-scale eddies growing at the jet boundary and propagating towards the jet axis with time/distance. %Regarding the role played by jet power and the ambient medium, \cite{ka09} derived a relation between jet power, $L_{\rm j}$, and mean ambient medium density, $\bar{n}_a$, that sets a separation between FRI and FRII jets, with FRIs located below $L_{\rm j}/\bar{n}_a \sim 10^{44 - 45} \, {\rm erg \ s^{-1} cm^3}$. A similar conclusion was also reached by \cite{tch16} from numerical simulations.

A second option, which might be complementary to turbulent entrainment from the jet boundary, is mass load by stellar mass loss \citep{ph83,ko94,BL96,lb02b,HB06}. \cite{ko94} showed that the interaction between the jet flow and the stellar winds from stars that may go through the jet cross-section as they orbit in the potential of the galaxy, can be treated as a hydrodynamical problem. The reason is that the gyroradii of the particles are much smaller than the size of the interaction region between the jet and the stellar wind. Thus, this situation can be modelled as a distributed source of mass injected into the jet. The first work dealing with this scenario was published by \citet{BL96}. In that paper, the authors studied the steady state structure of jets subject to mass load. They focused on low power, light and hot (non-magnetised) electron/proton jets, and concluded that the simulated jets could be efficiently decelerated by a population of old, low-mass stars within the typical FRI deceleration distance. A difference was noted between hotter and colder jets, because the former tend to cool down due to the entrainment of the colder wind particles, whereas the latter showed an increase in temperature due to the dissipation of kinetic energy associated with deceleration. 

\cite{HB06} performed an analytical study of the parameter space for stellar mass-loss rates from massive stars and jet powers that could end up in efficient jet deceleration. The authors derived the following useful relation by considering the deceleration distance as the distance at which the kinetic energy required to accelerate the loaded mass is approximately equal to the total initial jet kinetic power. The resulting expression, as given in \cite{pe14} is:  
\begin{eqnarray} \label{eq:hb}
l_{\rm d} \simeq \frac{1}{\gamma_{\rm j}} \! \! \left(\frac{L_{\rm j}}{10^{43} {\rm erg\,s}^{-1}}\right)
\!\! \left(\frac{\dot{M}}{10^{-11} {\rm M_\odot}{\rm yr}^{-1}}\right)^{-1} \times \nonumber \\ \left(\frac{n_s}{{\rm pc}^{-3}}\right)^{-1}\,  \!\! 
\! \left(\frac{R_{\rm j}}{10 {\rm pc}}\right)^{-2} \!\! \, 10^2\, {\rm kpc},
\end{eqnarray}

\noindent
where $\gamma_{\rm j}$ is the jet Lorentz factor, $L_{\rm j}$ is its kinetic power, $R_{\rm j}$ is the jet radius, $\dot{M}$ is the mean mass-loss rate of the stellar population in the galaxy, and $n_s$ is the number of stars per unit volume. This last parameter is obviously a decreasing function of distance to the galactic nucleus, so the expression stands as a lower limit to the deceleration distance. 

A number of works by Laing \& Bridle \citep[e.g.][]{lb02a,lb02b,lb14} have characterised the jet deceleration region as a typically extended region in which kinetic energy seems to be dissipated and the jet becomes X-ray bright \citep{lb14}. The authors conclude that entrainment is a gradual process that extends from the boundaries to the axis: the already slower outer layers of jets \citep[e.g., M87, CenA, NGC1052][]{wa18,mu14,ba16} could be decelerated on shorter scales than the faster spines. Furthermore, they also point out that stellar mass load can be more efficient in low power FRI jets (e.g., M84 and NGC193). In these sources no obvious transverse velocity gradients were found, which is more consistent with distributed mass load than with entrainment from the jet boundaries. \cite{wy13,wy15} have discussed the dynamical role of stellar mass-load in the jet in Centaurus~A, in the context of kinetic energy dissipation and particle acceleration, considering a detailed modelisation of the stellar population in the galaxy that includes a mixture of old and young stars. These works conclude that a large number of stars ($\sim 10^8$) could be located inside the jet, efficiently contributing to jet dynamics and non-thermal emission. On the contrary, in \citet{lb02a,lb02b}, the authors used a one-dimensional model of the source properties that required growing entrainment with distance to explain the jet and counter-jet properties in the more powerful ($10^{44}$\,erg\,s$^{-1}$) FRI jet 3C~31. They concluded that mass-entrainment by stellar winds could not be the main decelerating mechanism because the star number density is expected to fall with distance to the galactic centre, so the entrainment mechanism was more likely to be ambient gas mixing through the jet shear layer. 

\citet{pe14} presented a number of dynamic simulations based on the parameter space used by \citet{BL96}. The simulations showed that low power jets ($L_{\rm j} \leq 10^{42}\,{\rm erg/s}$) can be efficiently decelerated by a population of old, low-mass stars. The entrainment develops a runaway deceleration effect by increasing the jet thermal pressure, which triggers expansion and therefore an increased mass-load by the newly embedded stars. The jet expansion and deceleration causes a drop of the momentum flux per unit area, which brings the jet head close to stagnation. Nevertheless, there is a threshold of jet power, at a given mean stellar mass-loss rate, that can be estimated using Eq.~\ref{eq:hb}, over which stellar mass load has a marginal or negligible effect on the jet. 

A recent paper \citep{pe20} suggests that in the case of FRI jets with $L_{\rm j} \sim 10^{43}\,{\rm erg/s}$, the interaction of stars with the jet boundary as the enter or leave the jet could act as a trigger for turbulent mixing between the jet flow and the surrounding interstellar medium at tens to hundreds of parsecs \citep[see also][]{ta20}.The author proposed this solution for the triggering mechanism of FRI jet deceleration, based on the agreement of the time and spatial scales of this process with the observed deceleration scales \citep{lb14}.

Different studies have reported that the interaction between jets and stars can lead to efficient energy dissipation and high-energy emission \citep[e.g.,][]{bp97,ba10,ba12,br+12,aha17,AB13,wy13,wy15,pe+17,Vieyro:2017,tabr19}. Furthermore, the injection of protons in pair jets and the strong shocks generated have been suggested to be (and studied as) a possible source of proton acceleration and neutrino production \citep[e.g.,][]{br+12,mu18}. 

In this paper we present an extension of \citet{pe14} to magnetised jets with a jet power of $10^{43}\,~{\rm erg/s}$. The simulations presented here generate the steady state structure of the mass-loaded jets following the approach of \cite{ko15} \citep[see also ][]{fu18}, in order to study the structure of steady, narrow, axisymmetric relativistic flows using one-dimensional time-dependent simulations. This simplification has allowed us to run a large number of models, spanning several orders of magnitude in rest-mass density and internal energy, and three different values of the Lorentz factor. The stellar winds are included in the calculations as a source term in mass, in the same way as was done in \citet{pe14}. It is important to stress that our paper focuses on one of the two main strategies that have been followed to study this scenario, i.e., we study the possible global effect of mass-load by a distribution of stars along the jet flow channel, unlike other works that have focused on single jet-star interactions \citep[e.g.,][]{HB06,br+12,pe+17}.
 
The paper is organized as follows. A short description of the code and the setup of the simulations, together with the list of parameters used are presented in Section~\ref{sec:su}. The results are given in Section~\ref{sec:res}. We discuss these  results in Section~\ref{sec:disc}, and Section~\ref{sec:conc} collects the conclusions of this work.

\section{Numerical simulations}\label{sec:su}
%########################################

\subsection{The code}
%===============

The magnetohydrodynamical models have been computed following the approach developed by \cite{ko15}. This approach is designed to study the structure of steady, axisymmetric relativistic flows using one-dimensional time-independent simulations. The key point of the approach is the use of the {\it quasi-one-dimensional} approximation, which requires that the jet radius is much smaller than its length (i.e., narrow jets), and that the flow speed is close to the light speed $c$. Under these assumptions, together with that of negligible azimuthal velocity, the steady-state equations of relativistic magnetohydrodynamics (RMHD) can be treated as a one-dimensional (radial) problem in which the axial coordinate plays the role of the {\it temporal} coordinate. The code thus solves the one-dimensional time-dependent equations of RMHD plus a conservation equation for the tracer (jet mass fraction). 

The use of this approach allowed us to span a wide region of the parameter space, at the cost of giving a poor description of the jet boundary where the jet flow velocity becomes subrelativistic \citep[see, e.g.,][]{ma16,fu18}. However, in this paper we focus on jet deceleration due to mass-loading from stars inside the jet, a process in which the jet boundary plays no role. Besides, our approach is not suitable to describe the phase of strong jet deceleration once the flow velocity deviates significantly from the speed of light. Nevertheless, it is useful to discern the jet conditions under which the flow can or cannot be decelerated by stellar-wind mass-load.

\subsection{Set up} 
%=============

\subsubsection{Jet parameters}    
%-------------------------

   The majority of simulated models have a total power of $L_{\rm j} = 10^{43}$~erg/s, a typical value for FRI jets, and have a radius $R_{\rm j} = 1\,{\rm pc}$ at injection. The modelled jets are purely leptonic ($X_e = \rho_e/\rho = 1$). The jet density, Lorentz factor and axial component of the magnetic field are different among models but all are considered to be initially constant across the jet. The configuration of the toroidal field is also the same for all magnetised models
\begin{equation}\label{eq:bphi}
B^\phi(r) = \left\{ \begin{array}{ll}
\displaystyle{\frac{2 B_{\rm m, \, j}^\phi (r/R_{\rm m})}{1 + (r/R_{\rm m})^{2}}}, & 0 \leq r \leq R_{\rm j} \\ 
0,        & r > R_{\rm j},
               \end{array} \right.,
\end{equation}
with the same $R_{\rm m} = 0.37 R_{\rm j}$ (the radius at which the toroidal magnetic field reaches its maximum, $B^\phi_{\rm m, \, j}$), but different $B^{\phi}_{\rm m, \, j}$. Jets are injected in pressure equilibrium with the ambient medium, by computing the pressure distribution across the jet and matching them at the boundary as in \citet{ma15,ma16}. We have incorporated the relativistic gas equation of state of \cite{Sy57}, using the approximation given by \cite{cho10}. 

In order to compute the whole set of jet parameters, we started by fixing the jet rest-mass density, Lorentz factor, mean magnetic pitch angle\footnote{We define the pitch angle as $\phi_B = \tan^{-1}(B^\phi/B^z)$, so that $\phi_B=90^\circ$ stands for a purely toroidal field, whereas $\phi_B=0^\circ$ stands for a purely poloidal field.}, and ambient pressure. The jet densities, $\rho_{\rm j}$, span four orders of magnitude, from $10^{-32}$ to $10^{-28}$~g/cm$^{3}$; Lorentz factors, $\gamma_{\rm j}$, at injection range from 6 to 10, and mean magnetic pitch angles, $\phi_{\rm B, \, j}$, are $10^{\circ}$, $45^{\circ}$, and $75^{\circ}$. The ambient pressure at injection is $p_{\rm a, \, 0} = 10^{-7}$ dyn/cm$^2$. With these four parameters, and having fixed the jet power and radius, the remaining parameters of the jet can be derived; in particular, the mean gas pressure, $p_{\rm j}$, axial field, $B^z_{\rm j}$, toroidal field $B^{\phi}_{\rm j}$, and maximum toroidal field, $B^\phi_{\rm m, \, j}$.

The total power of the jet, including the rest-mass energy flux, is defined as
\begin{equation} 
\label{eq:power}
L_{\rm j} \, =\,   A_{\rm j} \left(\rho_{\rm j} h_{\rm j} \gamma_{\rm j}^2 + (B^\phi_{\rm j})^2\right) v_{\rm j},
\end{equation}
where $A_{\rm j}$ is the jet cross-section ($A_{\rm j} = \pi R_{\rm j}^2$) and $h_{\rm j}$ the specific enthalpy \citep{cho10},
\begin{eqnarray}
\label{eq:eos}
h_{\rm j} = \frac{5 c^2}{2 \xi} + (2 - \kappa) c^2 \left[
    \frac{9}{16} \frac{1}{\xi^2}  + \frac{1} {(2 - \kappa + \kappa \mu)^2}\right]^{1/2} \nonumber\\
    + \kappa c^2 \left[\frac{9}{16}\frac{1}{\xi^2}  + \frac{\mu^2}{(2 - \kappa + \kappa \mu)^2}\right]^{1/2},
\end{eqnarray}
with $\xi = \rho_{\rm j} c^2/p_{\rm j}$, $\kappa = n_{p^+}/n_{e^-}$ the ratio of proton and electron number densities, and $\mu = m_p/m_e$ the ratio of proton and electron masses. We note that in all the simulated models $\kappa = 0$ at injection. 

Furthermore, for the chosen configuration of the toroidal magnetic field, equilibrium with the ambient medium at injection demands that \citep{ma15}
\begin{equation}
p_{\rm j} = p_{\rm a, \, 0} - \frac{(B_{\rm j}^z)^2}{2},
\end{equation}
or, alternatively, 
\begin{equation}
\label{eq:transeq}
p_{\rm j} = p_{\rm a, \, 0} - \frac{(B_{\rm j}^\phi)^2}{2 \tan^2 \phi_{\rm B, \, j}}.
\end{equation}
Now, with $L_{\rm j}$, $\rho_{\rm j}$, $\gamma_{\rm j}$, $A_{\rm j}$, ${\phi_{\rm B, \, j}}$ and $p_{\rm a, \, 0}$ fixed, the system formed by Eqs.~\ref{eq:power}, \ref{eq:eos} (with $\kappa = 0$) and \ref{eq:transeq} is solved for $p_{\rm j}$ and the remaining parameters are derived.  The mean toroidal field is
\begin{equation}
B^{\phi}_{\rm j} = \sqrt{2 (p_{\rm a, \, 0} - p_{\rm j})} \tan \phi_{\rm B, \, j}.
\end{equation}
The (assumed constant) axial magnetic field is then $B^z_{\rm j} = B^{\phi}_{\rm j} / \tan \phi_{\rm B, \, j}$, and the maximum toroidal field is \citep[see][]{ma15},
\begin{equation}
B^\phi_{\rm m, \, j} = \frac{{B^{\phi}_{\rm j}} R_{\rm j}^2}{4 R_{\rm m} (R_{\rm j} - R_{\rm m} \arctan(R_{\rm j}/R_{\rm m}))}
\end{equation}
($\simeq 1.23 \, B^{\phi}_{\rm j}$ for $R_{\rm m} = 0.37 \, R_{\rm j}$).

The equilibrium profile for the gas pressure within the jet is then computed as in \cite{ma15}. Previous work with this code \citep{ma16} has shown that a contact discontinuity between the jet and the ambient medium can lead to the development of the pinching instability in axisymmetric flows and the disruption of the steady-state that we require for isolating the role of mass-load on jet dynamics. Therefore, a shear layer has been imposed in all the models by convolving the sharp jumps with the function ${\rm sech}\, r^m$ with $m = 8$ ($\Delta r_{\rm sl} \approx 0.2 R_{\rm j}$) in order to provide further stability to the solutions.
 
\subsubsection{The models}
%----------------------
Table~\ref{tab:jetpars} shows the relevant parameters used in a number of the different simulations discussed in this paper. In Appendix~A we show a table with the parameters of a number of intermediate models that were also analysed and contributed to the conclusions derived. 

The columns of Table~\ref{tab:jetpars} give the model name, jet rest-mass density, Lorentz factor, mean pitch angle, temperature, adiabatic exponent, $\Gamma_{\rm ad}$, specific enthalpy, axial magnetic field, maximum toroidal field, internal energy density, $e = \rho\, (h - c^2)/\Gamma_{\rm ad}$, magnetic energy density, $u_{\rm B}$, and the magnetization parameter, $\beta = u_{\rm B}/(2p)$, jet radius, power and ambient pressure at injection, respectively. Values of $h/c^2$ significantly larger than 1 imply models in which the internal energy dominates over the rest-mass energy density ({\it hot } models). According to the values in the table, models range between moderately hot (or {\it warm}; models J4\_\_) to hot. The values of $\beta$ indicate the ratio of the jet magnetic to internal energy densities. Values of $\beta$ are in the interval $\sim 0.1$ to $\sim 10$. 

Fixing the jet power at injection and ambient pressure, leaves three main parameters that may be changed to define the jet configuration, namely, the jet density, which we sweep in the range $10^{-32}$-$10^{-28}\,\rm{g\,cm^{-3}}$ for the main set of models, the Lorentz factor, ranging from 6 to 10, and the mean pitch angle of the magnetic field, which takes the values of $10^{\circ}$, $45^{\circ}$ and $75^{\circ}$. Imposing transversal equilibrium fixes the rest of the jet parameters, as explained in \citet{ma15,ma16}. 

We note here that there is little chance to find an ordered poloidal field at the simulated scales as implied by simple magnetic flux conservation \citep[see][]{bbr84}, and, from observational evidence, because the observed symmetry of the transverse brightness and polarization profiles is also inconsistent with global helical fields threading jets. Therefore, although the toroidal field component is probably ordered, the poloidal component \citep[comparable in magnitude at the base of the flaring region,][]{lb14} must have reversals: for instance, shear can contribute to the generation of a poloidal component across the jet's velocity gradients. Acknowledging these facts, we have simplified the set up of the magnetic field by including such an ordered structure. Nevertheless, the poloidal fields at injection given in Table~\ref{tab:jetpars} (1~mG) results in a poloidal flux of $3\times 10^{34}\,{\rm G cm^2}$, which compares to the estimate of $1 \times 10^{34}\,{\rm G cm^2}$ for a $10^9\,M_\odot$ by \citet{bbr84}, and is also within the order of magnitude given in \citet{za14}, which indicates that there is no major inconsistency between our adopted parameters and different theoretical approaches.

The names of the models provide information about their parameters. The second character, ranging from 4 to 9, gives the approximate (inverse) order of magnitude ratio between the jet density and the proton mass per cubic centimeter. Therefore, the larger the number, the smaller the density of the jet. The third character refers to the properties of the ambient medium (see Sec.~\ref{sss:amsw} and Table~\ref{tab:ambientpars}).

The last character of the model name provides information about both the jet Lorentz factor and the magnetic field structure: For each pitch angle, the letters stand for Lorentz factor 10, 7 and 6 (in alphabetical order), as indicated in Table~\ref{tab:jetpars}, with the Lorentz factor decreasing with the alphabetical order for a given pitch angle.

%with A, E, I, and X the letters for Lorentz factor 10, and and models without magnetic field, respectively.  

We have also run two sets of models, J91\_ and J92\_, with a lower power ($5\times10^{41}$~erg/s) and a larger radius (10 pc) to be compared with the purely hydrodynamical jet models A and D in \citet{pe14}, respectively. The parameters defining these models appear in the last part of Table~\ref{tab:jetpars}. They are injected with the same densities ($3\times10^{-33}\,{\rm g/cm^{3}}$ and $3\times10^{-35}\,{\rm g/cm^{3}}$, respectively) and Lorentz factor (3.2) as those in model A and D in \citet{pe14}. 

%%%%%%%%%%%%%%%%%%%%%%%%%%%%%%%%%%%%%%%%%%%%%%%%%%%%%%%%%%%%%%%%%%%%%%%%%%%%%%%%%%%%%%%%%%%%%%%%%
%
\begin{landscape}
\begin{table} 	
	\centering
	\caption{Simulation parameters.}
	\begin{tabular}{lcccccccccccccc} % four columns, alignment for each
		\hline
		Model & $\rho$ & $\gamma$ & $\phi_{\rm B}$ & $T$ & $\Gamma_{\rm ad}$ & $h$ & $B^{z}$  & $B_{\rm m}^\phi$ & $e$ & $u_{\rm B}$ & $\beta$ & $R_{\rm j}$ &$L_{\rm j}$ & $p_{\rm a, \, 0}$  \\
		 & [g/cm$^3$] &  & [$^\circ$]  &   [K] &  & [c$^2$]   &  [mG] & [mG] & [erg/cm$^3$] & [erg/cm$^3$] &  & [pc] & [erg/s] & [dyn/cm$^2$]  \\
	\hline
	        J4\_X & $10^{-28}$ &  10 & - & $5.8\times10^{8}$ & 1.62 & 1.24 & - & -& $ 1.44 \times 10^{-8}$ & - & - & 1 & $10^{43}$ & $10^{-7}$     \\
		J4\_Y &  $10^{-28}$ &  7 & - & $3.1\times10^{9}$ & 1.50 & 2.58 &  -  & -& $ 8.45 \times 10^{-8}$ & - & -& 1 & $10^{43}$ & $10^{-7}$    \\
		J4\_Z &  $10^{-28}$ &  6 & - & $4.6\times10^{9}$ & 1.46 & 3.47 & - & -& $ 1.56 \times 10^{-7}$ & - & -& 1 &$10^{43}$ & $10^{-7}$    \\
		J4\_A & $10^{-28}$ &  10 & 45 & $5.4\times10^{8}$ & 1.62 & 1.23 & $1.51$ & $1.92$& $1.26 \times 10^{-8}$ & $ 1.86\times 10^{-7}$ & 11.4 & 1 & $10^{43}$& $10^{-7}$    \\
		J4\_B &  $10^{-28}$ &  7 & 45 & $3.0\times10^{9}$ & 1.49 & 2.54 &  $1.17$  & $1.37$& $ 9.36 \times 10^{-8}$ &$ 1.11\times 10^{-7}$ & 1.19& 1 &$10^{43}$ & $10^{-7}$    \\
		J4\_C &  $10^{-28}$ &  6 & 45 & $4.6\times10^{9}$ & 1.45 & 3.49 & $0.87$  & $1.06$& $ 1.55 \times 10^{-7}$ & $ 6.21\times 10^{-8}$ & 0.44& 1 &$10^{43}$& $10^{-7}$    \\
		J4\_E & $10^{-28}$ &  10 & 10 & $5.8\times10^{8}$ & 1.62 & 1.26 & $1.51$  & $0.33$& $ 1.44 \times 10^{-8}$ &$ 1.84\times 10^{-7}$ & 10.4& 1 &$10^{43}$& $10^{-7}$    \\
		J4\_F &  $10^{-28}$ &  7 & 10 & $3.1\times10^{9}$ & 1.49 & 2.60 & $1.15$  & $0.25$& $ 9.45 \times 10^{-8}$ & $ 1.08\times 10^{-7}$ & 1.14& 1 &$10^{43}$& $10^{-7}$    \\
		J4\_G &  $10^{-28}$ &  6 & 10 & $4.6\times10^{9}$ & 1.46 & 3.52 & $0.87$  & $0.19$& $ 1.56 \times 10^{-7}$ & $ 5.94\times 10^{-8}$ & 0.42& 1 &$10^{43}$& $10^{-7}$    \\
		J4\_J &  $10^{-28}$ &  7 & 75 & $2.4\times10^{9}$ & 1.52 & 2.17 &  $1.28$  & $5.77$& $ 6.93 \times 10^{-8}$ & $ 1.66\times 10^{-7}$ & 2.30& 1 &$10^{43}$& $10^{-7}$    \\
		J4\_K &  $10^{-28}$ &  6 & 75 & $4.1\times10^{9}$ & 1.47 & 3.20 & $0.98$  & $4.40$& $ 1.35 \times 10^{-7}$ & $ 1.04\times 10^{-7}$ & 0.83& 1 &$10^{43}$& $10^{-7}$    \\

\hline
	        J8\_X  &  $10^{-32}$ & 10 & - & $1.8\times10^{13}$ & 1.33 & $1.21\times10^{4}$ & -  & -& $ 8.48\times 10^{-8}$ & - & -& 1 & $10^{43}$ & $10^{-7}$    \\
		J8\_Y  &  $10^{-32}$ & 7 & - & $3.8\times10^{13}$ & 1.33 & $2.56\times10^{4}$ & -  & -& $1.74\times 10^{-7}$  & - & -& 1 & $10^{43}$ & $10^{-7}$    \\
		J8\_Z  &  $10^{-32}$ & 6 & -  & $5.2\times10^{13}$ & 1.33 & $3.50\times10^{4}$ & - & -& $2.38\times 10^{-7}$  & - & -& 1 & $10^{43}$& $10^{-7}$    \\
	        J8\_A  &  $10^{-32}$ & 10 & 45 & $1.8\times10^{13}$ & 1.33 & $1.23\times10^{4}$ & $1.35$  & $1.65$ & $8.37\times 10^{-8}$ & $1.47\times 10^{-7}$ & 2.63& 1 & $10^{43}$& $10^{-7}$    \\
		J8\_B  &  $10^{-32}$ & 7 & 45 & $3.8\times10^{13}$ & 1.33 & $2.54\times10^{4}$ & $1.04$  & $1.28$ & $1.73\times 10^{-7}$  & $8.79\times 10^{-8}$ & 0.76& 1 & $10^{43}$& $10^{-7}$    \\
		J8\_C  &  $10^{-32}$ & 6 & 45 & $5.2\times10^{13}$ & 1.33 & $3.49\times10^{4}$ & $0.74$  & $0.91$ & $2.37\times 10^{-7}$  & $4.46\times 10^{-8}$ & 0.28& 1 & $10^{43}$& $10^{-7}$    \\
	        J8\_E  &  $10^{-32}$ & 10 & 10 & $1.8\times10^{13}$ & 1.33 & $1.26\times10^{4}$ & $1.33$  & $0.29$ & $8.48\times 10^{-8}$ & $1.45\times 10^{-7}$ & 2.56& 1 &$10^{43}$& $10^{-7}$    \\
		J8\_F  &  $10^{-32}$ & 7 & 10 & $3.8\times10^{13}$ & 1.33 & $2.58\times10^{4}$ & $1.03$  & $0.22$ & $1.74\times 10^{-7}$  & $8.53\times 10^{-8}$ & 0.74& 1 &$10^{43}$& $10^{-7}$    \\
		J8\_G  &  $10^{-32}$ & 6 & 10 & $5.2\times10^{13}$ & 1.33 & $3.52\times10^{4}$ & $0.73$  & $0.17$ & $2.38\times 10^{-7}$  & $4.30\times 10^{-8}$ & 0.27& 1 & $10^{43}$& $10^{-7}$    \\
	        J8\_I  &  $10^{-32}$ & 10 & 75 & $1.5\times10^{13}$ & 1.33 & $1.02\times10^{4}$ & $1.37$  & $6.46$ & $6.86\times 10^{-8}$ & $1.77\times 10^{-7}$ & 3.90& 1 & $10^{43}$& $10^{-7}$    \\
		J8\_J  &  $10^{-32}$ & 7 & 75 & $3.3\times10^{13}$ & 1.33 & $2.26\times10^{4}$ & $1.11$  & $5.08$ & $1.53\times 10^{-7}$  & $1.28\times 10^{-7}$ & 1.25& 1 &$10^{43}$& $10^{-7}$    \\
		J8\_K  &  $10^{-32}$ & 6  & 75 & $4.9\times10^{13}$ & 1.33 & $3.29\times10^{4}$ & $0.81$ & $3.71$ & $2.22\times 10^{-7}$  & $7.36\times 10^{-8}$ & 0.50& 1 & $10^{43}$& $10^{-7}$    \\

	\hline
        \hline
	       J91X & $3\times10^{-33}$ &  3.2 & -  & $3.2\times10^{11}$ & 1.33 & $2.13\times10^2$ & - & -& $ 4.28 \times 10^{-10}$ & - & -& 10 & $5 \times 10^{41}$ & $2.5\times10^{-10}$  \\
               J91D & $3\times10^{-33}$ &  3.2 & 45 & $3.2\times10^{11}$ & 1.34 & $2.05\times10^2$ & 0.015  & $0.018$& $4.12\times10^{-10}$ & $2.45\times10^{-10}$ & 0.89& 10 & $5 \times 10^{41}$ & $2.5\times10^{-10}$  \\
               J91H & $3\times10^{-33}$ &  3.2 & 10 & $3.2\times10^{11}$ & 1.34 & $2.13\times10^2$ & 0.015  & $3.16\times10^{-3}$ & $4.28\times10^{-10}$ & $2.14\times10^{-10}$ & 0.74& 10 & $5 \times 10^{41}$ & $2.5\times10^{-10}$  \\
%               J91L & $3\times10^{-33}$ &  3.2 & $1.34\times10^{11}$ & 1.34 & 91.4 & 0.019 & 75 & 0.089 & $1.81\times10^{-10}$ & $8.91\times10^{-10}$ & 7.26\\
\hline
	       J92X & $3\times10^{-35}$ &  3.2 & - & $3.2\times10^{13}$ & 1.33 & $2.14\times10^4$ & -  & -& $ 4.31 \times 10^{-10}$ & - & -& 10 & $5 \times 10^{41}$ & $2.5\times10^{-10}$  \\
               J92D & $3\times10^{-35}$ &  3.2 & 45 & $3.2\times10^{13}$ & 1.33 & $2.06\times10^4$ & $0.015$ & $0.018$& $4.15 \times 10^{-10}$ & $ 2.45\times 10^{-10}$ & 0.89& 10 & $5 \times 10^{41}$ & $2.5\times10^{-10}$  \\
               J92H & $3\times10^{-35}$ &  3.2 & 10 & $3.2\times10^{13}$ & 1.33 & $2.14\times10^4$ & $0.015$ & $3.16\times10^{-3}$& $4.30 \times 10^{-10}$ & $ 2.14\times 10^{-10}$ & 0.74& 10 & $5 \times 10^{41}$ & $2.5\times10^{-10}$  \\
 %              J92L & $3\times10^{-35}$ &  3.2 & $1.34\times10^{13}$ & 1.33 & $9.14\times10^3$ & $0.019$& 75 & $0.089$& $1.84 \times 10^{-10}$ & $ 8.91\times 10^{-10}$ & 7.26\\
	\hline
	\end{tabular}
		\label{tab:jetpars}
\end{table}
\end{landscape}
%
%%%%%%%%%%%%%%%%%%%%%%%%%%%%%%%%%%%%%%%%%%%%%%%%%%%%%%%%%%%%%%%%%%%%%%%%%%%%%%%%%%%%%%%%%%%%%%%%%%%%

%%%%%%%%%%%%%%%%%%%%%%%%%%%%%%%%%%%%%%%%%%%%%%%%%%%%%%%%%%%%%%%%%%%%%%%%%%%%%%%%%%%%%%%%%%%%%%%%%
%
\begin{table}
	\centering
	\caption{Kinetic ($F_k$), internal ($F_i$) and magnetic ($F_p$) energy fluxes as fractions of the total energy flux of the jet.}
	\begin{tabular}{lccc} % four columns, alignment for each
		\hline
		Model & $F_{\rm k}$ & $F_{\rm i}$ & $F_{\rm P}$ \\
\hline
	        J4\_X & 0.796 & 0.204 & - \\
		J4\_Y & 0.388 & 0.612 & - \\ 
		J4\_Z & 0.284 & 0.716 & - \\
		J4\_A & 0.796 & 0.188 & 0.016 \\
		J4\_B & 0.388 & 0.602 & $9.52 \times 10^{-3}$ \\
		J4\_C & 0.284 & 0.711 & $5.28 \times 10^{-3}$ \\
		J4\_E & 0.796 & 0.204 & $5.05 \times 10^{-4}$ \\
		J4\_F & 0.388 & 0.612 & $2.93 \times 10^{-4}$ \\
		J4\_G & 0.284 & 0.716 & $1.62 \times 10^{-4}$ \\
		J4\_J & 0.388 & 0.454 & 0.158 \\
		J4\_K & 0.284 & 0.624 & 0.092 \\
\hline
	        J8\_X & $7.96 \times 10^{-5}$ & 0.99992 & - \\
		J8\_Y & $3.88 \times 10^{-5}$ & 0.99996 & - \\
		J8\_Z & $2.84 \times 10^{-5}$ & 0.99997 & - \\
	        J8\_A & $7.96 \times 10^{-5}$ & 0.987 & 0.013 \\
		J8\_B & $3.88 \times 10^{-5}$ & 0.992 & $7.58 \times 10^{-3}$ \\
		J8\_C & $2.84 \times 10^{-5}$ & 0.996 & $3.81 \times 10^{-3}$ \\
	        J8\_E & $7.96 \times 10^{-5}$ & 0.9995 & $3.98 \times 10^{-4}$ \\
		J8\_F & $3.88 \times 10^{-5}$ & 0.9997 & $2.34 \times 10^{-4}$ \\
		J8\_G & $2.84 \times 10^{-5}$ & 0.9999 & $1.17 \times 10^{-4}$ \\
	        J8\_I & $7.96 \times 10^{-5}$ & 0.808 & 0.192 \\
		J8\_J & $3.88 \times 10^{-5}$ & 0.878 & 0.122 \\
		J8\_K & $2.84 \times 10^{-5}$ & 0.935 & 0.065 \\
\hline
\hline
	       J91X & $4.67\times10^{-3}$ & 0.995 & - \\
               J91D & $4.67\times10^{-3}$ & 0.957 & 0.038 \\         
               J91H & $4.67\times10^{-3}$ & 0.994 & 0.001 \\ 
 %             J91L & $3.54\times10^{-3}$ & 0.320 & 0.677 \\
\hline
	     J92X & $4.67\times10^{-5}$ & 0.999 & - \\
             J92D & $4.67\times10^{-5}$ & 0.962 & 0.038 \\ 
             J92H & $4.67\times10^{-5}$ & 0.999 & 0.001 \\ 		
%             J92L & $3.54\times10^{-5}$ & 0.323 & 0.677 \\
\hline

	\end{tabular}
			\label{tab:jetflux}
\end{table}
%
%%%%%%%%%%%%%%%%%%%%%%%%%%%%%%%%%%%%%%%%%%%%%%%%%%%%%%%%%%%%%%%%%%%%%%%%%%%%%%%%%%%%%%%%%%%%%%%%%%%%

Table~\ref{tab:jetflux} compiles the kinetic, internal and magnetic energy fluxes of all the models in terms of the total jet flux, $L_{\rm j}$. Note that except for the models with the higher rest mass densities and Lorentz factor (J4\_X, J4\_A, J4\_E) which are kinetically dominated, all the high-power models are dominated by the internal energy flux. Note also that the largest Poynting flux jets are those with the larger magnetic pitch angle, and among them, those with the larger Lorentz factor (J5\_I, J7\_I, J8\_I). Low power jet models (J9) are also dominated by the internal energy flux at injection. The magnetic field configuration chosen together with the condition of transversal equilibrium imposed to our models do not allow for Poynting-flux dominated jets within the domain of the space of parameters under study.

\subsubsection{Ambient medium and stellar winds}
%------------------------------------------
\label{sss:amsw}

%%%%%%%%%%%%%%%%%%%%%%%%%%%%%%%%%%%%%%%%%%%%%%%%%%%%%%%%%%%%%%%%%%%%%%%%%%%%%%%%%%%%%%%%%%%%%%%%%%%%
%
\begin{table}	
	\centering
	\caption{Different pressure core radii, mass injection rates and corresponding mean stellar mass-loss rates used in the simulations.}
	\begin{tabular}{lccc} 
		\hline
		Model  & $r_{\rm c}$ & $Q_0$ & $\dot{M}$  \\
                            & [kpc]    & [g y$^{-1}$pc$^{-3}$] &[M$_\odot$ yr$^{-1}$]  \\
                \hline
                J\_N\_     & 0.5    & $4.71 \times 10^{21}$ &  $10^{-13}$ \\
                J\_O\_    & 0.5     & $4.71 \times 10^{23}$ &  $10^{-11}$\\
                J\_P\_       & 0.2     & $4.71 \times 10^{23}$ & $10^{-11}$\\
                J\_Q\_    & 0.5     & $4.71 \times 10^{24}$ &$10^{-10}$\\
                J\_R\_       & 0.5    & $4.71 \times 10^{25}$ & $10^{-9}$ \\
                J\_S\_      & 0.2     & $4.71 \times 10^{25}$ & $10^{-9}$  \\
		\hline
	\end{tabular}
		\label{tab:ambientpars}
\end{table}
%
%%%%%%%%%%%%%%%%%%%%%%%%%%%%%%%%%%%%%%%%%%%%%%%%%%%%%%%%%%%%%%%%%%%%%%%%%%%%%%%%%%%%%%%%%%%%%%%%%%%%%

A pressure profile for the ambient medium is set up to reproduce typical galactic atmospheres and trigger the sideways expansion of the jet 
\begin{equation}
\label{eq:pprof}
p_{\rm a}(z) = p_{\rm a,\, 0} \left(1+ \left(\frac{z}{r_{\rm c}}\right)^2 \right)^{\alpha},
\end{equation}
where $r_{\rm c}$ is the inner core radius. This expression corresponds to that of an isothermal atmosphere with the same density profile as used in  \cite{pe14}. The mass loading of the jets from stellar winds is accounted for by adding a source term in the (relativistic) density conservation equation in the same way as described in \cite{pe14}. The mass load rate per unit volume follows
\begin{equation}
\label{eq:sprof}
Q(z) = Q_0 \left(1 + \left(\frac{z}{r_{c,s}}\right)^2 \right)^{\alpha_{\rm s}},
\end{equation}
where $r_{\rm c,\, s}$ is the stellar core radius. The composition of the injected gas is proton-electron, i.e., ionized hydrogen.

The last column of Table~\ref{tab:jetpars} shows the value of the ambient pressure at injection, which fixes the jet equilibrium configuration. All models were run with the same ambient pressure at injection $p_{\rm a, \, 0} = 10^{-7}$ dyn/cm$^2$, and the same stellar core radius, $r_{\rm c, \, s}= 500$~pc and $\alpha_{\rm s}= -1.095$ (see Eq.~\ref{eq:sprof}), but for the case of J9 models, where we have fixed the same ambient parameters as in \cite{pe14} for all models.\footnote{These parameters are $p_{\rm a, \, 0} = 2.5\times10^{-10}$~dyn/cm$^2$ and $r_{\rm c}=1.2$~kpc, $\alpha=-1.095$ in Eq.~\ref{eq:pprof}, and $Q_0 =4.71 \times 10^{21}~{\rm g \, yr^{-1}\,pc^{-3}}$, $r_{\rm c,\, s}=260$~pc, and $\alpha_s = -0.23$ in Eq.~\ref{eq:sprof}.} 

We change two parameters to set up the different simulations, namely, the galactic core radius in pressure, which takes the values $r_{\rm c} = 500$  and $ 200$ pc , and the fiducial mass-injection rate per unit volume, which is taken as $Q_0 = 4.71 \times 10^{23}~{\rm g \, yr^{-1}\,pc^{-3}}$, a value two orders of magnitude larger than the fiducial value in \cite{BL96},\footnote{The aim of this work is to study the scenarios in which jets could be decelerated by stellar winds alone. In this respect, the results obtained in \cite{pe14} pointed towards powerful mass-losses being required to decelerate jets with powers $\sim 10^{43}\,{\rm erg\,s^{-1}}$.} $Q_0 = 4.71 \times 10^{24}~{\rm g \, yr^{-1}\,pc^{-3}}$, and $Q_0 = 4.71 \times 10^{25}~{\rm g \, yr^{-1}\,pc^{-3}}$. The latter represents an extreme value of mass-load, corresponding to a mean stellar mass-loss of $\sim 10^{-9} \,{\rm M_\odot yr^{-1}}$ for a stellar density of $10\,{\rm pc^{-3}}$ \citep[see][]{Vieyro:2017}. These mass-loads correspond to mass injection rates which are much larger than expected for normal ellipticals. We know, from \cite{pe14}, that jet powers of $10^{43}\,{\rm erg\,s^{-1}}$ are probably too high for mass input from stars to have much effect for normal elliptical hosts. Although the aim of this work is to understand the basic physics of the process and to compare the models presented here with those in \cite{pe14}, we have added models with $Q_0 = 4.71 \times 10^{21}~{\rm g \, yr^{-1}\,pc^{-3}}$ \citep[as in][]{BL96} to discuss the effect of the expected mass load in typical elliptical galaxies on classical FRI jets. Finally, we also ran a set of models for which $Q_0 = 0~{\rm g \, yr^{-1}\,pc^{-3}}$ that were run as controls (the results of these models are not shown because they follow the expected behaviour and no significant features). 

The combinations, along with the indication of the corresponding character at the third position of the model name, are displayed in Table~\ref{tab:ambientpars}, and give rise to five possible ambient configurations through which all models in Table~\ref{tab:jetpars} have been run.

     In cylindrical coordinates, the numerical grid is $(r, z) \in [0, 20 R_{\rm j}] \times [0, 500 R_{\rm j}]$, with $R_{\rm j} = 1\,{\rm pc}$ in the high-power jets, and $R_{\rm j} = 10\,{\rm pc}$ in the low-power ones. The number of cells of the simulation box is $1600 \times 10000$, which implies a transversal resolution of $80\,{\rm cells}/R_{\rm j}$, and an axial resolution of $20\,{\rm cells}/R_{\rm j}$. Axisymmetry forces reflection boundary conditions on the jet axis. The outer boundary corresponds to the jet surface (defined as the contour line with tracer $f = 0.5$). At the boundary, the ambient pressure is set according to Eq.~\ref{eq:pprof} at each step. Moreover, the ambient density and the ambient flow speed are reset, respectively, to the values of the rest mass density in the lab frame and the jet flow velocity in the jet cell contiguous to the boundary, in order to avoid shocks being generated at the boundary between the jet and the ambient medium \citep[see][]{ko15}.
     
     The physical size of the numerical grids has been chosen to obtain a compromise between numerical resolution, a number of simulations that covers a large region of the parameter space, and a distance within which most of the stellar mass load is expected. On the one hand, beyond 500\,pc, the expected decrease in ambient density causes the jet to expand, thereby including more stars in its volume and encouraging deceleration. On the other hand, the expected drop in stellar number density would make deceleration by stellar winds more difficult, and, in addition, a hot jet is still expected to accelerate as it expands or by buoyancy. In conclusion, we expect the stronger deceleration by stellar winds to take place where the stellar density is larger, i.e., at the galactic core. 
     
\subsection{Summary on model names}

  A given model presented in this paper is thus defined in Tables~\ref{tab:jetpars} and \ref{tab:ambientpars}, with the second character of the model's name related to the jet rest-mass density at injection, and the fourth character to both the Lorentz factor and pitch angle. These parameters can be found in Table~\ref{tab:jetpars}. In contrast, the third character of a model's name is related to the ambient properties, which can be found in Table~\ref{tab:ambientpars}. The full information of the jet can thus be obtained by checking both tables. 

As an example, the jet properties of models J4OA and J8RA can be found in Table~\ref{tab:jetpars} under the J4\_A and J8\_A names (the empty space is left for the ambient character), which are shared by all models that have those same characters but are run within different ambient properties; and these can be found in Table~\ref{tab:ambientpars} with the corresponding letter (J\_O\_ and J\_R\_ in each case).       
                       
\section{Results} \label{sec:res}
%##########################

\begin{figure*} 
	\includegraphics[trim=0.cm 0.7cm 0.cm 0.cm,width=\textwidth]{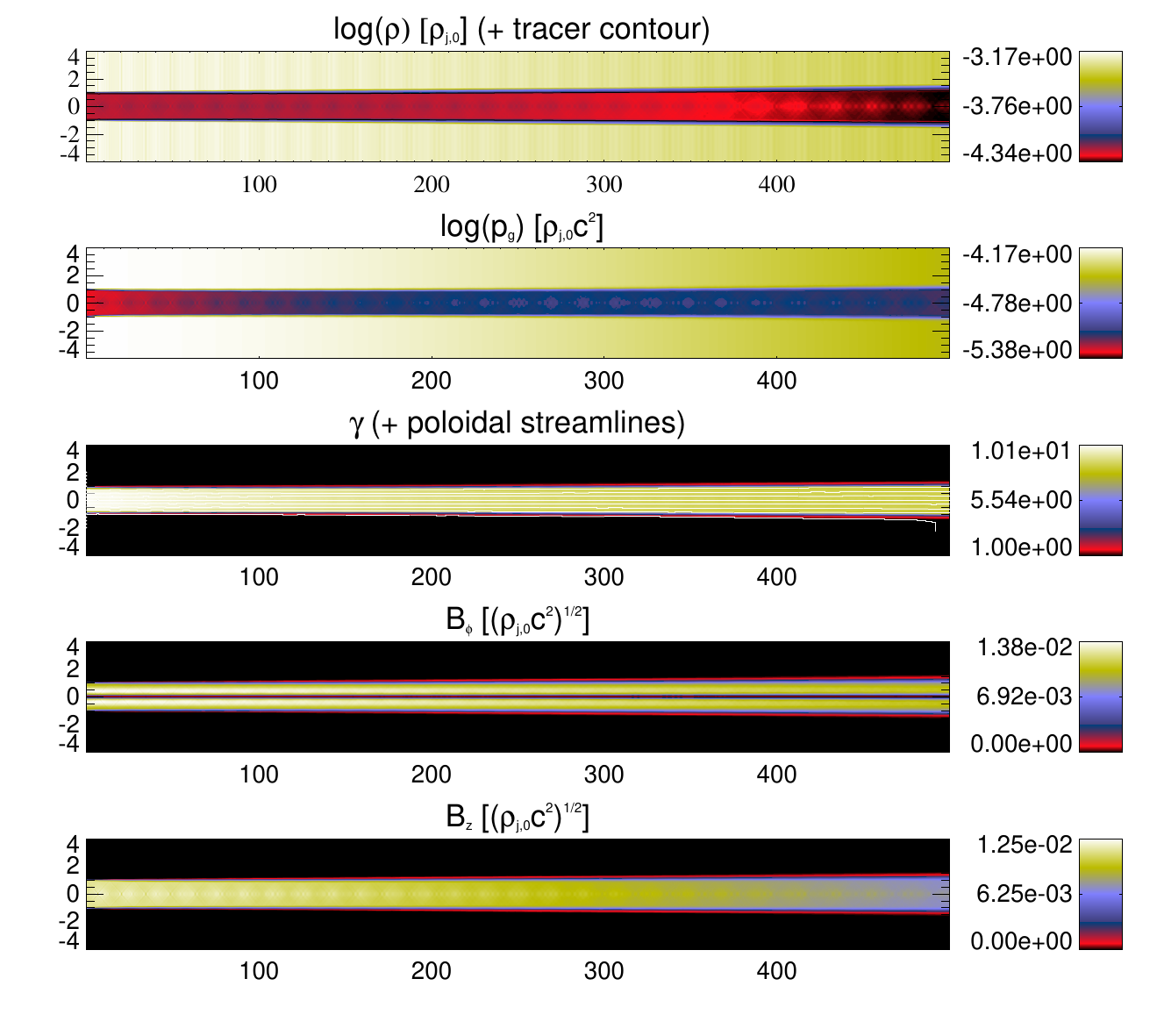} \,
\caption{2D maps of the logarithm of rest-mass density and gas pressure, Lorentz factor (including streamlines), toroidal and axial field components for model J4OA. The units for all plots are code units (in which unit density is $\rho_{\rm j,0} = 1.$, unit distance is the jet radius $R_{j,0} = 1.$ and $c=1.$).}
\label{fig:40A}
\end{figure*}

As an example, Fig.~\ref{fig:40A} show, from top to bottom, the 2D distributions of logarithm of the rest-mass density, gas pressure, Lorentz factor (plus poloidal streamlines), toroidal and axial field components, which define the steady-state solutions for model J4OA (see Appendix~\ref{app:maps} for the case of model J4OC). These images are representative of the results obtained for each model, which will be discussed later in the paper. Note that the ambient medium is set \emph{a priori} in these simulations and it is only used as a boundary condition for the computed transversal equilibrium structure.

We interpret our results in the context of the interplay between jet expansion (as they propagate through a pressure decreasing atmosphere), the subsequent cooling and acceleration (according to Bernoulli's law), the magnetic tension (which tends to avoid the jet expansion), and the loading of mass (which increases as the jet expansion and tends to decelerate the jet). Since all our models have dynamically unimportant Poynting fluxes we do not expect magnetic acceleration to play an important role.

%%%%%%%%%%%%%%%%%%%%%%%%%%%%%%%%%%%%%%%%%%%%%%%%%%%%%%%%%%%%%%%%%%%%%%%%%%%%%%%%%%%%%%%%%%%%%%%%%%%%
%
%
%%%%%%%%%%%%%%%%%%%%%%%%%%%%%%%%%%%%%%%%%%%%%%%%%%%%%%%%%%%%%%%%%%%%%%%%%%%%%%%%%%%%%%%%%%%%%%%%%%%%

\subsection{Low mass-load in low power jets. Comparison with RHD simulations}
%===========================================================
\label{ss:lmllpj}

In \citet{pe14} the authors found that a difference of one order of magnitude in the value of the fiducial mass-load rate, $Q_0$, could completely change the dynamics of the jets. In that paper, models C and D had the same initial conditions but different value of $Q_0$, which was one order of magnitude larger in model C. The hydrodynamical simulations showed that model D propagated with fairly constant radius and a minor deceleration, despite the entrainment, which only forced the cooling of the flow with distance (see Fig.~11 in that paper). In this model, the velocity in the jet dropped from $0.95c$ ($\gamma \approx 3.2$) to $0.83c$ ($\gamma \approx 1.8$) within the inner tens of parsecs after injection and oscillated around this value due to successive expansions and recollimations. In contrast, model C expanded and was efficiently decelerated and ultimately disrupted by the growth of a pinching Kelvin-Helmholtz instability.

%%%%%%%%%%%%%%%%%%%%%%%%%%%%%%%%%%%%%%%%%%%%%%%%%%%%%%%%%%%%%%%%%%%%%%%%%%%%%%%%%%%%%%%%%%%%%%%%%%%%
%
\begin{figure*} 
	\includegraphics[trim=2.cm 12cm 2.cm 4.cm,width=0.45\textwidth]{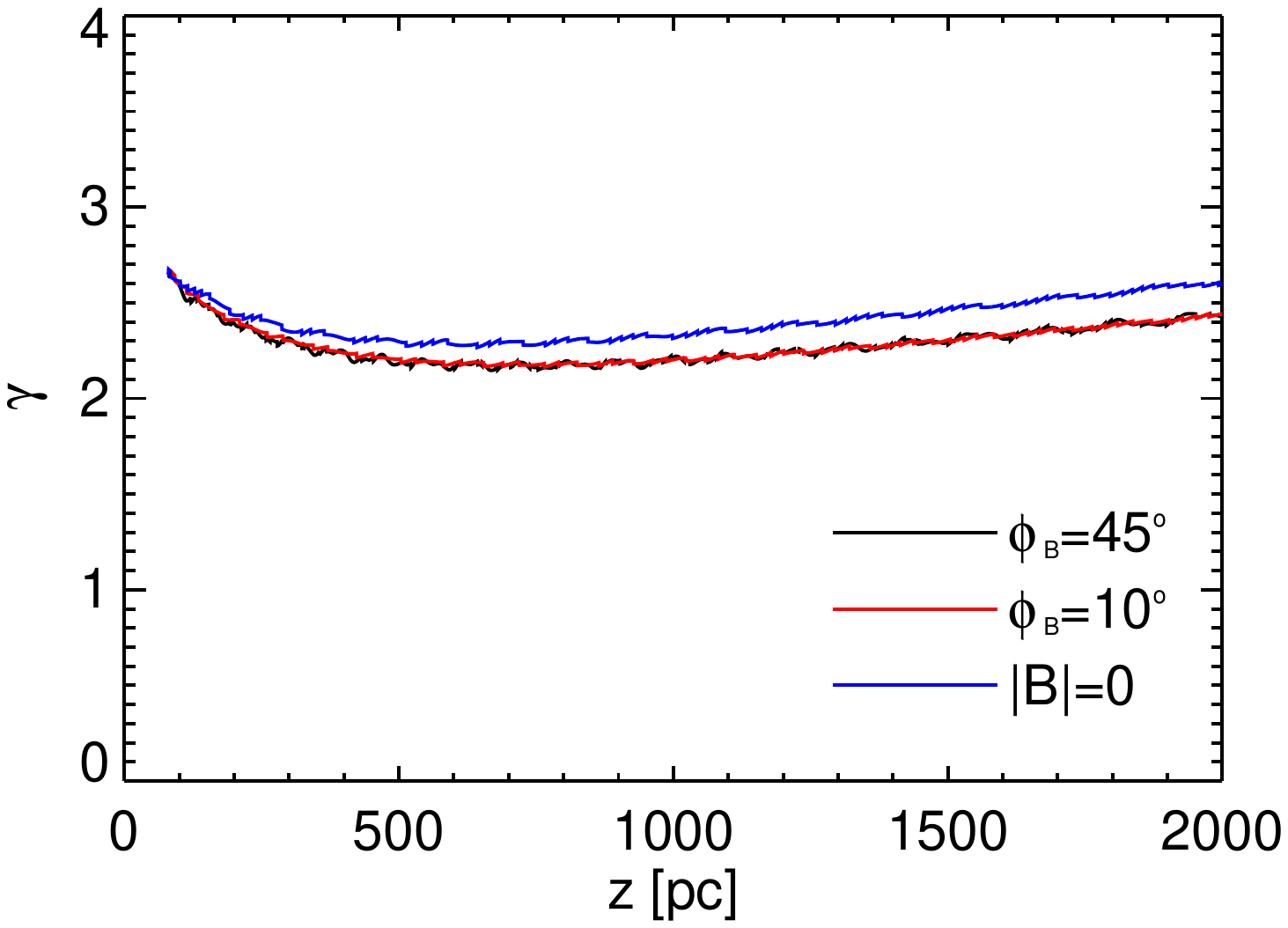} \,
	\includegraphics[trim=2.cm 12cm 2.cm 4.cm,width=0.45\textwidth]{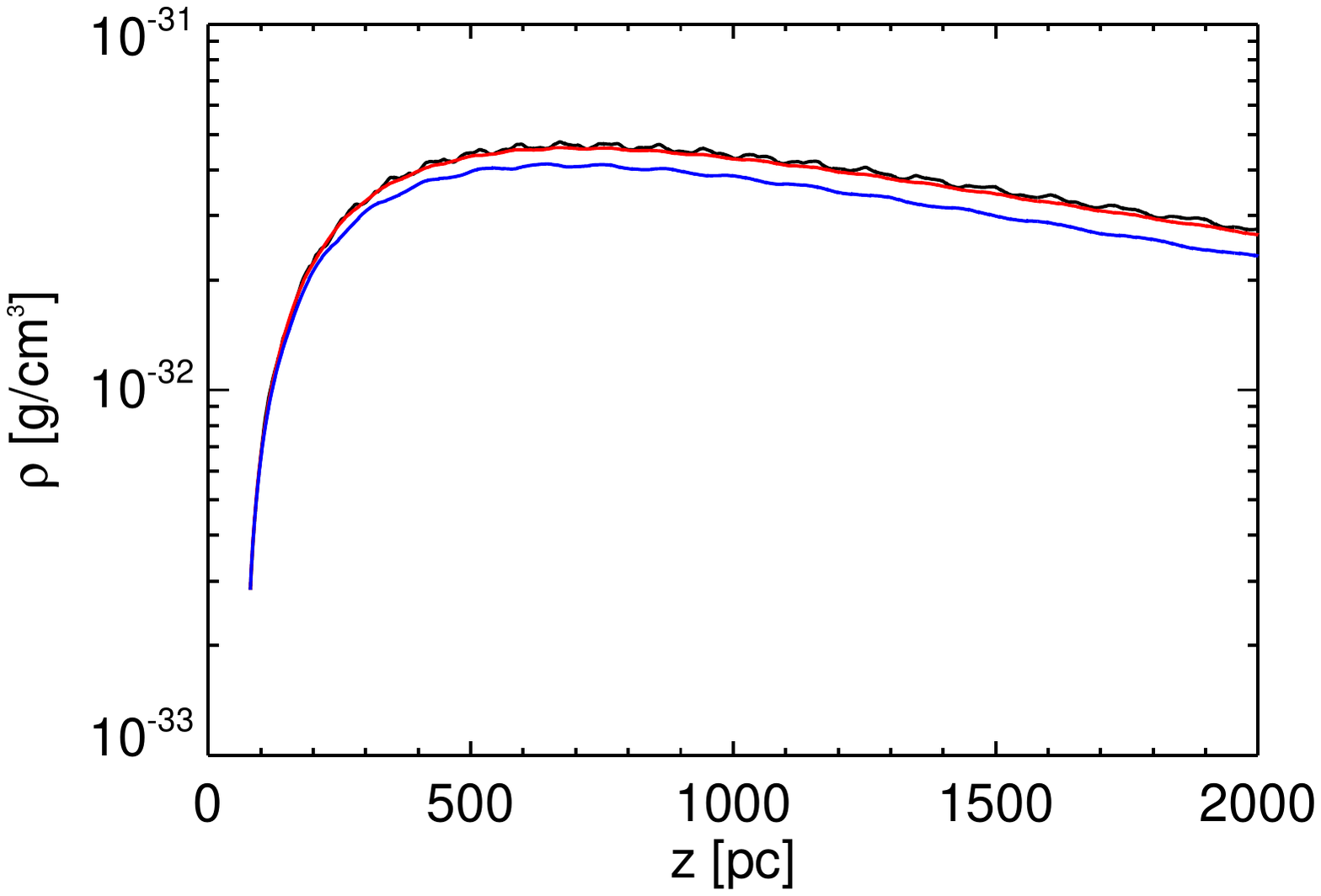} \\
	\includegraphics[trim=2.cm 12cm 2.cm 4.cm,width=0.45\textwidth]{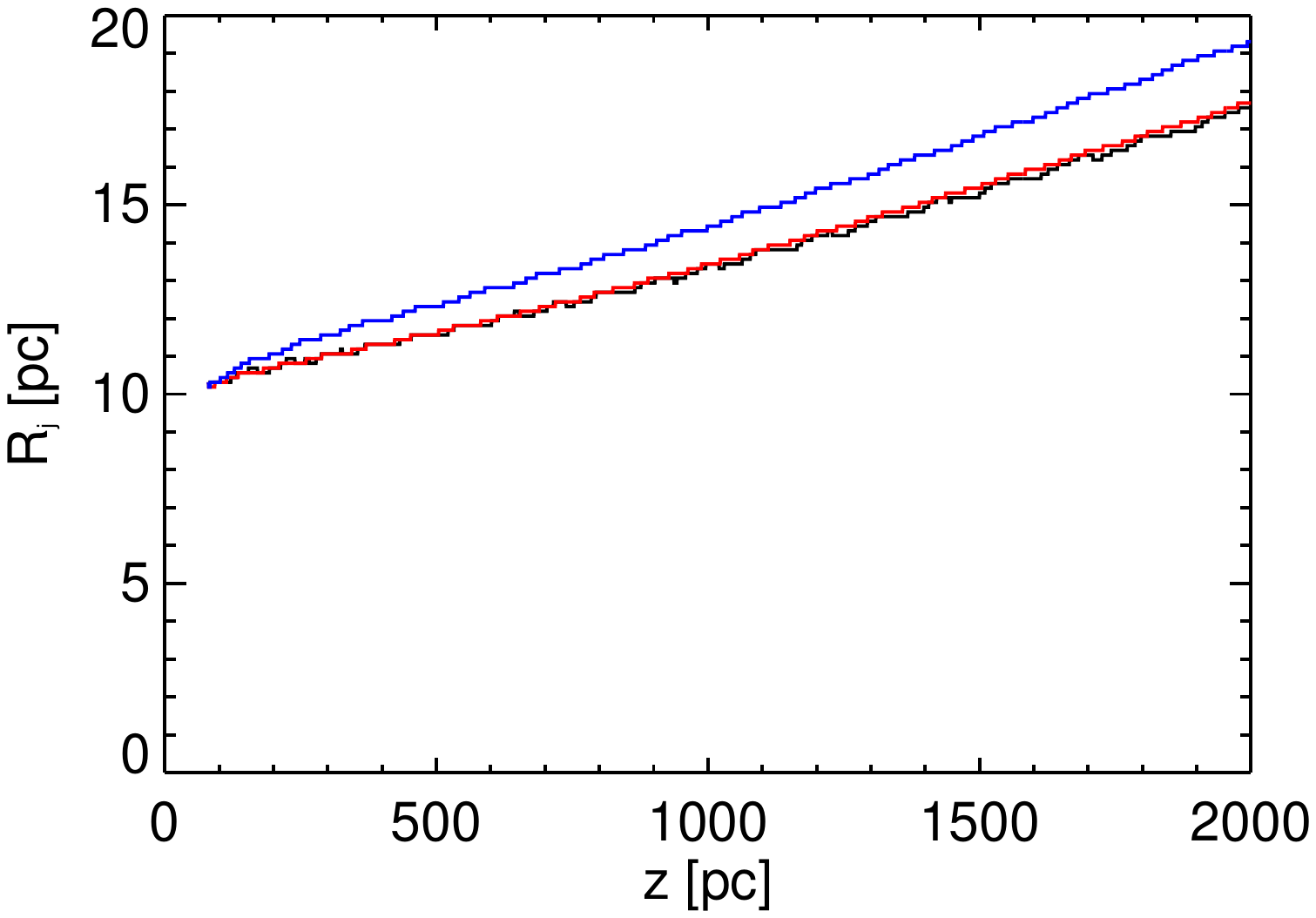} \,
   \includegraphics[trim=2.cm 12cm 2.cm 4.cm,width=0.45\textwidth]{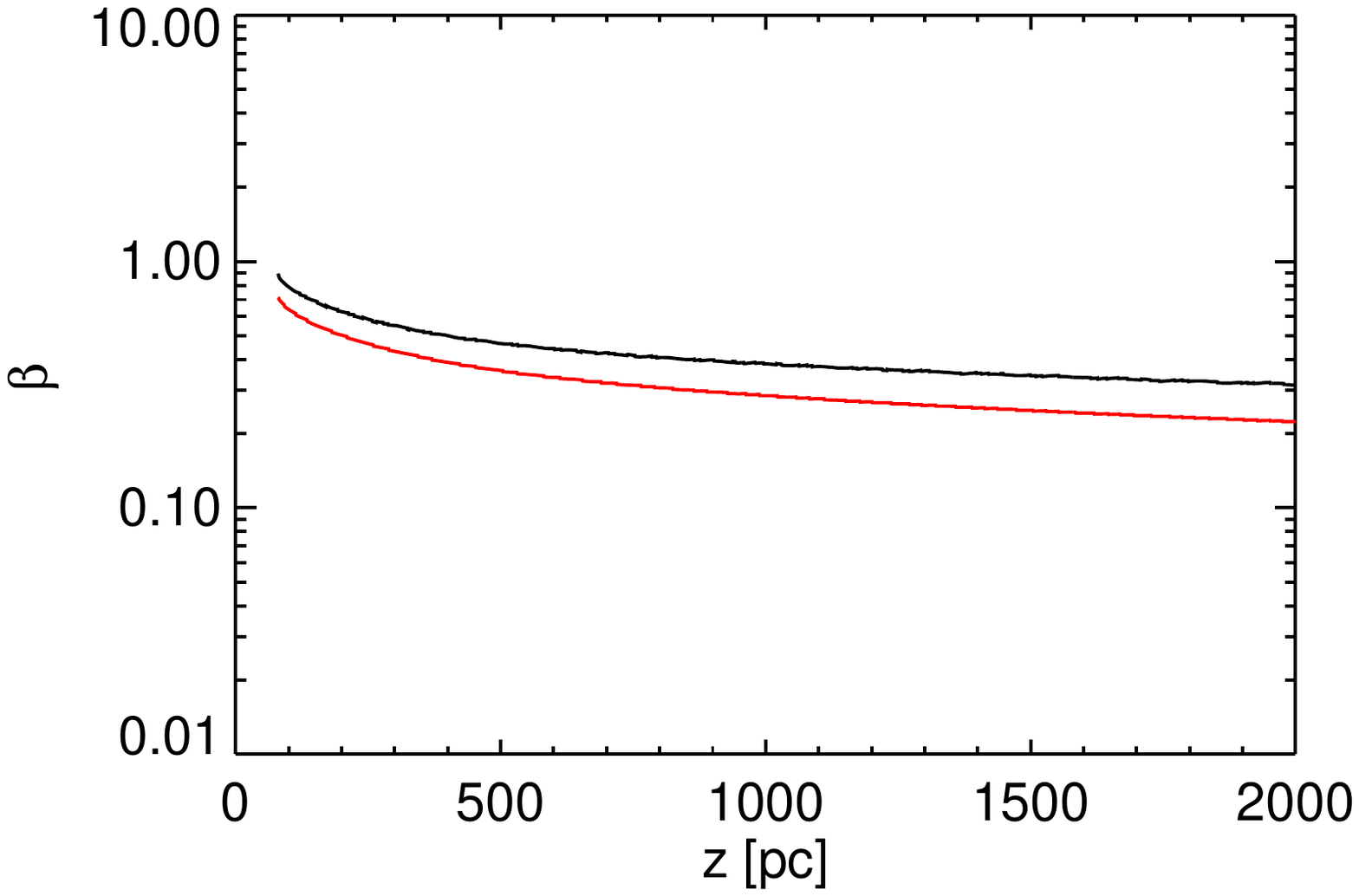} 
\caption{Mean jet Lorentz factor (top left), rest-mass density (top right), radius (bottom left) and $\beta$ parameter (bottom right) versus distance for models J91. In each plot, the black, red, and blue lines represent the jets with mean pitch angles $\phi_B=45^\circ$ (model J91D), $10^\circ$ (J91H), and hydro jets (J91X), respectively.} 
\label{fig:J91}
\end{figure*}
%
%%%%%%%%%%%%%%%%%%%%%%%%%%%%%%%%%%%%%%%%%%%%%%%%%%%%%%%%%%%%%%%%%%%%%%%%%%%%%%%%%%%%%%%%%%%%%%%%%%%%

Models J91 and J92 have been run with specific ambient parameters (see Table~\ref{tab:ambientpars}) to be compared with the purely hydrodynamical models presented in \citet{pe14}. {In particular, they were run with the same mass load as model D in that paper}.  

Figure~\ref{fig:J91}  shows the evolution with distance of the jet radius, and the averages of the jet rest-mass density, flow Lorentz factor and magnetization, for magnetised jets with pitch angles $\phi_B=45^\circ$ (model J91D) and $10^\circ$ (J91H), and for pure hydrodynamical jets (J91X). Models with $\phi_B=75^\circ$ could not be run because no equilibrium solution could be found for them. The plotted values are obtained as volume weighted means across the jet, bounded by the surface with jet mass fraction $f=0.5$, which is the reason why, for instance, the mean values of the jet Lorentz factors at injection are below their values on axis. In these very-low density models, moreover, the jet density is dominated by the mass-loading from the first numerical cell. This explains the discrepancy between the first rest-mass density values in the plots and the injection values given in Table~\ref{tab:jetpars}. The plots show a mild deceleration of the flow within the inner 500~pc, and an acceleration farther downstream. The jets expand with relatively small opening angles ($\approx 0.5^\circ$). The jet density grows very fast up to 500~pc, stalls at about this distance and slowly falls thereafter. Finally, the magnetisation falls in all the models due to expansion, which forces a drop in magnetic pressure, in parallel to a slowed down decrease in thermal pressure (due to the dissipation of kinetic energy and the injection of particles) induced by the mass load. The corresponding plots for models J92 are not shown because of their similarity to those for J91. 

A comparison between the quasi-one-dimensional (q1D, from now on) equilibrium models J91 and J92 and the aforementioned model D shows several differences: 

\begin{enumerate}[i)]
\item in model D the jet radius oscillates around the initial value, whereas it grows monotonically in models J91/J92; 
\item the jet velocity remains fairly constant in model D (in models J91/92, the jet flow accelerates once the mass-load drops beyond $r_s$); 
\item pressure in model D oscillates around some equilibrium value developing a pinched structure along the jet; 
\item the mean jet density in model D grows to values that are one order of magnitude larger than in q1D models. 
\end{enumerate}

These changes in behaviour result from differences in the way in which jets and ambient medium interact in the two types
of simulation. While hydrodynamical model D evolved through a dynamic ambient medium formed by the shocked jet gas (the jet's cocoon), and this medium was over-dense and over-pressured with respect to the jet \citep[see, e.g., Figs.~8 and 11 in][]{pe14}, q1D models J91/J92 evolve in pressure equilibrium with their environment (see Section~\ref{sec:su}). As a consequence, the ambient pressure gradient directly affects the jets in the q1D simulations, whereas the dynamical jets evolve completely embedded within their cocoons, with a fairly homogeneous pressure. This enables us to explain differences i)-iii) above. The higher densities attributed to model D with respect to q1D models J91/J92 (point iv) above) are an artefact of the definition of the jet surface 
%$v^z > 0.4c%
which in the case of model D comprised a wider section of the jet/ambient shear layer than in models J91/J92. It is interesting to remark that the steady-state simulations can be more representative of late time evolution, after the cocoon has evolved.
  
  Figure~\ref{fig:J91} shows that the purely hydrodynamical q1D jets, J91X (blue lines), undergo a slightly stronger acceleration than their magnetised counterparts as they propagate. At the same time, the rest-mass density is also slightly smaller in those models than in the magnetised ones. The relative extra acceleration and drop in rest-mass density of the hydrodynamical model are related to the extra expansion as compared to the magnetised counterparts (see Table\ref{tab:jetpars}).
    
  We do not find strong differences among the magnetised models, which is common to all the simulations presented in this work. The reason is the relatively small dynamical significance of the magnetic fields with respect to either the initial fluxes of kinetic and internal energy (see Tables~\ref{tab:jetpars} and \ref{tab:jetflux}) or because of the rapid increase of kinetic energy due to mass-load in hot, dilute jets. The latter option implies that the jet becomes rapidly dominated by the kinetic energy flux, making the magnetic energy flux negligible. 
  
%As we will discuss in the next sections, this is mainly due to the relation between the different energy fluxes. 

\subsection{Moderate mass-load}
%=======================

In this section we focus on the role of the magnetic field and its pitch angle on jets undergoing a moderate mass-load (characterized by a mass-load rate at injection of $Q_0 = 4.71 \times 10^{23}$ g yr$^{-1}$ pc$^{-3}$) by a stellar population with a mean mass-loss of $10^{-11}\,{\rm M_\odot\,yr^{-1}}$. With this aim, we present the results obtained for models J4O and J8O (four orders of magnitude lighter), which are representative limits of the parameter space we have studied. We consider magnetized models with different magnetic pitch angles ($\phi_B=10^\circ,\,45^\circ$, and $75^\circ$) and equivalent models without magnetic fields, as well as models with different Lorentz factors. Finally we compare our results with those obtained for models J4P and J8P, in which a change in the ambient pressure profile is introduced, to study the role of jet expansion induced by an earlier drop in ambient pressure. In order to facilitate the comparison among models, we use the same colour lines for the models with the same pitch angle in each of the plots.

In the absence of mass-loading, hydrodynamical jets embedded in an atmosphere with decreasing pressure expand, rarify, accelerate and cool down following Bernoulli's principle. When a jet is mass-loaded, its pressure tends to decrease more slowly as a result of the dissipation of kinetic energy and the injection of particles, which together force a faster expansion. 
Mass-load both increases the jet density and decelerates the jet flow, while temperature increases due to kinetic energy dissipation and decreases due to the injection of cold particles, so the resulting trends will depend on the initial distribution of parameters, with cold, fast jets possibly being heated and hot jets being cooled. The picture is completed when we consider the role of the magnetic field. In magnetised jets, the magnetic tension in the radial direction (caused by the toroidal component of the magnetic field) prevents rapid expansion. On the one hand, this favours a slower conversion of the internal energy into kinetic energy but, on the other hand, it delays the deceleration by mass-loading. Finally, significant magnetic acceleration would also possible in strongly magnetised jets under the appropriate conditions. 

%%%%%%%%%%%%%%%%%%%%%%%%%%%%%%%%%%%%%%%%%%%%%%%%%%%%%%%%%%%%%%%%%%%%%%%%%%%%%%%%%%%%%%%%%%%%%%%%%%%%%
%
\begin{figure*}
\includegraphics[trim=0.cm 0.cm 0.cm 1.5cm,width=\textwidth]{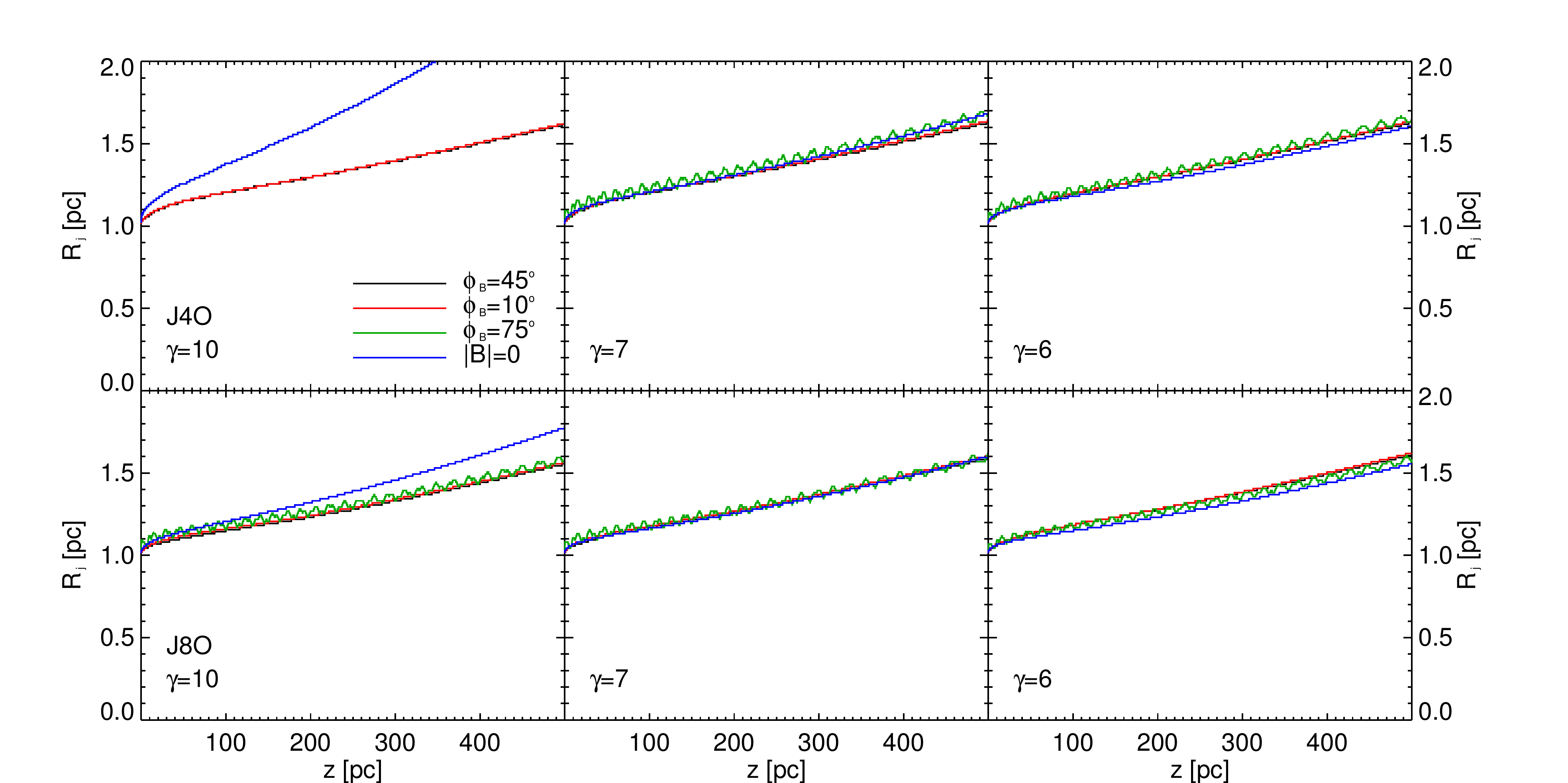} 
\caption{Jet radius versus distance for models J4O\_ (top panels) and J8O\_ (bottom panels). Each panel contains all the models with the same initial Lorentz factor. The left panels show the models with $\gamma = 10$ (models A, I, E, X), the central panels show those with $\gamma = 7$ (models B, J, F, Y), and the right panels show those with $\gamma = 6$ (models C, K, G, Z). In each plot, the black, red, green and blue lines represent the jets with mean pitch angles $\phi_B=45^\circ$ (models A, B, C), $10^\circ$ (E, F, G), $75^\circ$  (I, J, K), and purely hydro models, respectively.}
\label{fig:radi}
\end{figure*}
%
%%%%%%%%%%%%%%%%%%%%%%%%%%%%%%%%%%%%%%%%%%%%%%%%%%%%%%%%%%%%%%%%%%%%%%%%%%%%%%%%%%%%%%%%%%%%%%%%%%%%%

%%%%%%%%%%%%%%%%%%%%%%%%%%%%%%%%%%%%%%%%%%%%%%%%%%%%%%%%%%%%%%%%%%%%%%%%%%%%%%%%%%%%%%%%%%%%%%%%%%%%
%
\begin{figure*} 
\includegraphics[trim=0.cm 0.cm 0.cm 1.5cm,width=\textwidth]{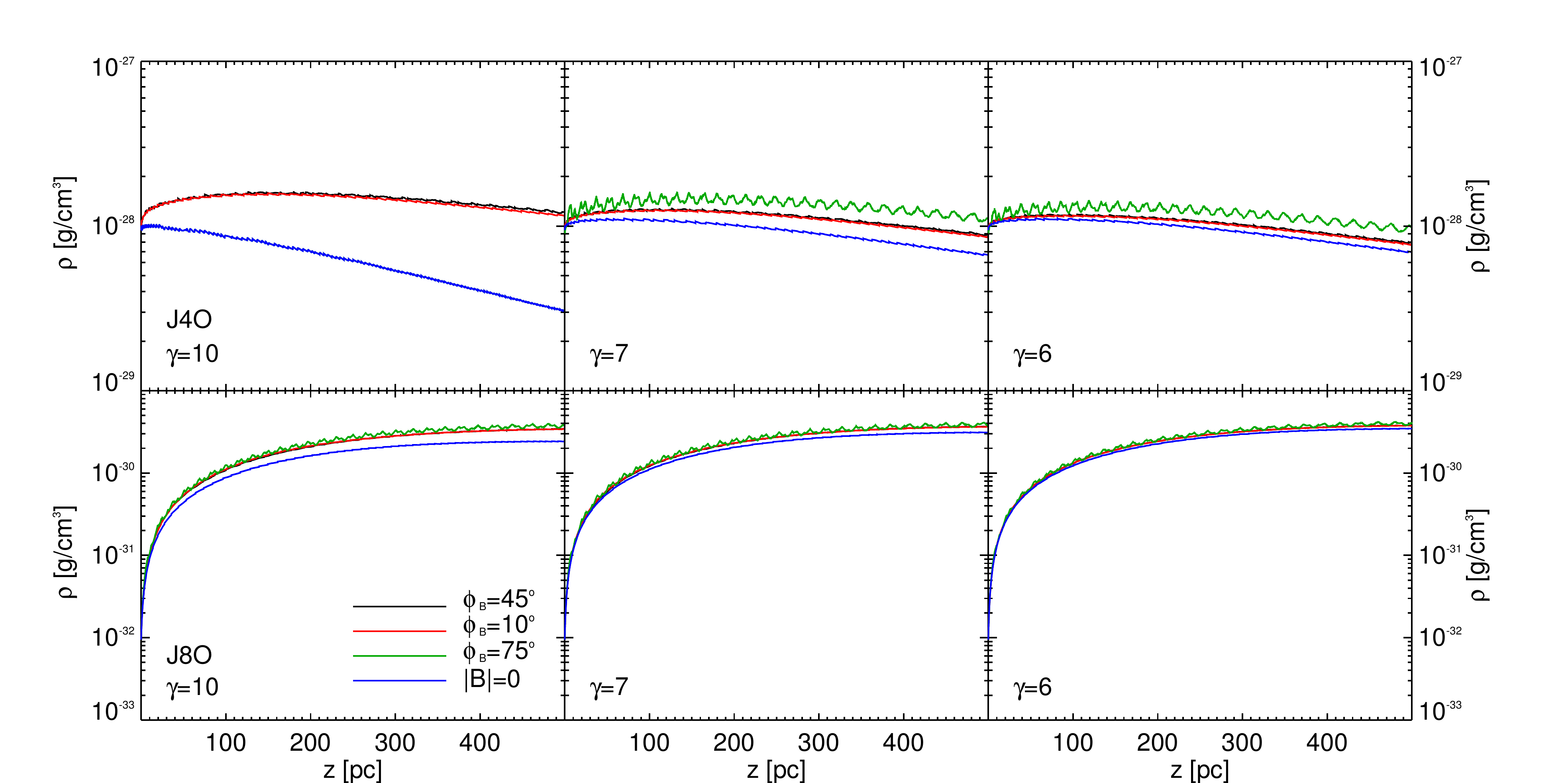} 
\caption{Mean jet rest-mass density versus distance for models J4O\_ (top panels) and J8O\_ (bottom panels). See the caption of Fig.~\ref{fig:radi}.}
\label{fig:densty}
\end{figure*}
%
%%%%%%%%%%%%%%%%%%%%%%%%%%%%%%%%%%%%%%%%%%%%%%%%%%%%%%%%%%%%%%%%%%%%%%%%%%%%%%%%%%%%%%%%%%%%%%%%%%%%

\begin{figure*} 
\includegraphics[trim=0.cm 0.cm 0.cm 1.5cm,width=\textwidth]{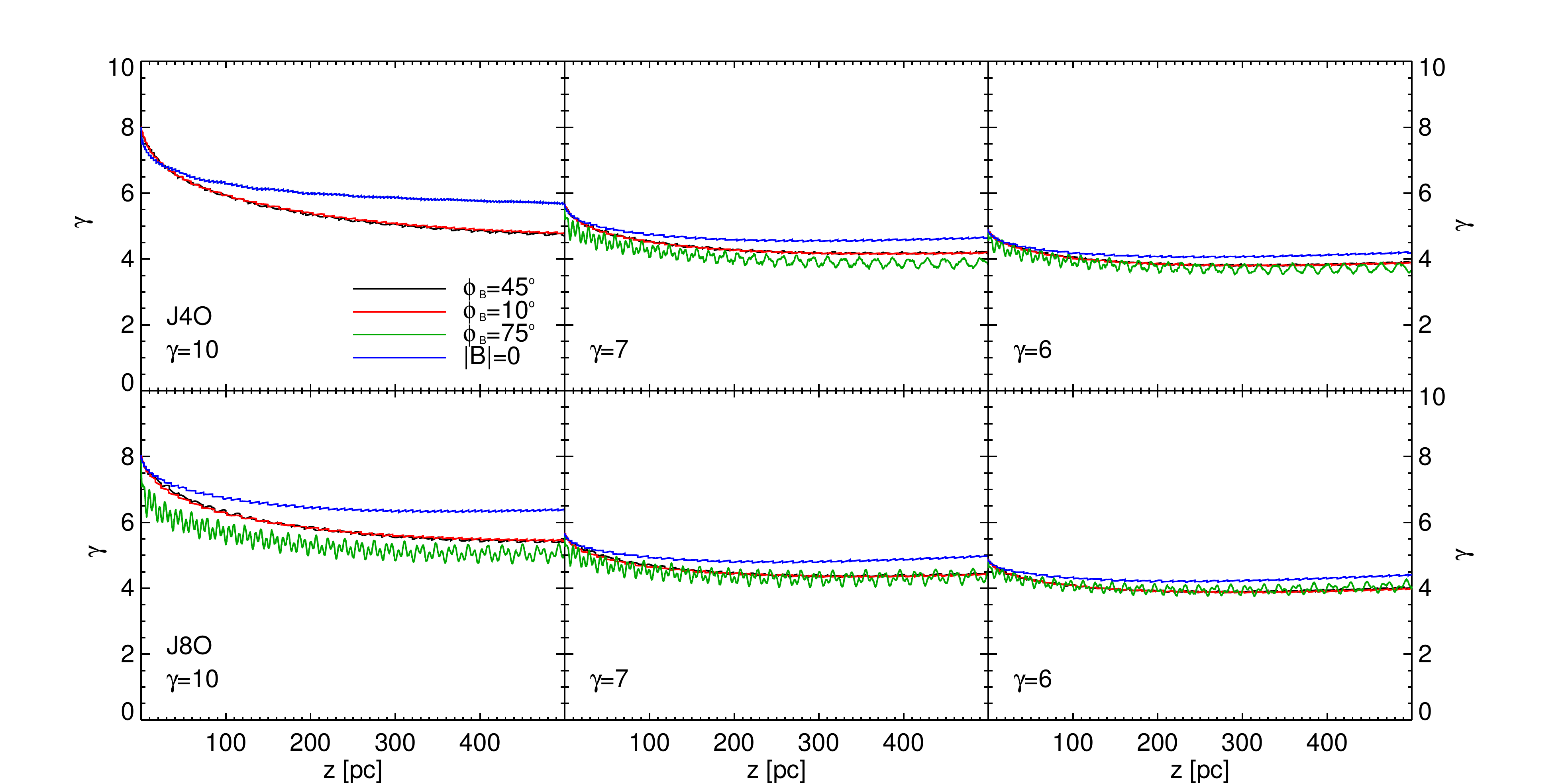} 
\caption{Mean jet Lorentz factor versus distance for models J4O\_ (top panels) and J8O\_ (bottom panels). See the caption of Fig.~\ref{fig:radi}.}
\label{fig:lor}
\end{figure*}

\begin{figure*} 
\includegraphics[trim=0.cm 0.cm 0.cm 1.5cm,width=\textwidth]{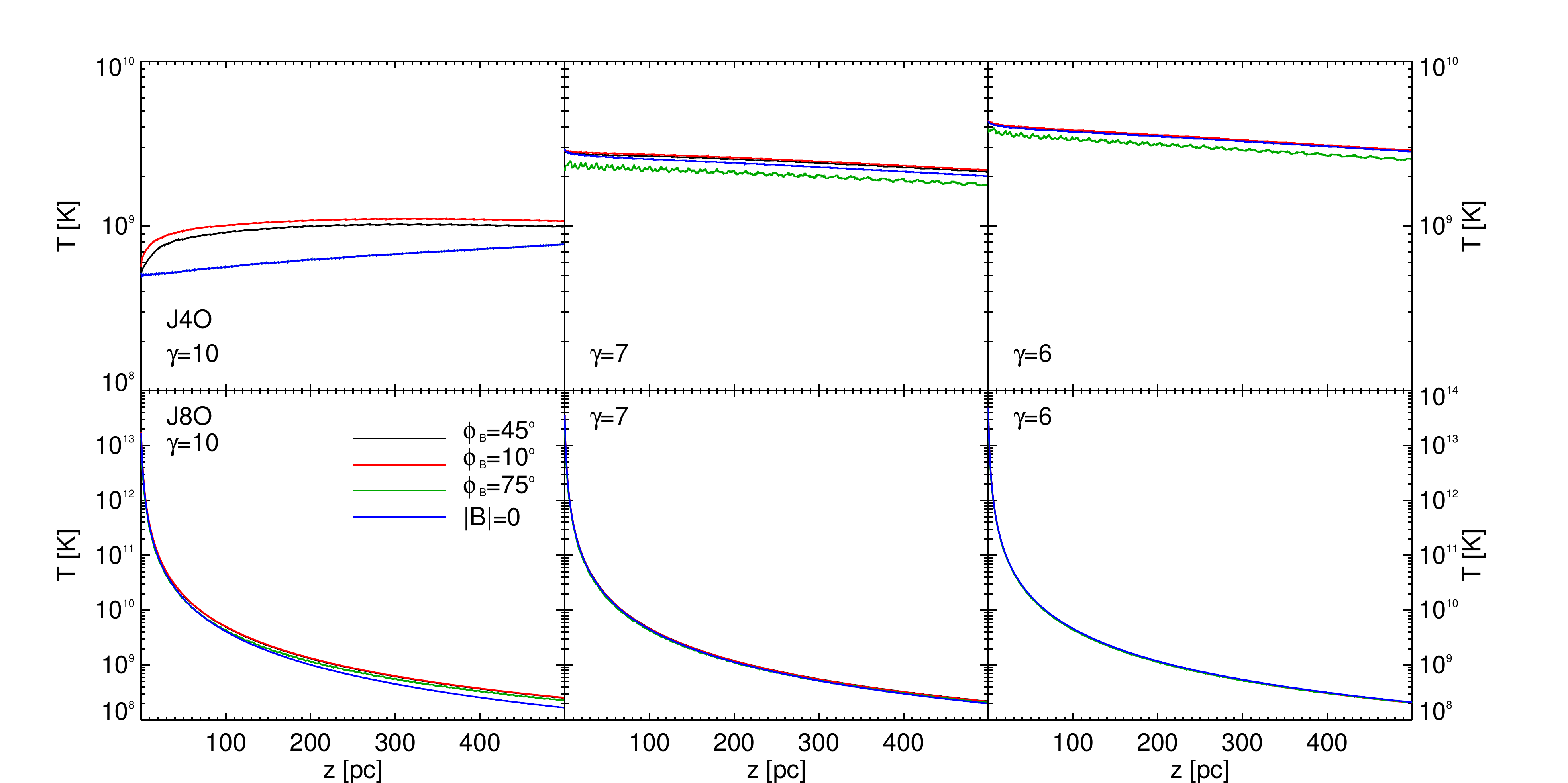} 
\caption{Jet temperature versus distance for models J4O\_ (top panels) and J8O\_ (bottom panels). See the caption of Fig.~\ref{fig:radi}.}
\label{fig:tem}
\end{figure*}

Figures~\ref{fig:radi}-\ref{fig:tem} show, respectively, the evolution of the jet radius, rest-mass density, flow Lorentz factor and temperature along the simulated grid (0.5 kpc long) for models J4O and J8O. The radii of the jets as a function of distance are shown in Fig.~\ref{fig:radi}.  All the jets expand as they propagate. The role of the magnetic tension in counterbalancing jet expansion is visible for the magnetised, $\gamma=10$ jets (models J4OA, J4OE, J8OA, J8OE, J8OI) when compared with the corresponding, purely hydrodynamical ones (J4OX, J8OX, left column of Fig.~\ref{fig:radi}). The enhanced expansion seen in model J4OX is the result of the efficient conversion of kinetic energy into internal energy in this non-magnetised, kinetically dominated jet model. 

The evolution of jet radius, rest-mass density and flow Lorentz factor along the jets are related by mass conservation (including the loaded mass). Rest-mass densities for models J4O and J8O are shown in Fig.~\ref{fig:densty}. Magnetised models J4O undergo an increase in the rest-mass density along the first tens of parsecs from the mass loading followed by a gentle decrease governed by the jet expansion (and the decrease of the mass-load rate with distance). In the case of model J4OX, the jet expansion is so fast that the density decreases right from the jet base. Finally, the density in models J8O is so small that it is dominated by the injected mass and keeps growing as the jet propagates.

Independently of the jet's initial density, $\gamma=10$ magnetised models tend to decelerate slightly, whereas $\gamma=6,\,7$ models tend to a constant Lorentz factor (see Fig.~\ref{fig:lor}). For J4O models, this can be understood if the mass load has a marginal or negligible effect on jet deceleration in these cases. For instance, in the case of the $\gamma = 10$ models J4OA and J4OE, the kinetic flux dominates over both the internal energy and the magnetic energy fluxes at injection (Table~\ref{tab:jetflux}) and mass entrainment seems to dissipate part of the kinetic energy flux, causing the mean jet temperature to increase slightly, as shown in Fig.~\ref{fig:tem} (top left panel). In contrast, in the case of J8OA and J8OE models, the slight decrease in the jet Lorentz factor can be assigned to the strong mass-load, compensated by expansion and cooling. In models with smaller Lorentz factors at injection (models J4OB, C, F, G, J, K; top-central and top-right panels in Figs.~\ref{fig:radi}-\ref{fig:tem}) the dominating energy flux is that of the internal energy (Table~\ref{tab:jetflux}), which allows the jet to keep an almost constant Lorentz factor even if considerably mass loaded, at the expense of a drop in the internal energy flux. 

Although the injected mass-load does not decelerate the jets, the increase of density in J8O models (see  Fig.~\ref{fig:densty}, bottom panels) implies a strong change in jet composition, leading to a baryonically dominated flow, with leptonic mass fraction changing from $X_e=1$ at injection to $X_e=10^{-3}-10^{-2}$. On the contrary, J4O models barely change their composition, with $X_e\simeq 0.95$ at $z=500\,{\rm pc}$. 

The figures also reveal that for the largest pitch angle ($\phi_B=75^\circ$, green lines) the toroidal field triggers small-scale oscillations (pinching) at the jet boundary, which can be observed in all the plots. The reason for this is that the equilibrium solutions found for this pitch angle are already neighbouring non-equilibrium configurations in the parameter space \citep[see][]{ma15}.

%%%%%%%%%%%%%%%%%%%%%%%%%%%%%%%%%%%%%%%%%%%%%%%%%%%%%%%%%%%%%%%%%%%%%%%%%%%%%%%%%%%%%%%%%%%%%%%%%%%%
%
%
%%%%%%%%%%%%%%%%%%%%%%%%%%%%%%%%%%%%%%%%%%%%%%%%%%%%%%%%%%%%%%%%%%%%%%%%%%%%%%%%%%%%%%%%%%%%%%%%%%%%

%%%%%%%%%%%%%%%%%%%%%%%%%%%%%%%%%%%%%%%%%%%%%%%%%%%%%%%%%%%%%%%%%%%%%%%%%%%%%%%%%%%%%%%%%%%%%%%%%%%%%
%
%[trim=4cm 7.5cm 10cm 10.5cm,width=0.2\textwidth]
%
%%%%%%%%%%%%%%%%%%%%%%%%%%%%%%%%%%%%%%%%%%%%%%%%%%%%%%%%%%%%%%%%%%%%%%%%%%%%%%%%%%%%%%%%%%%%%%%%%%%%%

\begin{figure} 
\includegraphics[trim=7.5cm 0.5cm 7.5cm 1.5cm,width=\columnwidth]{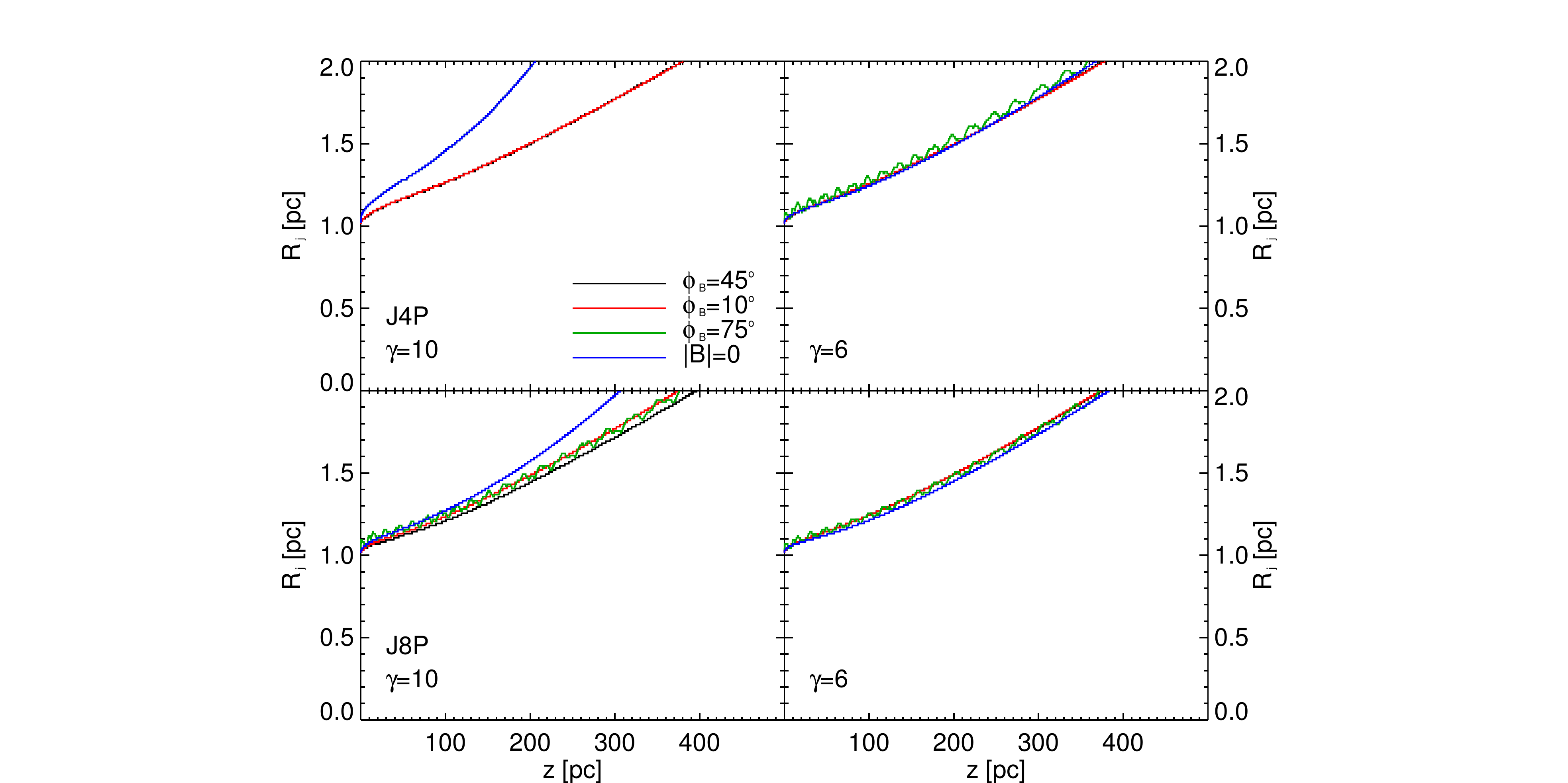} 
\caption{Jet radius for models J4P\_ (top panels) and J8P\_ (bottom panels) with Lorentz factor 10 (left column), and 6 (right column). In each plot, the black, red, green and blue lines represent the jets with mean pitch angles $\phi_B=45^\circ$, $10^\circ$, $75^\circ$, and purely hydro models, respectively.}
\label{fig:rad2}
\end{figure}

\begin{figure} 
\includegraphics[trim=7.cm 0.5cm 7.cm 1.5cm,width=\columnwidth]{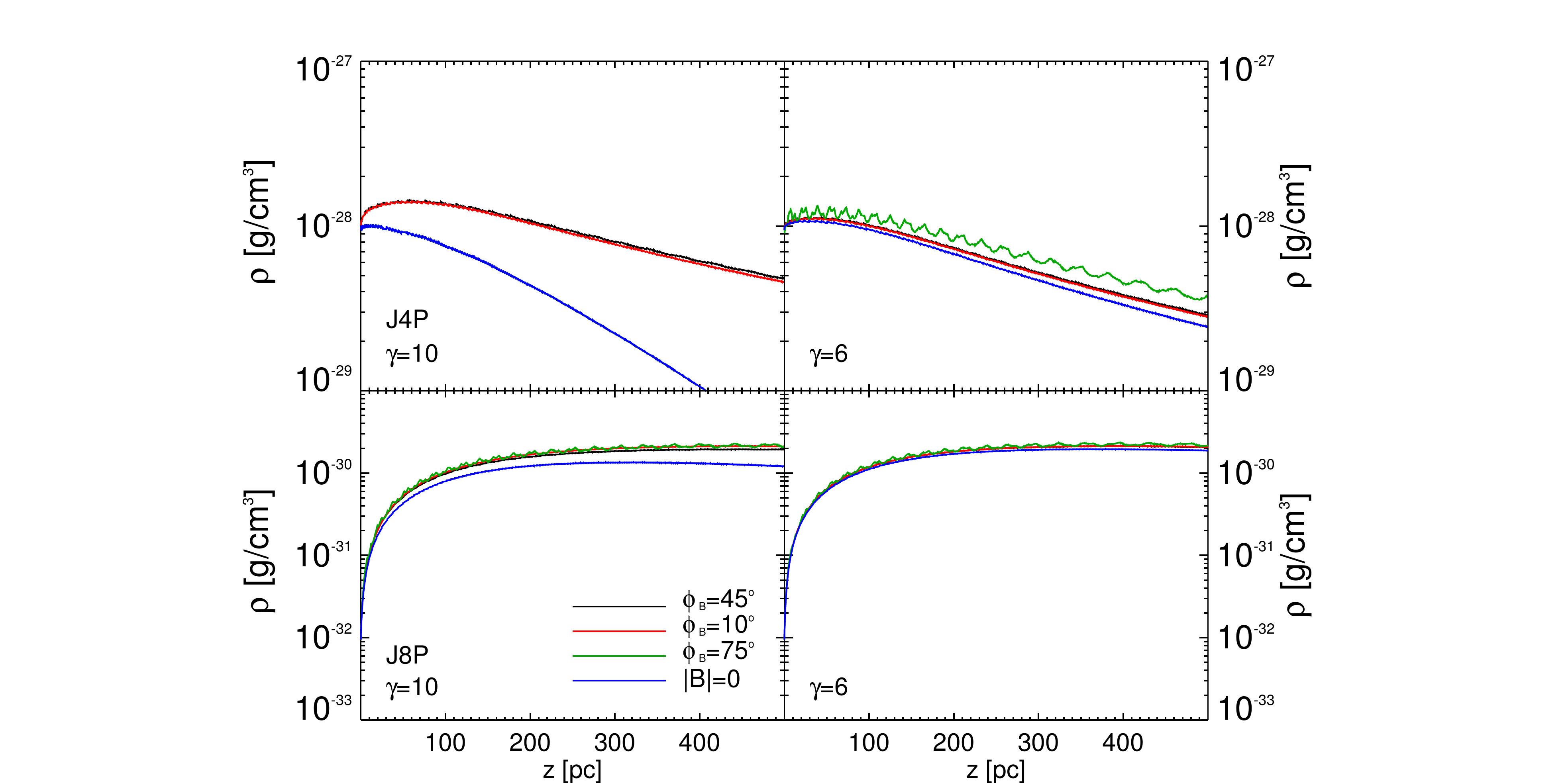} 
\caption{Rest-mass density versus distance for models J4P\_ (top panels) and J8P\_ (bottom panels) with Lorentz factor 10 (left column), and 6 (right column). In each plot, the black, red, green and blue lines represent the jets with mean pitch angles $\phi_B=45^\circ$, $10^\circ$, $75^\circ$, and purely hydro models, respectively.}
\label{fig:rho2}
\end{figure}

\begin{figure} 
\includegraphics[trim=7.5cm 0.5cm 7.5cm 1.5cm,width=\columnwidth]{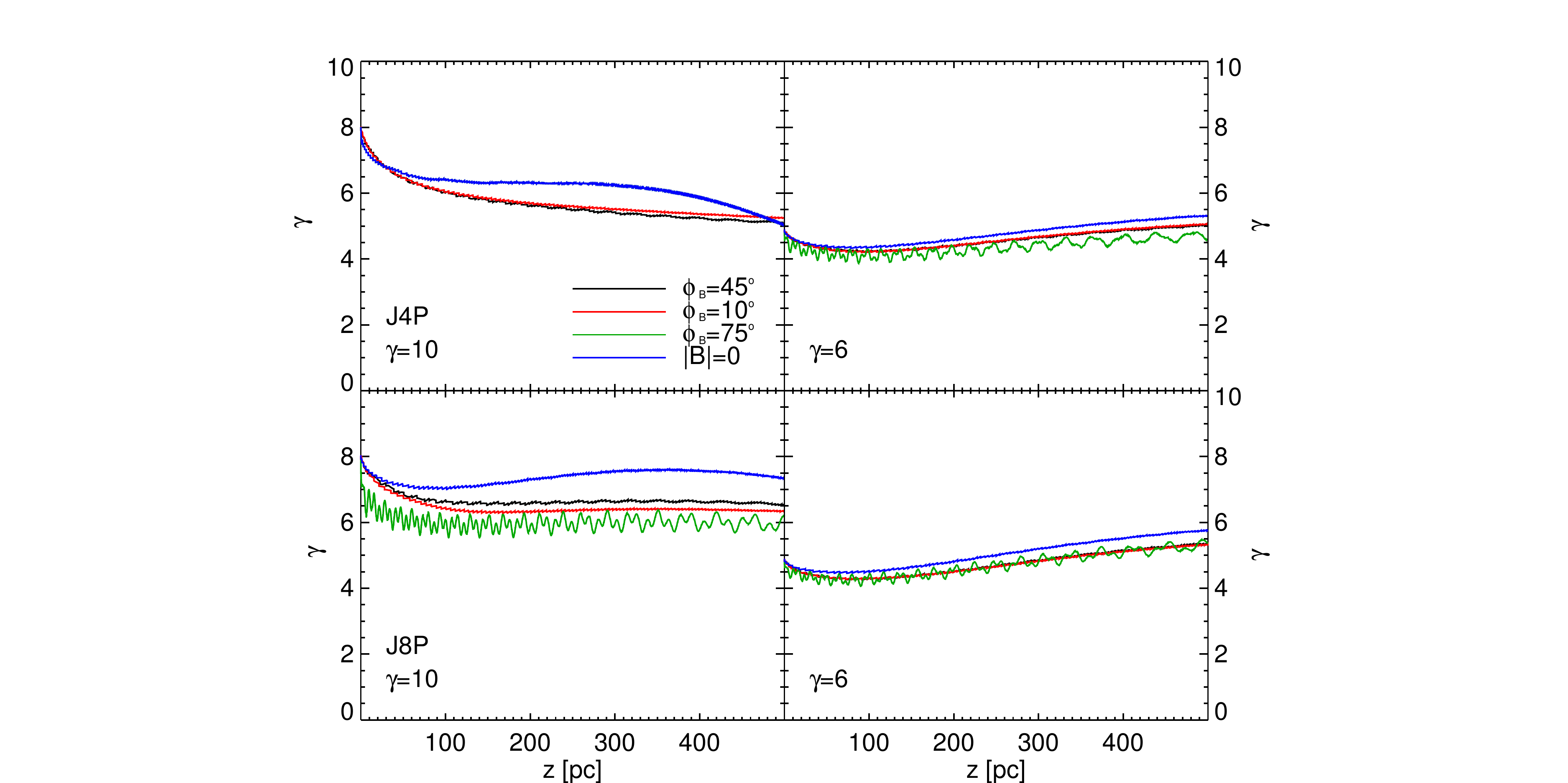} 
\caption{Mean jet Lorentz factor versus distance for models J4P\_ (top panels) and J8P\_ (bottom panels) with Lorentz factor 10 (left column), and 6 (right column). In each plot, the black, red, green and blue lines represent the jets with mean pitch angles $\phi_B=45^\circ$, $10^\circ$, $75^\circ$, and purely hydro models, respectively.}
\label{fig:lor2}
\end{figure}

A change in the pressure profile was introduced in order to check the role of an extended mass load region ($r_{\rm c, \, s}=500\,{\rm pc}$) within a falling pressure profile from $r_{\rm c} = 200\,{\rm pc}$ (J\_P\_ models). The results of models J4P (top panels) and J8P (bottom panels) are displayed in Figs.~\ref{fig:rad2}-\ref{fig:lor2}. The drop in pressure favours jet expansion, which is counter-balanced by magnetic tension in magnetised jets, but not in purely hydro jets, which therefore show the largest differences in evolution (see Fig.~\ref{fig:rad2}). Now, in the case of denser jets, the wider local opening angle of models J4P with respect to models J4O favours the flow acceleration in the former. However this acceleration is compensated by the increased mass-load and the enhanced dissipation of kinetic energy in $\gamma = 10$ models (see top panels of Fig.~\ref{fig:lor2}). In the case of dilute, hot models J8P, the increased opening angle favours the conversion of internal into kinetic energy via the Bernoulli effect and thus, jet acceleration, despite the increase in jet density (see bottom panels of Fig.~\ref{fig:lor2}). 

%%%%%%%%%%%%%%%%%%%%%%%%%%%%%%%%%%%%%%%%%%%%%%%%%%%%%%%%%%%%%%%%%%%%%%%%%%%%%%%%%%%%%%%%%%%%%%%%%%%%%
% 
%
%%%%%%%%%%%%%%%%%%%%%%%%%%%%%%%%%%%%%%%%%%%%%%%%%%%%%%%%%%%%%%%%%%%%%%%%%%%%%%%%%%%%%%%%%%%%%%%%%%%%%

%%%%%%%%%%%%%%%%%%%%%%%%%%%%%%%%%%%%%%%%%%%%%%%%%%%%%%%%%%%%%%%%%%%%%%%%%%%%%%%%%%%%%%%%%%%%%%%%%%%%%
%
%
%%%%%%%%%%%%%%%%%%%%%%%%%%%%%%%%%%%%%%%%%%%%%%%%%%%%%%%%%%%%%%%%%%%%%%%%%%%%%%%%%%%%%%%%%%%%%%%%%%%%%

%%%%%%%%%%%%%%%%%%%%%%%%%%%%%%%%%%%%%%%%%%%%%%%%%%%%%%%%%%%%%%%%%%%%%%%%%%%%%%%%%%%%%%%%%%%%%%%%%%%%%
%
%
%%%%%%%%%%%%%%%%%%%%%%%%%%%%%%%%%%%%%%%%%%%%%%%%%%%%%%%%%%%%%%%%%%%%%%%%%%%%%%%%%%%%%%%%%%%%%%%%%%%%%

\subsection{Intermediate and strong mass-loads}\label{sec:isload}
%=========================

In this section, we present the results of the simulations of jets with mass-load rate at injection $Q_0 = 4.71 \times 10^{24}~{\rm g \, yr^{-1}\,pc^{-3}}$ (equivalent to a mean stellar mass-loss rate of $10^{-10}\,{\rm M_\odot\,yr^{-1}}$ for a stellar density of $10\,{\rm pc^{-3}}$) in Appendix~\ref{app:int}, Figures~\ref{fig:lorQ}-\ref{fig:temQ}, on the one hand, and also for jet models with $Q_0 = 4.71 \times 10^{25}~{\rm g \, yr^{-1}\,pc^{-3}}$ (equivalent to a mean stellar mass-loss rate of $10^{-9}\,{\rm M_\odot\,yr^{-1}}$ for a stellar density of $10\,{\rm pc^{-3}}$). 

As the mass-load rate increases, a stronger deceleration in the jet flow is expected. At this point it is interesting to recall that our approach to obtain the stationary models is not valid once the flow reaches subrelativistic speeds (e.g., $\gamma \leq 2$) and therefore the subsequent jet properties are not reliable. Among the models in this set, model J4QX (purely hydrodynamical, fast and light) suffers a fast expansion and a sudden deceleration leading to a flow Lorentz factor $\gamma < 2$ beyond $z \approx 200$ pc. This result has to be taken as evidence for efficient jet deceleration. However, no valid conclusion about the subsequent evolution can be derived, beyond acknowledging that the jet is efficiently decelerated and tends to becoming sub-relativistic and transonic.

% Radius
Models Q expand faster than corresponding models O. At $z = 500$ pc, all Q models have a jet radius well above 2 initial jet radii whereas O models have a radius below this value (with the exception of the pure hydrodynamical model J4OX). This is probably caused by the dissipation of kinetic energy (enhanced in the intermediate mass-load models), which produces heating, an increase of internal jet pressure, and then expansion. Moreover, there is a big difference between the local jet radius of the hydrodynamical models Q and their corresponding magnetised counterparts caused by the lack of a collimating, toroidal field in the purely hydrodynamical models. This is more striking in the case of the fastest $\gamma = 10$ cases, J4QX, J8QX, where the kinetic energy budget available for dissipation is larger.

% Rest-mass density
The rest-mass density profiles of models 4Q are almost identical to those of models 4O (with the only exception of the purely hydrodynamical model J4QX). In models 4Q the rest mass density increases from the injection point dominated by the load of mass, reaches a maximum at $z \leq 100$ pc and then decrases dominated by the jet expansion. In the lighter models 8Q (as it was in models 8O) the rest mass density profile is dominated by the mass load and keeps growing until $z = 500$ pc despite the jet expansion. The only difference between the rest mass density profiles of models 8Q and 8O is quantitative and reflects the difference in one order of magnitude between the mass-load rates of the two sets of models.

% Lorentz factor
Unlike models O (with a moderate mass-load rate), models Q experience a strong deceleration down to Lorentz factors $\simeq 3$ with the already commented exception of the purely hydrodynamical model J4QX, which reaches $\gamma \approx 2$ at $z \approx 200$ pc, and to a less extent, J8QX, reaching $\gamma \approx 2$ at $z \approx 400$ pc. The strong deceleration experienced by models Q in comparison with models O is due to the larger mass-load rate of the former. The even stronger deceleration seen in the fast models J4QX and J8QX is the result of the enhanced dissipation of kinetic energy in these models as compared with the magnetised models or the purely hydrodynamical ones with smaller Lorentz factors at injection. In the case of the magnetised models, the presence of a collimating magnetic field prevents the expansion of the jet and thefefore reduces the mass-loading. In the case of hydro jets with smaller Lorentz factors, the kinetic energy dissipation is not so strong as to increase the jet pressure because of the smaller relative initial kinetic energy budget, forcing further jet expansion and more entrainment. 

%\subsection{Strong mass-load}
%=====================

\begin{figure*} 
\includegraphics[trim=0.cm 0.7cm 0.cm 0.cm,width=\textwidth]{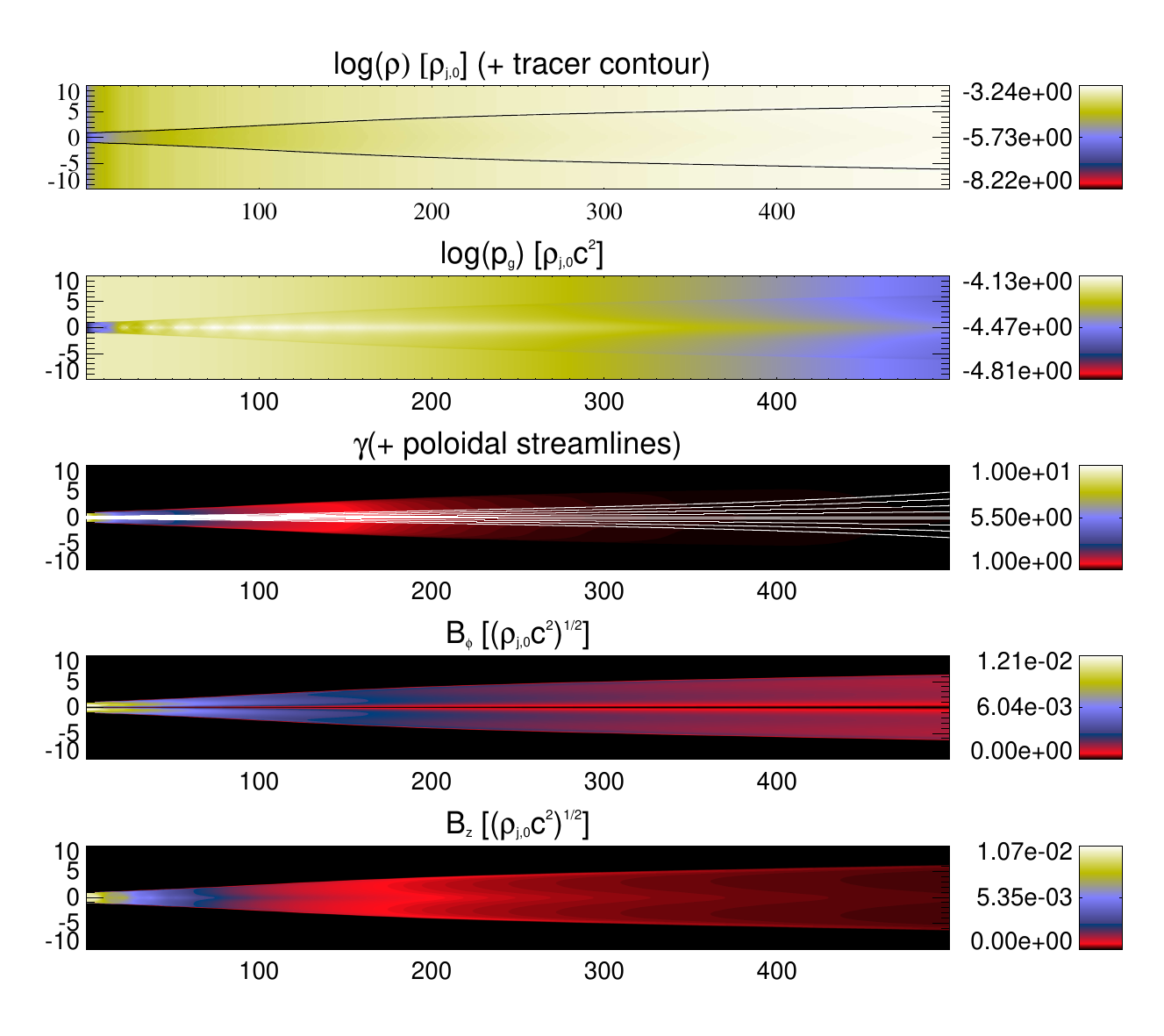} \,
\caption{Same as Fig.~\ref{fig:40A} for the case of model J8RA. The units for all plots are code units (in which $\rho_{\rm j,0} = 1.$, $R_{j,0} = 1.$ and  $c=1.$).} 
\label{fig:8RA}
\end{figure*}

Figure~\ref{fig:8RA} shows the 2D distributions corresponding to the steady-state models J8RA (Lorentz factor 10 at injection; see the Appendix~\ref{app:maps} for the case of J8RC, with Lorentz factor 6 at injection). The linear plots showing the mean jet Lorentz factor, rest-mass density, jet radius and temperature versus distance for simulations J4R\_ and J8R\_ are shown in Appendix~\ref{app:int}, Figures~\ref{fig:lorR}-\ref{fig:temR}. In contrast to the corresponding models O shown in Figures~\ref{fig:40A} and \ref{fig:radi}-\ref{fig:tem}, models R show larger opening angles and stronger decelerations. In fact all these heavily mass-loaded model jets reach the limiting Lorentz factor 2 at around 100 pc, and significantly closer to the injection point, $z \approx 30$ pc, in the case of the pure hydrodynamical, $\gamma = 10$, models. In the magnetised models, the sideways expansion of the jet finds the opposition of the radial component of the magnetic tension hence reducing the expansion rate and, correspondingly the entrained mass. The result is a delayed deceleration.

%%%%%%%%%%%%%%%%%%%%%%%%%%%%%%%%%%%%%%%%%%%%%%%%%%%%%%%%%%%%%%%%%%%%%%%%%%%%%%%%%%%%%%%%%%%%%%%%%%
%
%
%%%%%%%%%%%%%%%%%%%%%%%%%%%%%%%%%%%%%%%%%%%%%%%%%%%%%%%%%%%%%%%%%%%%%%%%%%%%%%%%%%%%%%%%%%%%%%%%%%%

\section{Discussion}\label{sec:disc}
%##############################

%\subsection{Comparison with RHD, dynamic simulations}
%=========================================

% In Sec.~\ref{ss:lmllpj} we compared our q1D simulations of steady, magnetised and unmagnetised jets with low mass-load (models J91 and J92) with previous 2D si<mulations of dynamical unmagnetised jets \citep{pe14}. The comparison shows the relevance of the ambient pressure in the evolution of jets. Hydrodynamical 2D jets evolved surrounded by an overpressured cocoon which prevents jet expansion, whereas the q1D solutions, independently of their magnetisation, are obtained in pressure equilibrium with the ambient medium. The fall in ambient pressure in the case of q1D models causes the expansion of the jet and the subsequent flow acceleration through the Bernoulli effect. 

%Focusing in the comparison of magnetised and unmagnetised q1D models, we conclude that the similarity between both kind of models is the consequence that models J91 and J92 are internal energy dominated and the magnetic tension does not play a fundamental role because the jet expansion is moderate (the opening angle is $2.7^\circ$). However, changing the initial parameters of the magnetised jets to make them Poynting flux dominated would prevent a fair comparison with the purely hydrodynamical 2D simulations.
  
\subsection{Magnetisation}
%==================

\begin{figure} 
\includegraphics[trim=7.cm 0.5cm 7.cm 1.5cm,width=\columnwidth]{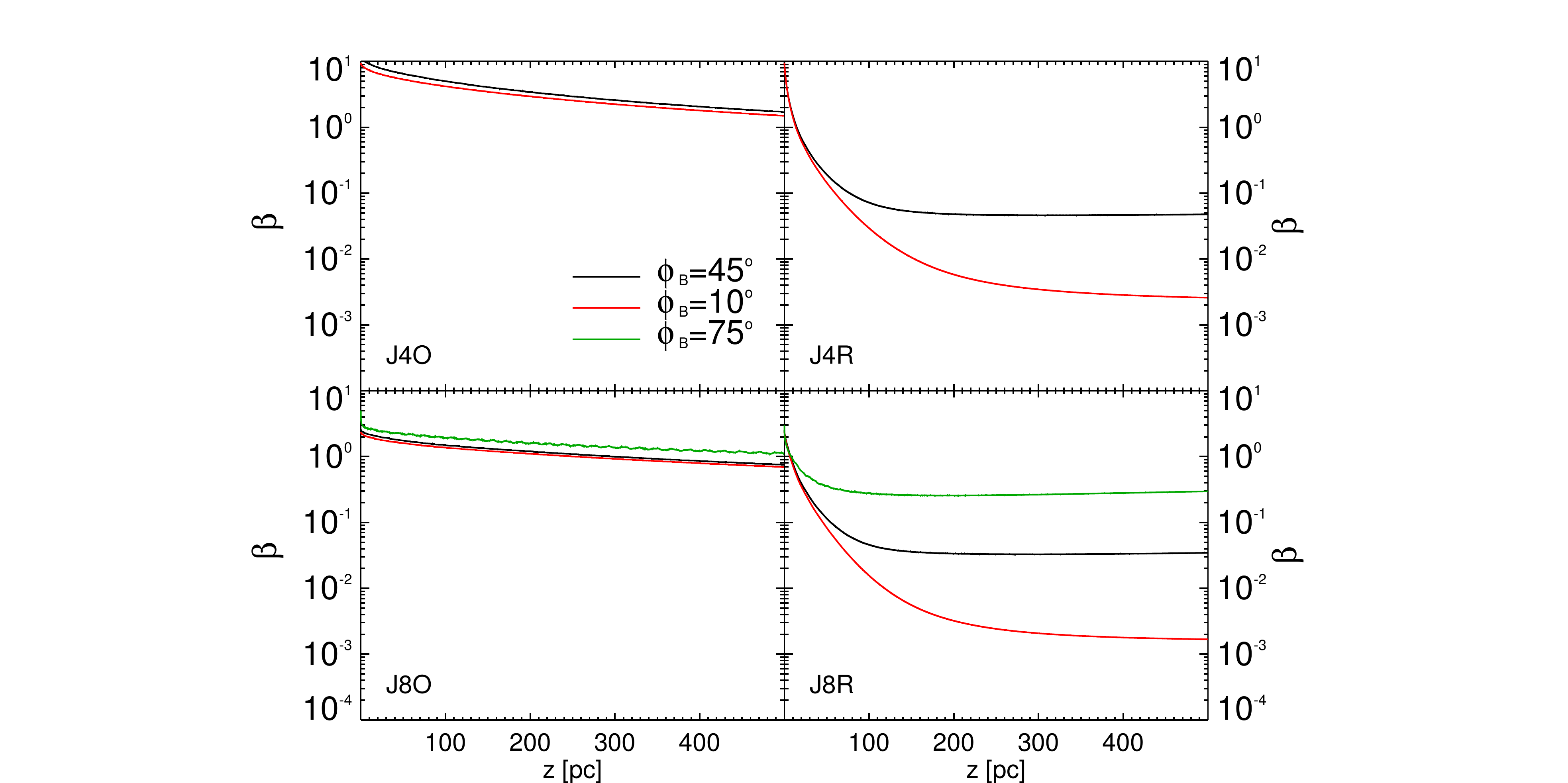} 
\caption{Mean magnetisation parameter, $\beta$, for models J4O and J4R, from left to right (top panels), and J8O and J8R (bottom panels) with Lorentz factor 10. In each plot, the black, and red lines represent the jets with mean pitch angles $\phi_B=45^\circ$ and $10^\circ$.}
\label{fig:beta1}
\end{figure}

Figure~\ref{fig:beta1} shows the mean magnetisation, $\beta$, for models J4 (top panels) and J8 (bottom panels) with Lorentz factor 10 for moderate (models O; left panels) and heavy mass-load rates (models R: right panels)\footnote{Models with Lorentz factors 7 and 6 show equivalent behaviours, albeit with different initial values of $\beta$.}. The magnetisation falls faster in models where the deceleration is also faster (i.e., those with a higher mass load rate) and, within them, those with a smaller magnetic pitch angle. Let us try to understand these trends with the help of the following expression for $\beta$ 
\begin{equation} 
\label{eq:beta}
\beta = \frac{(B^{\phi})^2}{2 p} \left( \frac{1}{\gamma^2}+\frac{1}{\tan^2 \phi_B}\right),
\end{equation}
where the first term in the brackets accounts for the toroidal magnetic field in the jet reference frame and the second term for the contribution of the axial field. According to it, there are obvious differences in the evolution of the magnetisation between jets in which $\gamma \rightarrow 1$ due to deceleration, and those that accelerate or keep a constant velocity, and between those with large and small widening of the magnetic pitch angle along the evolution. Before proceeding, let us recall that the information derived from strongly decelerated jet simulations should be used with caution when $\gamma$ approaches one.
 
%%%%%%%%%%%%%%%%%%%%%%%%%%%%%%%%%%%%%%%%%%%%%%%%%%%%%%%%%%%%%%%%%%%%%%%%%%%%%%%%%%%%%%%%%%%%%%%%%%%%
%
%
%%%%%%%%%%%%%%%%%%%%%%%%%%%%%%%%%%%%%%%%%%%%%%%%%%%%%%%%%%%%%%%%%%%%%%%%%%%%%%%%%%%%%%%%%%%%%%%%%%%%

If the jet is strongly decelerated, the first term in brackets grows to 1 and the second term tends to zero as the jet expands and $\phi_B \rightarrow 90^{\circ}$. Thus, the expected trend will be that $\beta$ tends to ${(B^{\phi})^2}/{(2 p)}$ in those cases, which results in 
\begin{equation}
\beta \approx \frac{(B^{\phi}_0)^2 \,R_{\rm j,0}^2}{2 p \, R_{\rm j}^2}
\end{equation}
(where $B^{\phi}_0$ and $R_{\rm j, 0}$ are the average toroidal magnetic field and the jet radius at $z = 0$, respectively). Note that $p$ falls with distance unless dissipation is significant,  thus partially compensating the increase in $R_{\rm j}$ with distance.  

In the case of jets that are not decelerated by mass entrainment and thus $\gamma^2 \gg 1$, the dominating term in Eq.~(\ref{eq:beta}) is $1/\tan^2 \phi_B$.\footnote{In our simulations, $1/\gamma^2$ ranges between 0.01 and 0.03 at injection, whereas $1/\tan^2 \phi_B$ ranges between 0.3 and 32 for the selected initial pitch angles.} Therefore, we have
\begin{eqnarray}
\beta & \approx & \frac{(B^{\phi})^2}{2 p \, \tan^2 \phi_B} \, = \, \frac{(B^{\phi})^2}{2 p \, (B^{\phi}/B^z)^2} = \nonumber \\
  & = & \frac{(B^z)^2}{2 p} \approx  \frac{(B^{z}_0)^2 \, R_{\rm j,0}^4}{2 p \, R_{\rm j}^4}, 
\end{eqnarray}
where $B^z_0$ is the axial magnetic field at injection. This is valid while the Lorentz factor of the jet is relatively large and jet expansion is not strong (when $R_j$ grows and $\gamma$ tends to one, the relation between terms in Eq.~\ref{eq:beta} would change). The stronger dependence on $R_{\rm j}$ ($\beta \propto 1/(p\, R_{\rm j}^4)$) explains the continuous fall of the magnetisation with distance observed in the left-column panels of Fig.~\ref{fig:beta1}, despite the fall of $p$ with distance.

\begin{figure} 
\includegraphics[trim=7.cm 0.5cm 7.cm 1.5cm,width=\columnwidth]{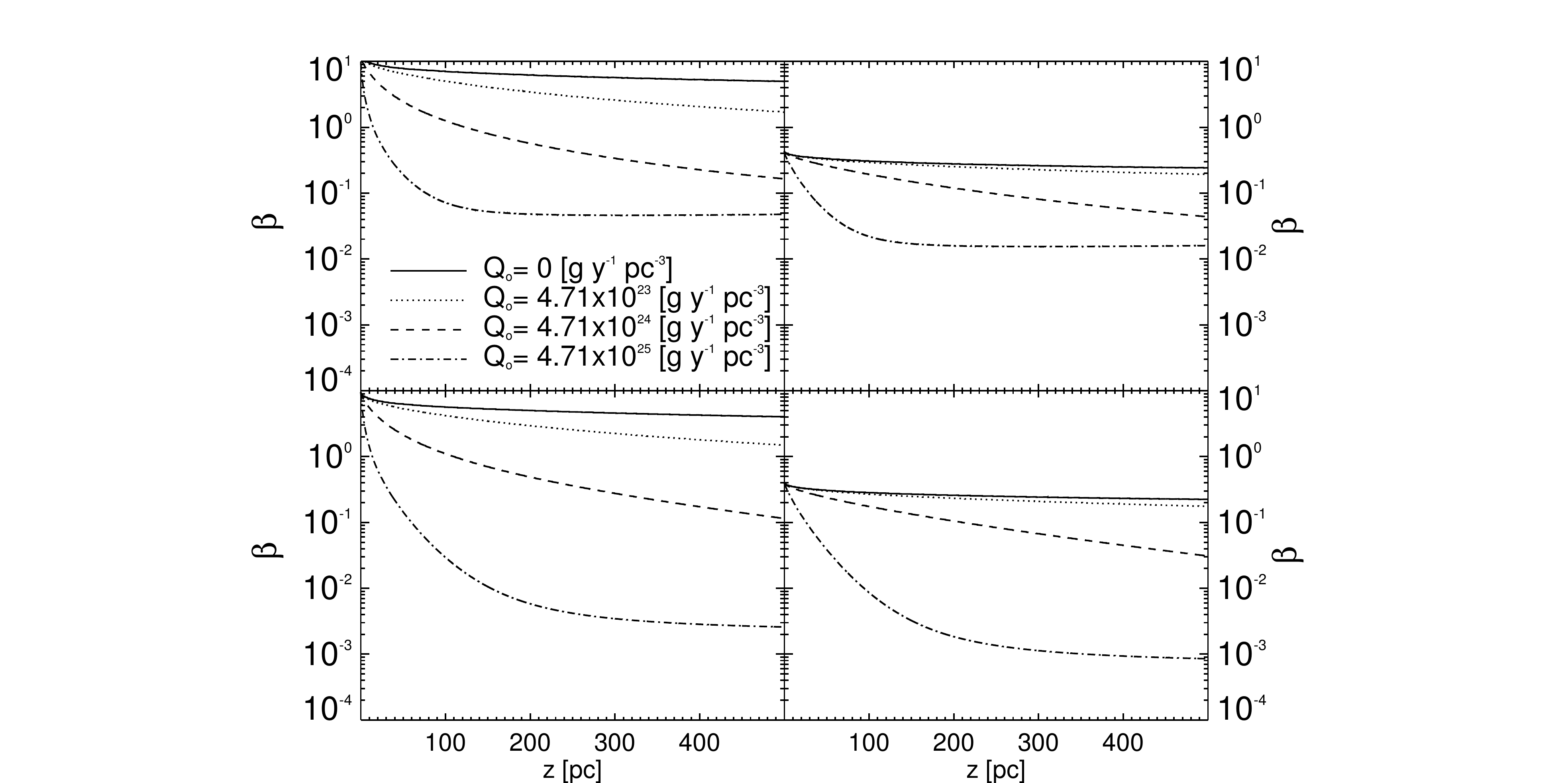} 
\caption{Mean magnetisation parameter, $\beta$, for models J4, with Lorentz factor 10 (left panels) and Lorentz factor 6 (right panels), and pitch angles of $45^{\circ}$ (top panels) and  $10^{\circ}$ (bottom panels). In each plot, solid, dotted, dashed and dash-dotted lines represent the models without mass-load, moderate, intermediate and heavy mass-load, respectively, for each Lorentz factor.}
\label{fig:beta3}
\end{figure}

\begin{figure} 
\includegraphics[trim=7cm 0.5cm 7cm 0.5cm,width=\columnwidth]{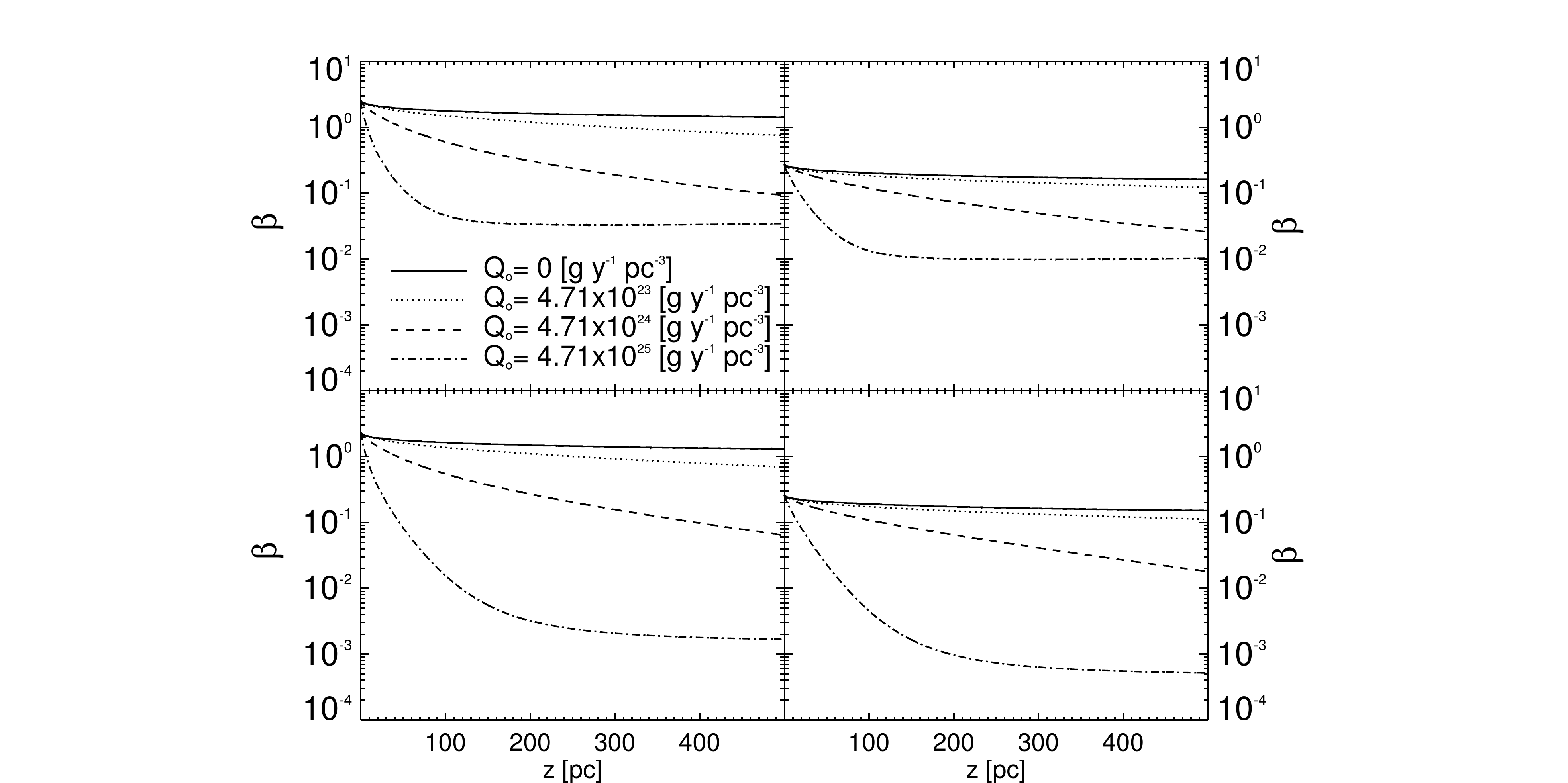} 
\caption{Mean magnetisation parameter, $\beta$, for models J8, with Lorentz factor 10 (left panels) and Lorentz factor 6 (right panels), and pitch angles of $45^{\circ}$ (top panels) and  $10^{\circ}$ (bottom panels). In each plot, solid, dotted, dashed and dash-dotted lines represent the models without mass-load, moderate, intermediate and heavy mass-load, respectively, for each Lorentz factor.}
\label{fig:beta4}
\end{figure}

As stated above, a direct consequence of the relative increase in $p$ that follows entrainment in cold and fast jets (e.g., because of the dissipation of kinetic energy) is the fall of $\beta$ to smaller values. Figures~\ref{fig:beta3} and \ref{fig:beta4} show this effect by comparing the evolution of the magnetization for J4 and J8 models, respectively, with Lorentz factors 10 (left panels) and 6 (right panels) and pitch angles $\phi_B=45^\circ$ (top panels) and $10^\circ$ (bottom panels), for no mass-load (purple lines), moderate mass-load (solid black or red lines) and strong mass-load (dashed black or red lines). The figures show that the magnetisation falls faster in the Lorentz factor 10 jets, because dissipation is more relevant in this case. The effect is visible even for the case of moderate mass-load values. 

Altogether, we can conclude that mass-load, even if relatively small, contributes to the  fall of magnetisation at large distances. 

%%%%%%%%%%%%%%%%%%%%%%%%%%%%%%%%%%%%%%%%%%%%%%%%%%%%%%%%%%%%%%%%%%%%%%%%%%%%%%%%%%%%%%%%%%%%%%%%%%%%
%
%
%%%%%%%%%%%%%%%%%%%%%%%%%%%%%%%%%%%%%%%%%%%%%%%%%%%%%%%%%%%%%%%%%%%%%%%%%%%%%%%%%%%%%%%%%%%%%%%%%%%%

%%%%%%%%%%%%%%%%%%%%%%%%%%%%%%%%%%%%%%%%%%%%%%%%%%%%%%%%%%%%%%%%%%%%%%%%%%%%%%%%%%%%%%%%%%%%%%%%%%%%
%
%
%%%%%%%%%%%%%%%%%%%%%%%%%%%%%%%%%%%%%%%%%%%%%%%%%%%%%%%%%%%%%%%%%%%%%%%%%%%%%%%%%%%%%%%%%%%%%%%%%%%%

\subsection{Magnetic acceleration} \label{sec:mac}
%==================================

\begin{figure}
\includegraphics[trim=7.cm 0.5cm 7.cm 1.5cm,width=\columnwidth]{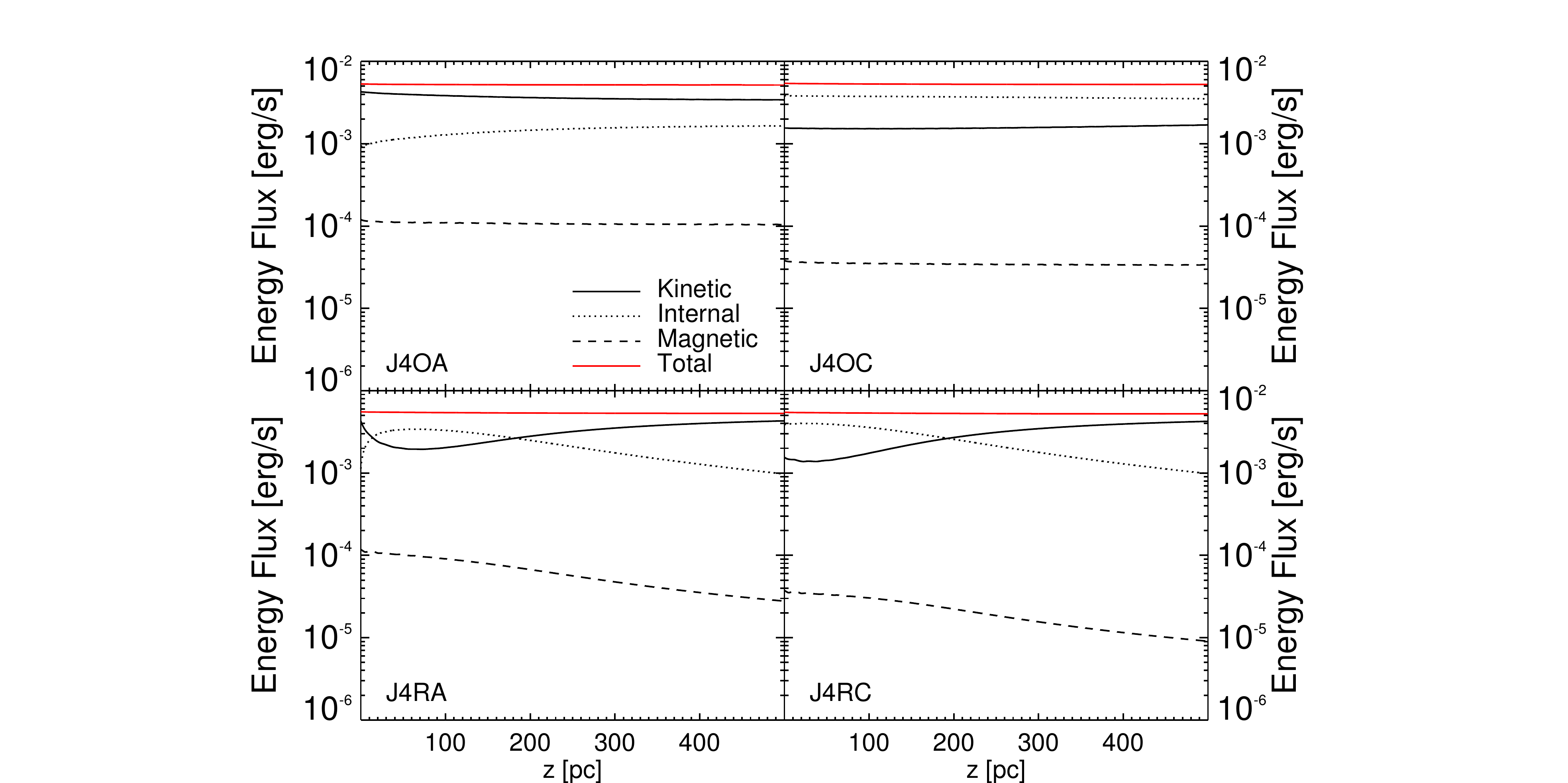} 
\caption{Energy fluxes for models J4OA (top left), J4OC (top right), J4RA (bottom left) and J4RC (bottom right), i.e., J4 models with Lorentz factors 10 (left) and 6 (right) and with moderate (top) and heavy (bottom) mass-load. The red line stands for the total flux, the dotted line for the internal energy flux, the solid line for the kinetic energy flux and the dashed line for the magnetic energy flux.}
\label{fig:acc1}
\end{figure}

\begin{figure*} 
\includegraphics[trim=0.cm 0.cm 0.cm 1.5cm,width=\textwidth]{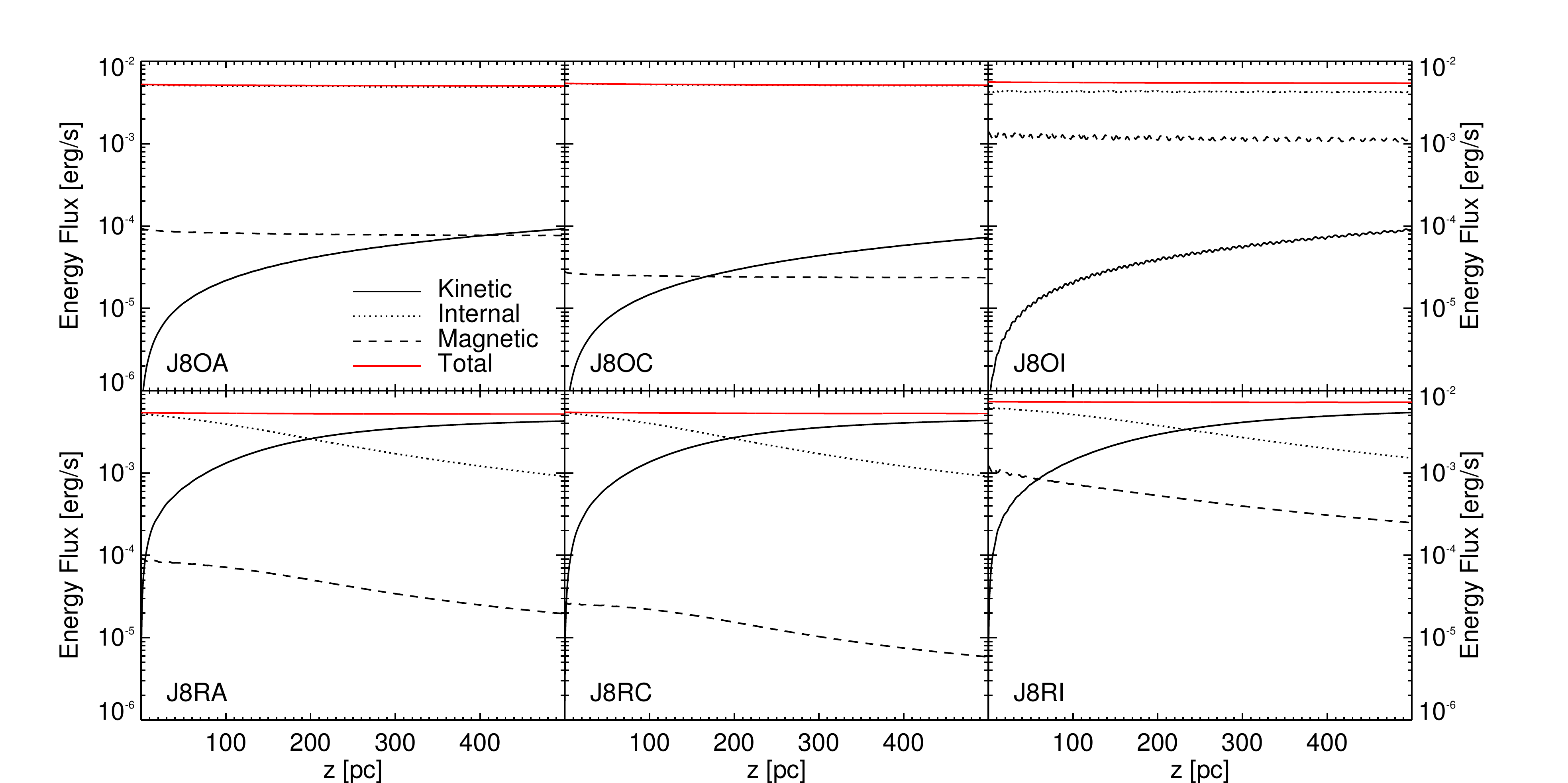} 
\caption{Energy fluxes for models J8OA (top left), J8OC (top centre), J8OI (top right), J8RA (bottom left), J8RC (bottom centre), and J8RI (bottom right) i.e., J8 models with Lorentz factors 10 (left and right) and 6 (centre), pitch angles of $45^{\circ}$ (left and centre) and $75^{\circ}$ (right), and with moderate (top) and heavy (bottom) mass-load. The red line stands for the total flux, the dotted line for the internal energy flux, the solid line for the kinetic energy flux and the dashed line for the magnetic energy flux.}
\label{fig:acc2}
\end{figure*}

An interesting effect is observed in a number of the simulated models, related to the conversion of Poynting flux into kinetic energy flux. In those models where the radius does not grow linearly with $z$, we observe a drop in Poynting flux and an extended deceleration distance as compared to purely hydro jets. Figures~\ref{fig:acc1} and \ref{fig:acc2} show, respectively, the energy fluxes for a selection of models J4 and J8 with different flow Lorentz factors, magnetic pitch angles and mass load rates. From the evolution of the radius with distance for the cases shown in these figures (see Figs.~\ref{fig:radi} and \ref{fig:radR}), we observe that a non-linear expansion of the jet (such as the one seen in models J4RC, J8RA, J8RC, J8RI) relates to a drop in the magnetic (Poynting) flux. However, because the jets showing non-linear expansions are heavily mass-loaded, the flow does not accelerate. Instead, the magnetic field acts in two different ways to extend the deceleration zone. On the one hand, the magnetic tension slows the radial expansion, thus allowing the high internal energy in these models (J8) to be slowly transferred into kinetic energy, via Bernoulli, as opposed to hydro jets (for which a faster expansion implies a faster transfer of the internal energy -and a faster mass load). On the other hand, the drop in magnetic energy necessarily implies an increase in kinetic energy (in ideal RMHD), thus delaying the deceleration of the jet. Nevertheless, we recall that this effect is seen in the more strongly decelerated models and the results should be treated with caution when $\gamma$ approaches one.

%%%%%%%%%%%%%%%%%%%%%%%%%%%%%%%%%%%%%%%%%%%%%%%%%%%%%%%%%%%%%%%%%%%%%%%%%%%%%%%%%%%%%%%%%%%%%%%%%%%%
%
%
%%%%%%%%%%%%%%%%%%%%%%%%%%%%%%%%%%%%%%%%%%%%%%%%%%%%%%%%%%%%%%%%%%%%%%%%%%%%%%%%%%%%%%%%%%%%%%%%%%%%

%%%%%%%%%%%%%%%%%%%%%%%%%%%%%%%%%%%%%%%%%%%%%%%%%%%%%%%%%%%%%%%%%%%%%%%%%%%%%%%%%%%%%%%%%%%%%%%%%%%%
%
%
%%%%%%%%%%%%%%%%%%%%%%%%%%%%%%%%%%%%%%%%%%%%%%%%%%%%%%%%%%%%%%%%%%%%%%%%%%%%%%%%%%%%%%%%%%%%%%%%%%%%

\subsection{The dynamical role of mass-load}  \label{sec:bern}
%====================================

Mass conservation in a steady relativistic flow (with $v \simeq c$) is given approximately by $\rho \gamma R_{\rm j}^2 c =$ constant in the absence of mass-load. If there is some mass-loading, the equivalent expression can be written as:
\begin{equation} \label{eq:mcons}
\rho \gamma R_{\rm j}^2 v  = \rho_0 \gamma_0 R_{\rm j, 0}^2 c + Q R_{\rm j}^2 \Delta z, 
\end{equation}
where subscript $0$ stands for injection values, the left hand side of the equation gives the mass flux at some point downstream, $Q$ represents the mean mass-load rate, and $\Delta z$ is the distance over which the jet is mass-loaded.

%On the one hand, if $\rho_0 \gamma_0 R_{\rm j,0}^2 c \leq Q R_{\rm j}^2 \Delta z$ then the mass-load dominates the jet mass flux. On the other hand, if  $\rho_0 \gamma_0 R_{\rm j,0}^2 c \gg Q R_{\rm j}^2 \Delta z$, the mass-load is negligible compared with the initial mass flux. 
  
Therefore the (approximate) minimum value of rest-mass density in the jet's reference frame for its mass flux not to be dominated by the mass-load, i.e., for the jet not to be decelerated to sub-relativistic speeds, can be obtained by equating the two terms on the r.h.s. of Equation~\ref{eq:mcons}, and expressed
%, for $\Delta z \gg$, as:
as 
\begin{eqnarray} \label{eq:mcrit}
\rho_0 \,\gamma_0 \, > \,6.7 \times 10^{-27}\, \left(\frac{\dot{M}}{10^{-11} {\rm M_\odot}{\rm yr}^{-1}}\right)\, \, 
\left(\frac{N}{10\,{\rm pc^{-3}}}\right) \, \times\nonumber \\ \left(\frac{\Delta z}{1\, {\rm kpc}}\right)^3 \left( \frac{\tan \alpha }{\tan 1^\circ }\right)^2 \left(\frac{R_{\rm j, 0}}{1\, {\rm pc}}\right)^{-2} \frac{1}{c} \, {\rm g\,cm^{-3}},
\end{eqnarray}
where we have separated $Q$ as the multiplication of the mean stellar wind mass-loss $\dot{M}$ times the stellar density $N$ (for $Q = 2\times 10^{23}\,{\rm g\,yr^{-1}\,pc^{-3}}$, $\alpha=\tan^{-1}(R_{\rm j}/\Delta z)$ is the jet half opening angle (where we assume that $R_{\rm j, \, 0} \ll R_{\rm j}$), and we consistently use $v_0 \simeq c$. 

If we apply this formula to the simulated models, setting all fractions to one except for $Q=\dot{M} N$, we find that the minimum $\rho_0\,\gamma_0$ is $\simeq 10^{-26} \,{\rm g \,cm^{-3}}$ for models O, while it is $\simeq 10^{-24} \,{\rm g \,cm^{-3}}$ for models R, with models Q between them. The relation is thus able to predict that mass load will be relevant in J4 jets and critical in J8 jets (see Figs.~\ref{fig:densty}, \ref{fig:rho2}, \ref{fig:rhoQ}, and \ref{fig:rhoR}), which have values of $\rho_0\,\gamma_0\,\sim\, 10^{-27}$ and $\sim 10^{-31}\,{\rm g\,cm^{-3}}$, respectively.    

Although this relation does not include the effects of magnetic tension, dissipation of kinetic energy, the sudden increase of the jet opening angle once mass-loading starts to be relevant, or the drop in stellar density with distance, it represents a good a priori estimate of the expected role that stellar mass-load can play on jet dynamics for given initial conditions.

We can also calculate the distance at which the entrained mass equals the initial jet mass flux (note that this would imply the effective deceleration of the jet only if the momentum and energy fluxes are dominated by particles). Comparing the two terms on the right-hand-side of Eq.~(\ref{eq:mcons}), we can state that mass-load will start to be relevant in terms of mass flux after a distance $\Delta z = l_{\rm m}$ such that
\begin{equation} \label{eq:mcmp}
\dot{M} N l_{\rm m}^3 \tan^2\alpha_{\rm j} \simeq \rho_{0} \gamma_{0} R_{\rm j, 0}^2 c.   
\end{equation}
Now, writing the initial mass flux in terms of the jet power (assuming $v_0 \simeq c$):
\begin{equation} \label{eq:mpow} 
\rho_{0} \gamma_{0} \pi R_{\rm j, 0}^2 c \simeq \frac{L_{\rm j}}{\gamma_{0} h_0},
\end{equation}    
and taking into account that the specific enthalpy is $h_0 \geq c^2$, substitution in Eq.~(\ref{eq:mcmp}) allows us to get an upper bound on $l_{\rm m}$ \citep[cf.][]{pe14},
\begin{eqnarray} \label{eq:mcrit}
l_m \, \leq\,390 \, \left(\frac{1}{\gamma_{0}(\tan(\alpha))^2}\right)^{1/3}\,  
\left(\frac{L_{\rm j}}{10^{43}\,{\rm erg\,s^{-1}}}\right)^{1/3} \, \times\nonumber \\ \left(\frac{\dot{M}}{10^{-11}\, {\rm M_\odot\,yr^{-1}}}\right)^{-1/3} \left( \frac{N}{0.1\,{\rm pc^{-3}}}\right)^{-1/3}   \, {\rm pc}.
\end{eqnarray}

%   We can also calculate the distance at which the entrained mass equals the initial jet mass flux (note that this would imply the effective %deceleration of the jet only if the momentum and energy fluxes are dominated by particles). Comparing the two terms on the right-hand-side of %Eq.~(\ref{eq:mcons}), we can state that mass-load will start to be relevant in terms of mass flux when
%\begin{equation} \label{eq:mcmp}
%\dot{M} N (\Delta z)^3 \tan^2\alpha_{\rm j} \simeq \rho_{0} \gamma_{0} R_{\rm j, 0}^2 c.   
%\end{equation}
%Writing the initial mass flux in terms of the jet power (for $v_0 \simeq c$):
%\begin{equation} \label{eq:mpow} 
%\rho_{0} \gamma_{0} \pi R_{\rm j, 0}^2 c \simeq \frac{L_{\rm j}}{\gamma_{0} h_0},
%\end{equation}    
%which, considering that specific enthalpy is $h_0 \geq c^2$, and substituting $\Delta z$ by $l_m$, and $R_{\rm j}$ by $\Delta z \,\tan(\alpha)$, %brings us to the distance at which both terms become equal:
%  \begin{eqnarray} \label{eq:mcrit}
%  l_m \, \leq\,390 \, \left(\frac{1}{\gamma_{0}(\tan(\alpha))^2}\right)^{1/3}\,  
%\left(\frac{L_{\rm j}}{10^{43}\,{\rm erg\,s^{-1}}}\right)^{1/3} \, \times\nonumber \\ \left(\frac{\dot{M}}{10^{-11}\, {\rm %M_\odot\,yr^{-1}}}\right)^{-1/3} \left( \frac{N}{0.1\,{\rm pc^{-3}}}\right)^{-1/3}   \, {\rm pc},
%\end{eqnarray}
 
Although this expression is not very sensitive to variations in the parameters within an order of magnitude, it
can vary with the jet opening angle, $\alpha$. For an opening angle of $1^\circ$, $l_m \simeq 8.4/\gamma_0$~kpc, whereas $\alpha=1.5^\circ$ brings $l_m$ to $\simeq 1$~kpc. In the case of a jet in free expansion, $\alpha \simeq 1/\gamma_0$, so if $\gamma_0 \simeq 10$, then $l_m \leq 200$~pc. Again, Eq.~(\ref{eq:mcrit}) does not take into account the drop in $N$ with distance, but a large drop is not expected within the inner hundreds of parsecs for the 'core' ellipticals typically hosting FRI jets, which have roughly constant stellar densities on these scales \citep{la07,ru05}.

\subsection{Summary of jet dynamics}
%===========================

\begin{figure*} 
	\includegraphics[width=\textwidth]{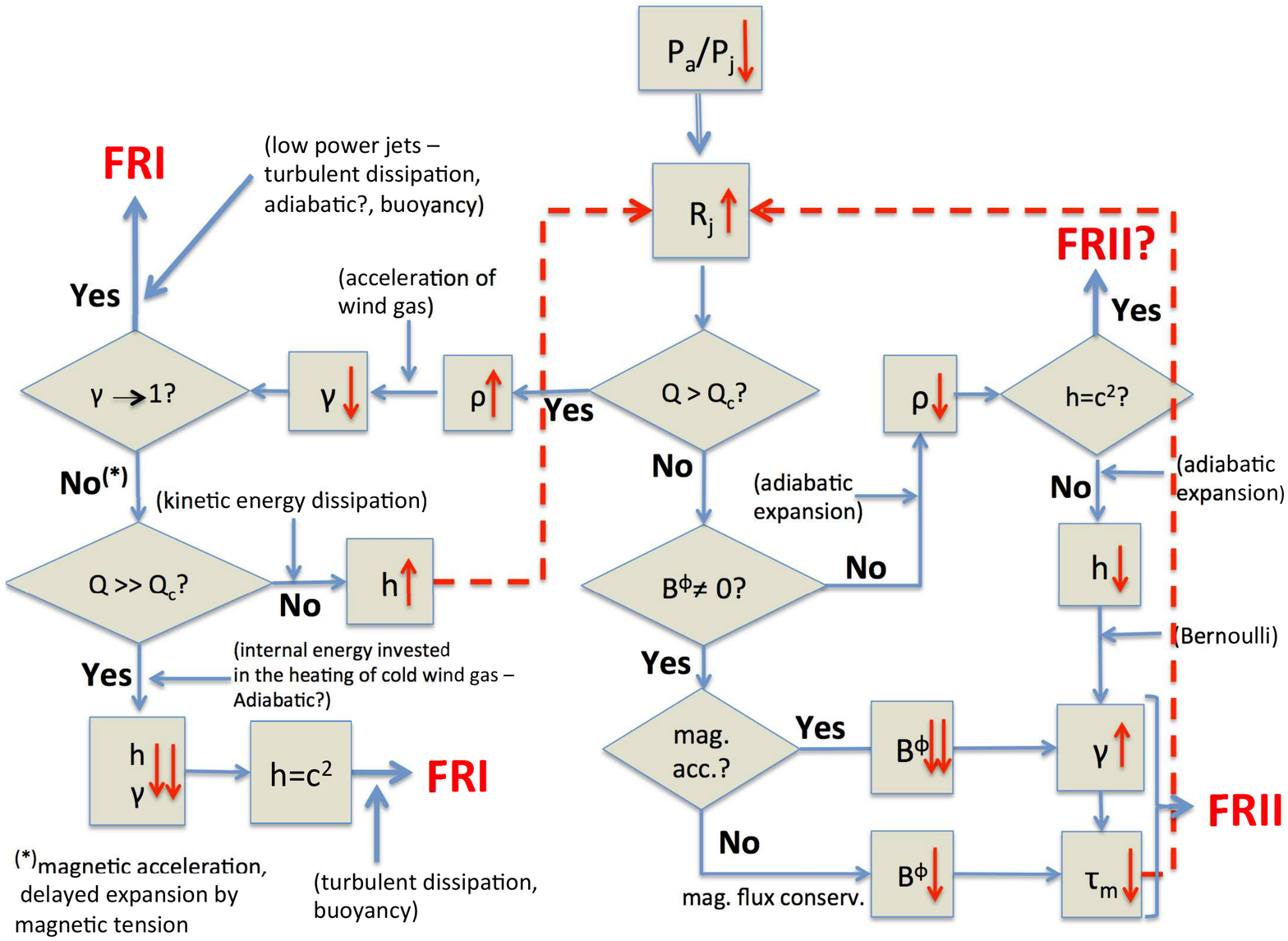}
	\caption{Flow diagram of jet evolution. The rhombi indicate options for the evolution of the flow that depend on the local jet properties, whereas the squares indicate natural consequences of the described processes. (*) Although it has also been indicated in the diagram, it is relevant to stress that the presence of a toroidal field would slow down the processes described along the left branch, by delaying jet expansion (thus reducing the entrainment of starts and/or extending thermal acceleration) or via magnetic acceleration (see the text).}
	\label{fig:flow}
\end{figure*}

Figure~\ref{fig:flow} shows a flow diagram that summarizes the different evolutionary paths discussed in this paper for mass-loaded jets. The diagram applies to both FRI and FRII relativistic jets and takes into account the effects of the magnetic field in the evolution. In this context, we identify FRI jets as those that expand with large opening angles and decelerate to sub-relativistic speeds ($\gamma \approx 1$) whereas FRII jets remain significantly relativistic and well collimated.

A drop in the ambient pressure $p_a$ (or, equivalently, an initial jet overpressure) [$p_a/p_j \downarrow$ in the diagram] induces the expansion of the jet, which translates into an increase of the jet cross-section [$R_j \uparrow$]. The subsequent evolution depends on whether  or not the jet is significantly mass-loaded 
%(with respect to its initial mass flux;  Sect.~\ref{sec:bern}) 
[$Q > Q_c ?$].  

From Eq.~(\ref{eq:mcons}) it is easy to show that
\begin{equation}
	\label{eq:WW0}
	\frac{\rho}{\rho_0} = \frac{Q \bar{R}_{\rm j}^2 \Delta z}{R_{\rm j}^2 \gamma v \rho_0} + \frac{R_{\rm j, \, 0}^2 \gamma_0 c}{R_{\rm j}^2 \gamma v},
\end{equation}
where $\bar{R}_{\rm j}$ stands for the mean value of the radius along $\Delta z$ (if $R_{\rm j, \, 0} \ll R_{\rm j}$, then $\bar{R}_{\rm j} \approx R_{\rm j}/2$). Thus, the comparison of the two terms on the right-hand-side of the equation will determine the ratio $\rho/\rho_0$, namely, whether the rest-mass density of the jet decays along $\Delta z$ as a result of the increase in radius, or grows as a result of the mass load. The comparison between these terms leads to the definition of a critical mass load:
\begin{equation}\label{eq:Qc}
	Q_c \,=\, \frac{R_{\rm j, \, 0}^2}{\bar{R_{\rm j}}^2} \, \frac{\rho_0\,\gamma_0 \,c}{\Delta z}.
\end{equation} 

If the mass load is above this critical value [left sector of the diagram], the jet rest-mass density rises with distance [$\rho \uparrow$], and the jet is decelerated [$\gamma \downarrow]$. This is what we clearly observe in models with higher values of $Q_0$ (most of the Q models and all the R ones). These jets can lead to FRIs once they have been sufficiently decelerated [top-left corner of the diagram].

On the contrary, if $Q \ll Q_c$ [central/right sector of the diagram], the jet follows a well-known adiabatic expansion process in which density and pressure drop [$\rho \downarrow$], and the Bernoulli effect takes place as long as there is internal energy available (i.e., the jet enthalpy, $h$, is larger than $c^2$ [$h = c^2?$]), producing acceleration [$\gamma \uparrow$]. This is seen in J4O and J5O models, although the effects of mass-load are also noticeable in these cases because the mass-load is not very much smaller than $Q_c$ (see Sect.~\ref{sec:bern}). Jets without significant mass load (and in the absence of other disrupting instabilities, e.g., magnetic instabilities) can keep extracting internal energy and accelerating by the Bernoulli process, and remain collimated to end up as an FR~II [top-right corner of the diagram].

Using the energy conservation equation,
\begin{equation}
	\pi R_{\rm j, \, 0}^2 \rho_0 h_0\,\gamma_0^2 \,c \,=\, \pi R_{\rm j}^2 \rho h\,\gamma^2 \,v,
\end{equation}    
leads to the following ratio for the jet enthalpy at injection and at some point downstream of the flow:
\begin{equation}
	\frac{h_0}{h}\, = \frac {\rho\,R_{\rm j}^2\,\gamma^2 v}{\rho_0\,R_{\rm j, \, 0}^2\,\gamma_0^2 c},
\end{equation}    
which, considering Eqs.~\ref{eq:WW0} and \ref{eq:Qc}, can be rewritten as:
\begin{equation}
	\frac{h_0}{h}\, = \, \frac{\gamma}{\gamma_0}\,\left( 1 + \frac{Q}{Q_c} \right),
\end{equation}
where the term in {parentheses} is always larger than 1. The equation reduces to Bernoulli if $Q \ll Q_c$. In this case, the jet accelerates at the cost of reducing the specific internal energy. However if the mass loading is large [left sector of the diagram], the internal energy can be replenished {[$h \uparrow$]} due to the dissipation of kinetic energy at the entrainment sites. This possibility dominates and is observed in denser, colder, and faster jets (e.g., see the upper left panels in Figs.~\ref{fig:lor} and \ref{fig:tem}, and \ref{fig:lorQ} and \ref{fig:temQ}). The increase of the internal energy implies an increase of pressure that also favours further expansion [left red broken line]. In cases in which mass load is relatively important (models Q and R), expansion has a runaway effect caused by the entrainment of more stars (or clouds) as the jet cross section grows. As a consequence, in jets that are initially hot and with $Q \geq Q_c$, the mass entrainment increases as the jet expands {[$Q \gg Q_c$]} {and decelerates, while the internal energy of the jet is invested in heating the entrained material [$h, \gamma \downarrow \downarrow$]}. {This kind of situation can lead also to FRI jets [bottom-left corner of the diagram].} The {self-sustained process of expansion/mass entrainment} can be delayed by a toroidal field, which slows down jet expansion through the magnetic tension {($\tau_m = (B^\phi)^2/\gamma^2 R_j$)}, and flow deceleration if magnetic acceleration takes place (see Sec.~\ref{sec:mac}).

If the jets are weakly magnetised [$B^\phi ?$]; bottom-right sector of the diagram, super-Alfv\'enic flows would overexpand beyond pressure equilibrium with the ambient medium, triggering recollimation shocks; or would continue expanding to longer distances, if the pressure gradient is shallow. In many of our simulations, we see that non-magnetised jets (models N, represented by blue lines in the line plots) show larger opening angles than magnetised jets. If the jets are strongly magnetised, they can undergo magnetic acceleration if differential expansion is induced \citep[see, e.g.][and references therein]{ko12} and generate a fast drop in the toroidal field [$B^\phi \downarrow \downarrow$] and a consequent relative increase in the Lorentz factor. The expansion also leads to a drop in the field, albeit slower [$B^\phi \downarrow$], due to conservation of the magnetic flux. In both cases, there is a relative drop of the magnetic tension [$\tau_m \downarrow$], which may allow further expansion [right red broken line].\footnote{We actually observe hints of magnetic acceleration in those jets with non-conical expansion (see Figs.~\ref{fig:acc1}-\ref{fig:acc2}, Figs.~\ref{fig:radi} and \ref{fig:radR} and Sect.~\ref{sec:mac}).} We can see that the situations depicted in the right sector of the diagram end up in jet acceleration, thus typically producing large-scale FRII morphologies.

%%%%%%%%%%%%%%%%%%%%%%%%%%%%%%%%%%%%%%%%%%%%%%%%%%%%%%%%%%%%%%%%%%%%%%%%%%%%%%%%%%%%%%%%%%%%%%%%%%%%%%%%
%
%
%%%%%%%%%%%%%%%%%%%%%%%%%%%%%%%%%%%%%%%%%%%%%%%%%%%%%%%%%%%%%%%%%%%%%%%%%%%%%%%%%%%%%%%%%%%%%%%%%%%%%%%%

It is relevant to stress that in this flow diagram we describe competing effects that can take place at the same time for a given jet and can either have opposite effects, or contribute in the same direction.

\subsection{Astrophysical implications}
%============================

  In \cite{lb14}, the authors modelled a number of FRI jets by fitting distributions of velocity, rest-frame emissivity and magnetic-field ordering to the brightness and polarization distributions distributions of the sources. A series of common (or frequent) features were found: 
\begin{enumerate}  
  \item an initial brightness and geometrical flaring beyond the dense galactic corona, 
  \item development of a wide velocity distribution and deceleration (from $v\simeq 0.8\,c$ to $v \leq 0.4\,c$ on axis),
  \item a region of high emissivity roughly coincident with the deceleration,
  \item recollimation to a conical expansion with roughly constant velocity farther downstream,
   \item a transformation of the magnetic field structure from a mixture of an ordered toroidal component and a disordered longitudinal one (with the longitudinal field dominating) at the jet base to mainly toroidal after deceleration. 
\end{enumerate}
The models of emissivity and field-component ratio are not consistent with adiabatic losses and flux freezing alone.  The authors conclude that ongoing particle acceleration is required, consistent with the short cooling times for observed X-ray synchrotron emission.   
   
   The brightness flaring starts at distances that range from $0.7$ to $2$~kpc, approximately, whereas the range of distances at which strong deceleration starts is more spread, although it happens within the inner 4~kpc for most of the sources. The position of the brightness flaring sets the distance from which the model fitted to the sources in
\cite{lb14} can give reasonable determinations of the physical parameters. This is, thus, beyond the region
that we have covered in this work \citep[we also note that our approximation does not allow us to run simulations with $v_{\rm j} \simeq 0.8\,c$, which is the typical fitted velocity at that distance in][]{lb14}. 
   
   The simulations that we show here would cover the inner region, in which the modelling of FRI jets by Laing \& Bridle is much less reliable due to the narrowness and faintness of both the jets and the counter-jets. Our results describe the jet structure in the region of the dense corona,   
close to the active nucleus. In this region, FRI jets are probably still relativistic, as revealed by the brightness asymmetries in VLBI images of those FRI jets that can be imaged \citep[see, e.g., the strong brightness asymmetry in the case of NGC~315,][]{lis18}. Therefore, the approximation of jet axial speed close to $c$ that is required by our simulations remains valid.

\begin{figure} 
\includegraphics[trim=7.cm 0.5cm 7.cm 1.5cm,width=\columnwidth]{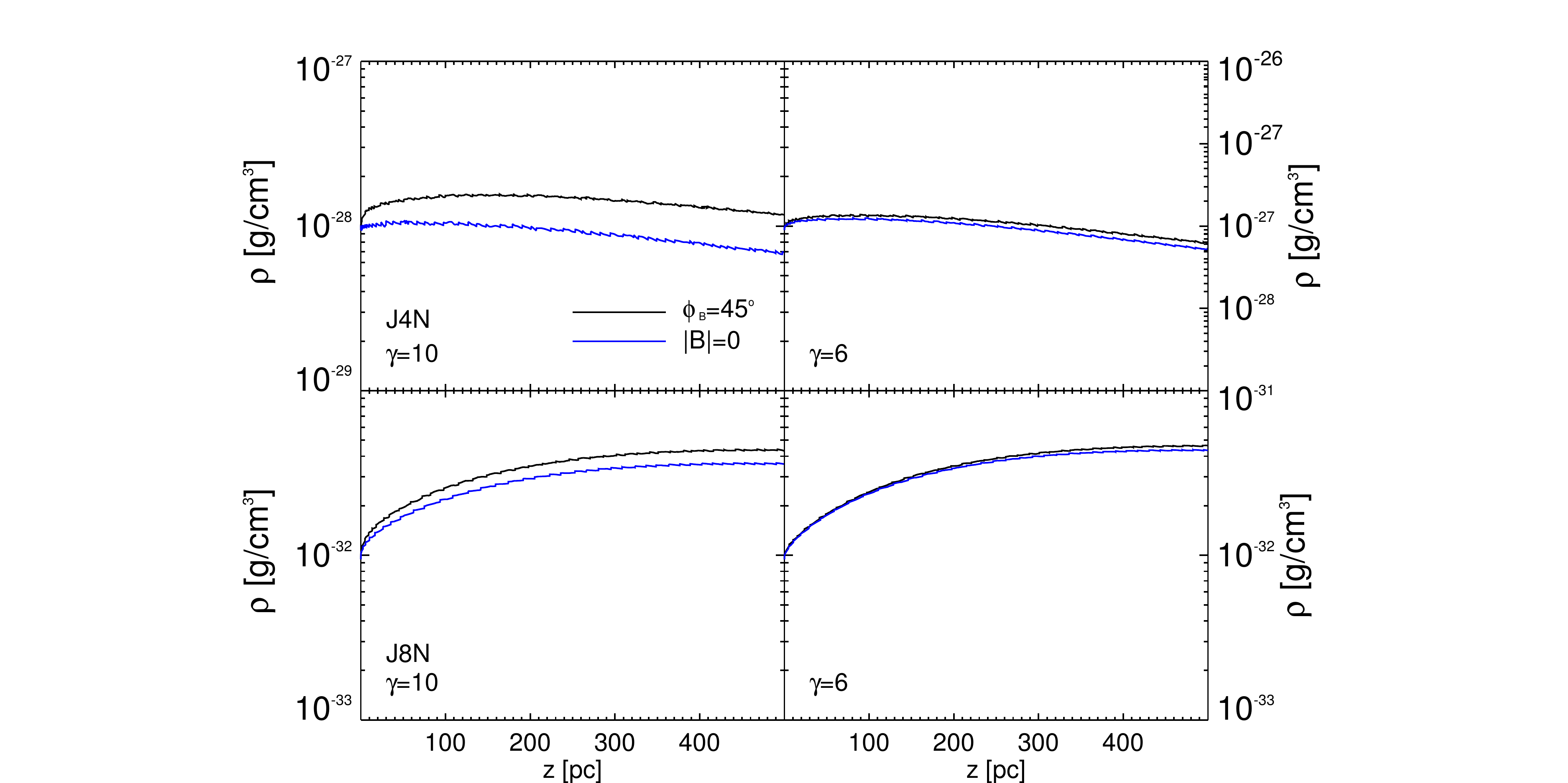} 
\caption{Rest-mass density versus distance for models J4N\_ (top panels) and J8N\_ (bottom panels) with Lorentz factor 10 (left column), and 6 (right column). In each plot, the black and blue lines represent the jets with mean pitch angles $\phi_B=45^\circ$, and purely hydro models, respectively.}
\label{fig:rho19}
\end{figure}

\begin{figure} 
\includegraphics[trim=7.5cm 0.5cm 7.5cm 1.5cm,width=\columnwidth]{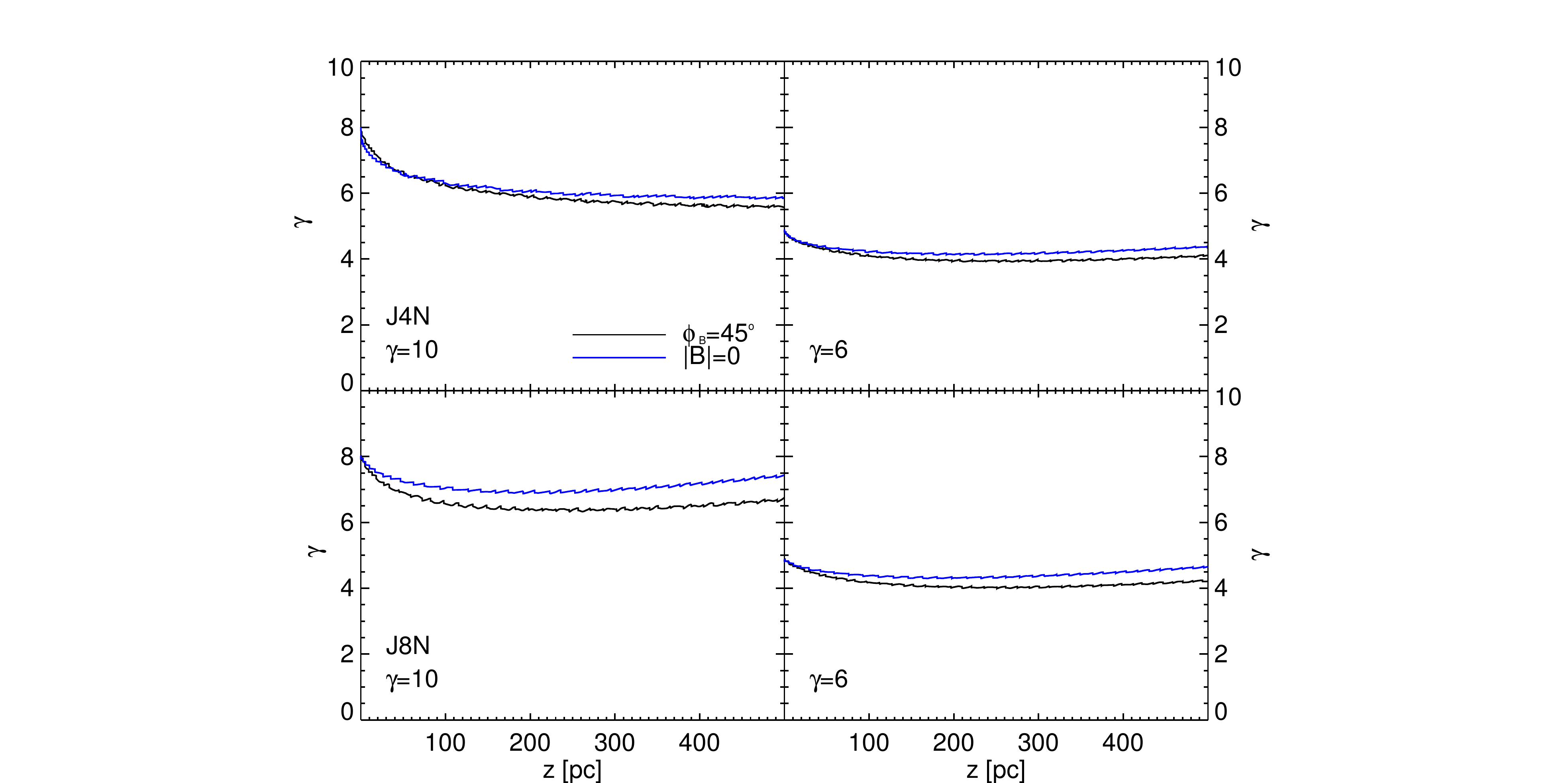} 
\caption{Mean jet Lorentz factor versus distance for models J4N\_ (top panels) and J8N\_ (bottom panels) with Lorentz factor 10 (left column), and 6 (right column). In each plot, the black and blue lines represent the jets with mean pitch angles $\phi_B=45^\circ$, and purely hydro models, respectively.}
\label{fig:lor20}
\end{figure}

Our simulations show that jets with powers $L_{\rm j} \simeq 10^{43}\,{\rm erg\,s^{-1}}$ can be decelerated from velocities that are close to $c$ to that given by the models at the onset of the flaring region, $\simeq 0.8\, c$ for sufficient mass-loss rates per star and stellar densities within the galactic core. These are nevertheless possibly too large with respect to the expected ones from typical stellar populations in these galaxies. In Figs.~\ref{fig:rho19} and \ref{fig:lor20} we show the evolution of rest-mass density and Lorentz factor obtained for a number of J4N and J8N models, i.e., those with the expected mass-load in elliptical galaxies (see Table~\ref{tab:ambientpars}. The results (also for the radius and the jet temperature, not shown) are qualitatively very similar to the equivalent J4O and J8O models and confirm that, although it can have a non-negligible effect on the jet parameters (mild deceleration and an increase in the rest-mass density are observed), stellar mass-load cannot, by itself, explain jet deceleration to mildly relativistic or sub-relativistic speeds as it happens in classical FRI jets \citep[as anticipated by][]{lb02b,pe14}. 

%
%%%%%%%%%%%%%%%%%%%%%%%%%%%%%%%%%%%%%%%%%%%%%%%%%%%%%%%%%%%%%%%%%%%%%%%%%%%%%%%%%%%%%%%%%%%%%%%%%%%%%

%%%%%%%%%%%%%%%%%%%%%%%%%%%%%%%%%%%%%%%%%%%%%%%%%%%%%%%%%%%%%%%%%%%%%%%%%%%%%%%%%%%%%%%%%%%%%%%%%%%%%
%

In Section~\ref{sec:bern}, we have shown that even well collimated jets can mass-load in quantities equivalent to their initial mass-fluxes at distances that are of the order of the flaring distances given in \cite{lb14}. Whether this load is also sufficient to trigger the emissivity flaring via the shocks and turbulence triggered by the jet-star interactions \citep[e.g.,][]{br+12,pe+17} remains to be studied. 
However, the visible drop in the jet Lorentz factor and the increase (or delayed decrease) in the jet rest-mass density or temperature (not shown) produced by this process even for the expected stellar populations in ellipticals (see Figs.~\ref{fig:rho19} and \ref{fig:lor20}), shows that it can efficiently contribute to the dissipation of the jet energy flux and to its radiative output at these scales \citep[e.g.,][]{wy15,Vieyro:2017}.

Furthermore, our simulations also show that, when mass load becomes sufficient, the jets undergo an expansion that resembles the geometrical flaring observed in FRI jets, even in slowly decaying ambient pressures (see Figs.~\ref{fig:radi} and \ref{fig:radR}). This effect would be exaggerated in the case of rapidly decreasing atmospheres outside the dense galactic cores, which is where it takes place in the observed FRI sources, according to \cite{lb14}. It is important to stress that once the jet inner spine is decelerated, the
entrainment triggered at the boundaries, probably by small-scale instabilities \citep[e.g.,][]{mk07,mk09,mm13,ma17,to17,gk18a,gk18b} can reach the axis in shorter distances than in the case of jets with large Lorentz factors.

Finally, adding the effect of mass-load by stars to the possible role of stars in triggering mixing with the ambient medium, as suggested in \citet{pe20}, could well be a sufficient combination of processes to explain jet deceleration in relatively powerful FRI sources, whereas stellar mass loss could be a sufficient mechanism for jet deceleration in less powerful jets \citep{pe14}.

\section{Conclusions} \label{sec:conc}
We have run RMHD simulations of stationary jets, with a source term in the mass equation that accounts for mass-loading by stellar winds within the host galaxy. We have simulated jets with a total power of $10^{43}\,{\rm erg s^{-1}}$ but with a wide range of internal energies and densities, and different magnetic field configurations, Lorentz factors and mean stellar populations (i.e., mean stellar wind mass loss). The models are injected in transversal equilibrium, which is updated for each spatial step along the jet axis, as the injected particles and the ambient medium pressure change the jet-to-environment conditions. 

  Our results show that: 
  \begin{itemize}
  \item Jets with $10^{43}\,{\rm erg s^{-1}}$ may undergo mild deceleration even for old stellar populations, which, in the frame of the suggestion that stars may also trigger jet-ISM mixing at the jet boundary \citep{pe20}, can altogether represent an efficient deceleration mechanism for classical FRI jets.
  \item Initially dilute and hot models cool and mass-load very rapidly, changing from an electron-positron dominated jet to an electron-proton dominated jet.
  \item Initially denser and faster jets can undergo heating caused by the dissipation of kinetic energy as they are mass-loaded.
  \item The toroidal field acts to collimate the flow, with a double effect with respect to non-magnetised jets:
     \begin{itemize}
         \item Bernoulli acceleration becomes more gradual, and
         \item it prevents the entrainment of further stars by preventing fast jet expansion.
     \end{itemize}
   \item The effects of entrainment are that:
    \begin{itemize}
         \item the magnetic field structure tends to be dominated by the toroidal field, by forcing jet expansion, and  
       \item the ratio of magnetic energy to internal energy densities drops with distance.
     \end{itemize}  
  \item A population of stars dominated by red giants would rapidly decelerate the jets to mildly relativistic speeds. However, such a population is not present in typical FRI host galaxies, so it is improbable that stellar winds alone can decelerate the jets.  
   \item In the case of strong mass-load, we have observed a correlation between differential expansion and a drop in magnetic energy flux that delays deceleration, i.e., that is invested into flow acceleration.
   \item The mass-load by stars in the dense corona of galaxies, where the stellar density is larger, could trigger the initial jet deceleration before the brightness and geometrical flaring start.
   \item In general, mass-load might contribute to the change of the jet properties and composition (loading protons) along its spatial evolution and can efficiently generate dissipation of the jet energy flux and cause part of the observed radiative output at these scales.
  \end{itemize}

   Future simulations should include a full three-dimensional, dynamic approach to the problem, introducing cooling terms to account for the possible non-adiabatic behaviour of jet evolution during deceleration \citep{lb14}. In addition, simulations of jet evolution under the constant perturbation of star crossings at the boundaries, as suggested by \citet{pe20}, should be performed, although the different scales that this scenario implies strongly complicate the set-ups. 

\section*{Acknowledgements}
This work has been supported by the Spanish Ministry of Science through Grants PID2019-105510GB-C31, PID2019-107427GB-C33 and AYA2016-77237-C3-3-P, and from the Generalitat Valenciana through grant PROMETEU/2019/071.
 
\section*{Data Availability}
The data underlying this article will be shared on reasonable request to the corresponding author.

%\bibliographystyle{mnras}

%%%%%%%%%%%%%%%%%%%%%%%%%%%%%%%%%%%%%%%%%%%%%%%%%%

%%%%%%%%%%%%%%%%% APPENDICES %%%%%%%%%%%%%%%%%%%%%

\appendix

\section{Supplementary material} \label{app:extra}

\subsection{Other simulated jets}
Tables~\ref{tab:jetpars2} and {tab:jetflux2} show the parameters of a number of extra simulations that were run and used in order to obtain a better understanding of the physics of the process studied than if only those discussed in the paper had been run. 
%%%%%%%%%%%%%%%%%%%%%%%%%%%%%%%%%%%%%%%%%%%%%%%%%%%%%%%%%%%%%%%%%%%%%%%%%%%%%%%%%%%%%%%%%%%%%%%%%
%
\begin{landscape}
\begin{table}	
	\centering
	\caption{Extra jet parameters.}
	\label{tab:jetpars2}
	\begin{tabular}{lcccccccccccccc} % four columns, alignment for each
		\hline
		Model & $\rho$ & $\gamma$ & $\phi_{\rm B}$ & $T$ & $\Gamma_{\rm ad}$ & $h$ & $B^{z}$  & $B_{\rm m}^\phi$ & $e$ & $u_{\rm B}$ & $\beta$ & $R_{\rm j}$ &$L_{\rm j}$ & $p_{\rm a, \, 0}$  \\
		 & [g/cm$^3$] &  & [$^\circ$]  &   [K] &  & [c$^2$]   &  [mG] & [mG] & [erg/cm$^3$] & [erg/cm$^3$] &  & [pc] & [erg/s] & [dyn/cm$^2$]  \\
	\hline
	\hline
		J5\_X &  $10^{-29}$ & 10 & -  & $1.8\times10^{10}$ & 1.37 & 12.6 & - & -& $ 7.61\times 10^{-8}$ & - & - & 1 & $10^{43}$ & $10^{-7}$  \\
		J5\_Y &  $10^{-29}$ & 7 & -  & $3.8\times10^{10}$ & 1.35 & 25.8 & - & -& $ 1.65\times 10^{-7}$  & - & -& 1 & $10^{43}$ & $10^{-7}$  \\
		J5\_Z &  $10^{-29}$ & 6 & - & $5.2\times10^{10}$ & 1.34 & 35.2 & -  & -& $ 2.29\times 10^{-7}$ & - & - & 1 & $10^{43}$ & $10^{-7}$  \\
		J5\_A &  $10^{-29}$ & 10 & 45 & $1.8\times10^{10}$ & 1.37 & 12.4 & $1.36$  & $1.65$& $ 7.50\times 10^{-8}$ & $ 1.48\times 10^{-7}$ & 2.66& 1 & $10^{43}$ & $10^{-7}$   \\
		J5\_B &  $10^{-29}$ & 7 & 45  & $3.8\times10^{10}$ & 1.35 & 25.6 & $1.04$ & $1.28$&  $ 1.64\times 10^{-7}$ & $ 8.82\times 10^{-8}$ & 0.77& 1 & $10^{43}$ & $10^{-7}$  \\
		J5\_C &  $10^{-29}$ & 6 & 45 & $5.2\times10^{10}$ & 1.34 & 35.1 & $0.74$ & $0.91$& $ 2.28\times 10^{-7}$ & $ 4.48\times 10^{-8}$ & 0.28& 1 & $10^{43}$ & $10^{-7}$   \\
		J5\_E &  $10^{-29}$ & 10 & 10 & $1.8\times10^{10}$ & 1.37 & 12.6 & $1.35$ & $0.29$& $ 7.61\times 10^{-8}$ & $ 1.45\times 10^{-7}$ & 2.59& 1 & $10^{43}$ & $10^{-7}$  \\
		J5\_F &  $10^{-29}$ & 7 & 10 & $3.8\times10^{10}$ & 1.35 & 25.8 & $1.03$ & $0.22$& $ 1.66\times 10^{-7}$ & $ 8.56\times 10^{-8}$ & 0.74& 1 & $10^{43}$ & $10^{-7}$  \\
		J5\_G &  $10^{-29}$ & 6 & 10 & $5.2\times10^{10}$ & 1.35 & 35.2 & $0.74$ & $0.17$ & $ 2.29\times 10^{-7}$ & $ 4.30\times 10^{-8}$ & 0.27& 1 & $10^{43}$ & $10^{-7}$  \\
		J5\_I &  $10^{-29}$ & 10 & 75 & $1.5\times10^{10}$ & 1.38 & 10.2 & $1.37$ & $6.60$& $ 5.98\times 10^{-8}$ & $ 1.78\times 10^{-7}$ & 3.95& 1 & $10^{43}$ & $10^{-7}$   \\
		J5\_J &  $10^{-29}$ & 7 & 75 & $3.3\times10^{10}$ & 1.35 & 22.6 & $1.13$ & $5.08$& $ 1.44\times 10^{-7}$  & $ 1.28\times 10^{-7}$ & 1.26& 1 & $10^{43}$ & $10^{-7}$  \\
		J5\_K &  $10^{-29}$ & 6 & 75 & $4.8\times10^{10}$ & 1.35 & 32.9 & $0.83$ & $3.71$& $ 2.13\times 10^{-7}$ & $ 7.40\times 10^{-8}$ & 0.50& 1 & $10^{43}$ & $10^{-7}$   \\
\hline
	        J7\_X   &  $10^{-31}$ & 10 & - & $1.8\times10^{12}$ & 1.33 & $1.25\times10^{3}$ & -  & -& $ 8.47\times 10^{-8}$ & - & -& 1 & $10^{43}$ & $10^{-7}$  \\
		J7\_Y   &  $10^{-31}$ & 7 & - & $3.8\times10^{12}$ & 1.33 & $2.58\times10^{3}$ & - &  -& $ 1.74\times 10^{-7}$ & - & -& 1 & $10^{43}$ & $10^{-7}$  \\
		J7\_Z &  $10^{-31}$ & 6 & - & $5.2\times10^{12}$ & 1.33 & $3.52\times10^{3}$ & - &  -& $ 2.38\times 10^{-7}$  & - & -& 1 & $10^{43}$ & $10^{-7}$  \\
		J7\_A   &  $10^{-31}$ & 10 & 45 & $1.8\times10^{12}$ & 1.33 & $1.24\times10^{3}$ & $1.35$  & $1.65$ & $ 8.36\times 10^{-8}$ & $ 1.47\times 10^{-7}$ & 2.63& 1 & $10^{43}$ & $10^{-7}$  \\
		J7\_B   &  $10^{-31}$ & 7 & 45 & $3.4\times10^{12}$ & 1.33 & $2.56\times10^{3}$ & $1.04$ &  $1.28$ & $ 1.73\times 10^{-7}$  & $ 8.79\times 10^{-8}$ & 0.76& 1 & $10^{43}$ & $10^{-7}$  \\
		J7\_C &  $10^{-31}$ & 6 & 45 & $5.2\times10^{12}$ & 1.33 & $3.51\times10^{3}$ & $0.74$ &  $0.91$ & $ 2.37\times 10^{-7}$  & $ 4.46\times 10^{-8}$ & 0.28& 1 & $10^{43}$ & $10^{-7}$  \\
		J7\_E   &  $10^{-31}$ & 10& 10 & $1.8\times10^{12}$ & 1.33 & $1.26\times10^{3}$ & $1.35$  & $0.29$ & $ 8.48\times 10^{-8}$ & $ 1.45\times 10^{-7}$ & 2.56& 1 & $10^{43}$ & $10^{-7}$  \\
		J7\_F  &  $10^{-31}$ & 7 & 10& $3.8\times10^{12}$ & 1.33 & $2.58\times10^{3}$ & $1.03$  & $0.22$ & $ 1.74\times 10^{-7}$  & $ 8.53\times 10^{-8}$ & 0.73& 1 & $10^{43}$ & $10^{-7}$  \\
	       J7\_G &  $10^{-31}$ & 6 & 10 & $5.2\times10^{12}$ & 1.33 & $3.52\times10^{3}$ & $7.28$ & $0.17$ & $ 2.38\times 10^{-7}$  & $ 4.28\times 10^{-8}$ & 0.27& 1 & $10^{43}$ & $10^{-7}$  \\
		J7\_I   &  $10^{-31}$ & 10 & 75  & $1.5\times10^{12}$ & 1.33 & $1.01\times10^{3}$ & $1.37$ & $6.46$ & $ 6.85\times 10^{-8}$ & $ 1.77\times 10^{-7}$ & 3.90& 1 & $10^{43}$ & $10^{-7}$  \\
		J7\_J   &  $10^{-31}$ & 7 & 75 & $3.3\times10^{12}$ & 1.33 & $2.26\times10^{3}$ & $1.11$  &  $5.08$ & $ 1.53\times 10^{-7}$  & $ 1.28\times 10^{-7}$ & 1.25& 1 & $10^{43}$ & $10^{-7}$  \\
		J7\_K &  $10^{-31}$ & 6 & 75 & $4.8\times10^{12}$ & 1.33 & $3.29\times10^{3}$ & $0.81$ &  $3.71$ & $ 2.22\times 10^{-7}$  & $ 7.36\times 10^{-8}$ & 0.50& 1 & $10^{43}$ & $10^{-7}$  \\
	\hline

	\end{tabular}
\end{table}
\end{landscape}

\begin{table}
	\centering
	\caption{Kinetic, internal energy and magnetic jet fluxes in terms of the total jet flux for the simulations in Table~\ref{tab:jetpars2}.}
		\label{tab:jetflux2}
	\begin{tabular}{lccc} % four columns, alignment for each
		\hline
		Model & $F_{\rm k}$ & $F_{\rm i}$ & $F_{\rm P}$ \\
\hline
		J5\_X & 0.080 & 0.920 & - \\
		J5\_Y & 0.039 & 0.961 & - \\
		J5\_Z & 0.028 & 0.972 & - \\
		J5\_A & 0.080 & 0.907 & 0.013 \\
		J5\_B & 0.039 & 0.954 & $7.60 \times 10^{-3}$ \\
		J5\_C & 0.028 & 0.968 & $3.82 \times 10^{-3}$ \\
		J5\_E & 0.080 & 0.920 & $4.00 \times 10^{-4}$ \\
		J5\_F & 0.039 & 0.961 & $2.32 \times 10^{-4}$ \\
		J5\_G & 0.028 & 0.971 & $1.20 \times 10^{-4}$ \\
		J5\_I & 0.080 & 0.728 & 0.192 \\
		J5\_J & 0.039 & 0.839 & 0.122 \\
		J5\_K & 0.028 & 0.907 & 0.065 \\
\hline
	        J7\_X & $7.96 \times 10^{-4}$ & 0.9992 & - \\
		J7\_Y  & $3.88 \times 10^{-4}$ & 0.9996 & - \\
		J7\_Z & $2.84 \times 10^{-4}$ & 0.9997& - \\
		J7\_A & $7.96 \times 10^{-4}$& 0.986 & 0.013 \\
		J7\_B & $3.88 \times 10^{-4}$ & 0.992 & $7.58 \times 10^{-3}$ \\
		J7\_C & $2.84 \times 10^{-4}$ & 0.996 & $3.81 \times 10^{-3}$ \\
		J7\_E & $7.96 \times 10^{-4}$ & 0.999 & $3.98 \times 10^{-4}$ \\
		J7\_F & $3.88 \times 10^{-4}$ & 0.9994 & $2.34 \times 10^{-4}$ \\
	        J7\_G & $2.84 \times 10^{-4}$ & 0.9996 & $1.17 \times 10^{-4}$ \\
		J7\_I & $7.96 \times 10^{-4}$ & 0.807 & 0.192 \\
		J7\_J & $3.88 \times 10^{-4}$ & 0.878 & 0.122 \\
		J7\_K & $2.84 \times 10^{-4}$ & 0.935 & 0.065 \\

	\end{tabular}
\end{table}
%
%%%%%%%%%%%%%%%%%%%%%%%%%%%%%%%%%%%%%%%%%%%%%%%%%%%%%%%%%%%%%%%%%%%%%%%%%%%%%%%%%%%%%%%%%%%%%%%%%%%%%%%%%%%%%%%%%%%%%%%%%%


\subsection{Two-dimensional maps} \label{app:maps}

\begin{figure*} 
	\includegraphics[width=\textwidth]{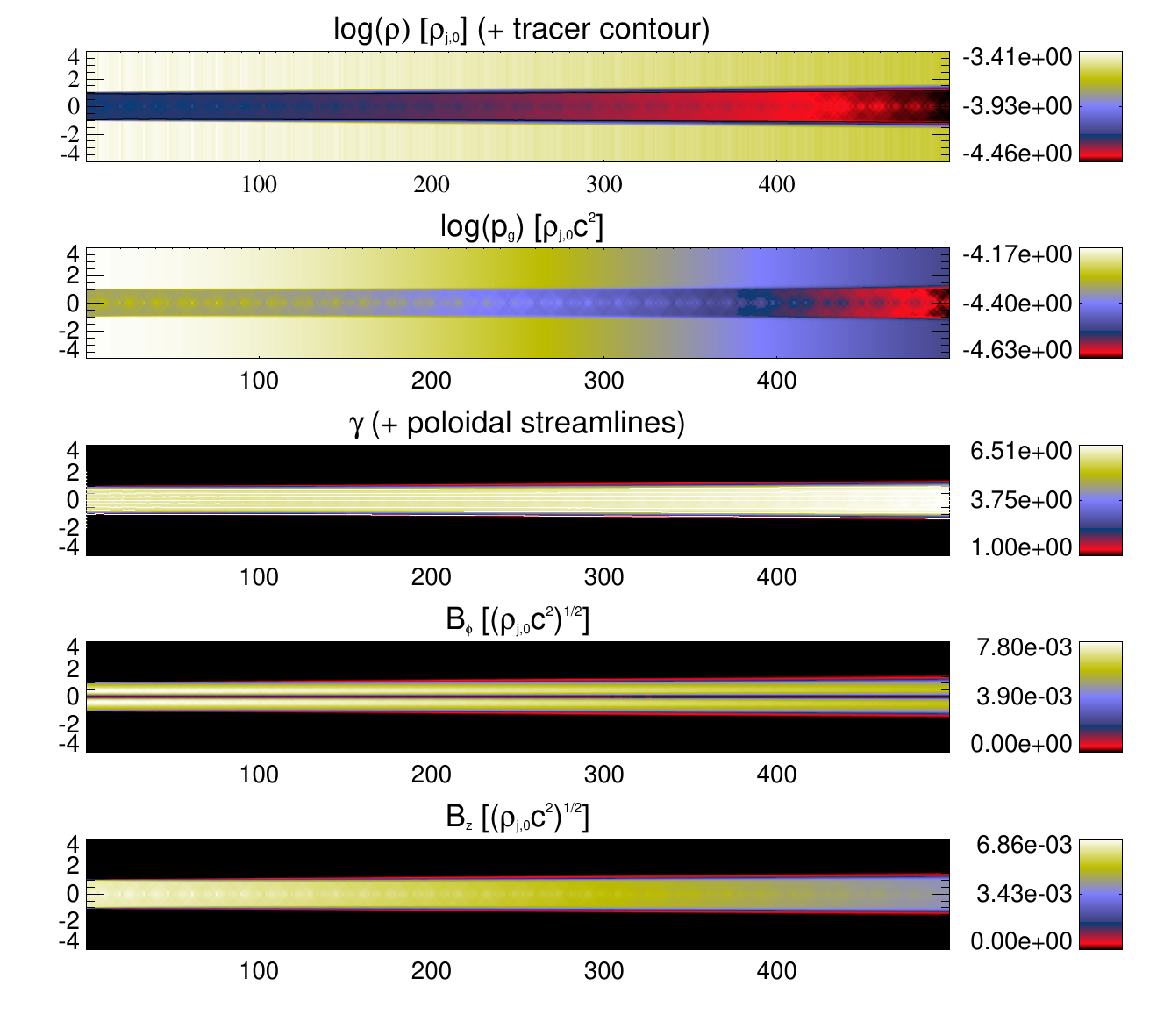} \,
\caption{Same as Fig.~\ref{fig:40A} for the case of model J4OC. The units for all plots are code units (in which $\rho_{\rm j,0} = 1$, $R_{j,0} = 1$ and $c=1$).}
\label{fig:40C}
\end{figure*}

\begin{figure*} 
\includegraphics[width=\textwidth]{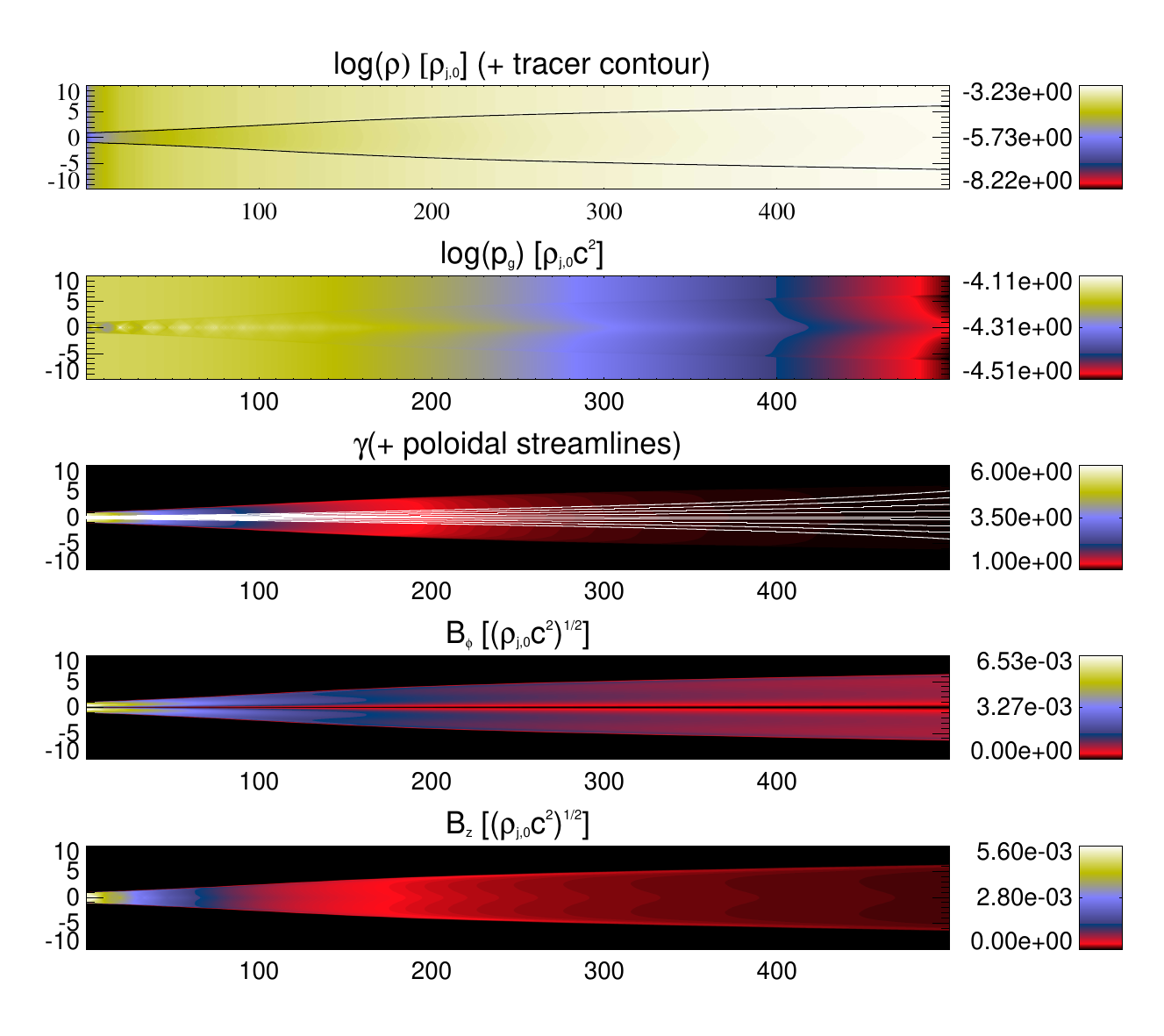} \,
\caption{Same as Fig.~\ref{fig:40A} for the case of model J8RC. The units for all plots are code units (in which $\rho_{\rm j,0} = 1.$, $R_{j,0} = 1.$ and $c=1.$).}
\label{fig:8RC}
\end{figure*}

The figures shown here (\ref{fig:40C} and \ref{fig:8RC}) show the two dimensional distributions of physical variables in simulations J4OC and J8RC, which complement Figs.~\ref{fig:40A} and \ref{fig:8RA} shown in the paper.

%%%%%%%%%%%%%%%%%%%%%%%%%%%%%%%%%%%%%%%%%%%%%%%%%%%%%%%%%%%%%%%%%%%%%%%%%%%%%%%%%%%%%%%%%%%%%%%%%%%%
%
%
%%%%%%%%%%%%%%%%%%%%%%%%%%%%%%%%%%%%%%%%%%%%%%%%%%%%%%%%%%%%%%%%%%%%%%%%%%%%%%%%%%%%%%%%%%%%%%%%%%%%

%%%%%%%%%%%%%%%%%%%%%%%%%%%%%%%%%%%%%%%%%%%%%%%%%%%%%%%%%%%%%%%%%%%%%%%%%%%%%%%%%%%%%%%%%%%%%%%%%%%
%
%
%%%%%%%%%%%%%%%%%%%%%%%%%%%%%%%%%%%%%%%%%%%%%%%%%%%%%%%%%%%%%%%%%%%%%%%%%%%%%%%%%%%%%%%%%%%%%%%%%%%

\subsection{Intermediate and strong mass-load} \label{app:int}
In this Appendix we show the Figures of mean Lorentz factor, rest-mass density, radius and temperature for models J4Q\_, J8Q\_ (intermediate mass-load models) and J4R\_ and J8R\_  (strong mass-load models) with Lorentz factors 10 and 6. These results are discussed in the paper (Section~\ref{sec:isload}).

 \begin{figure*} 
    \includegraphics[width=\textwidth]{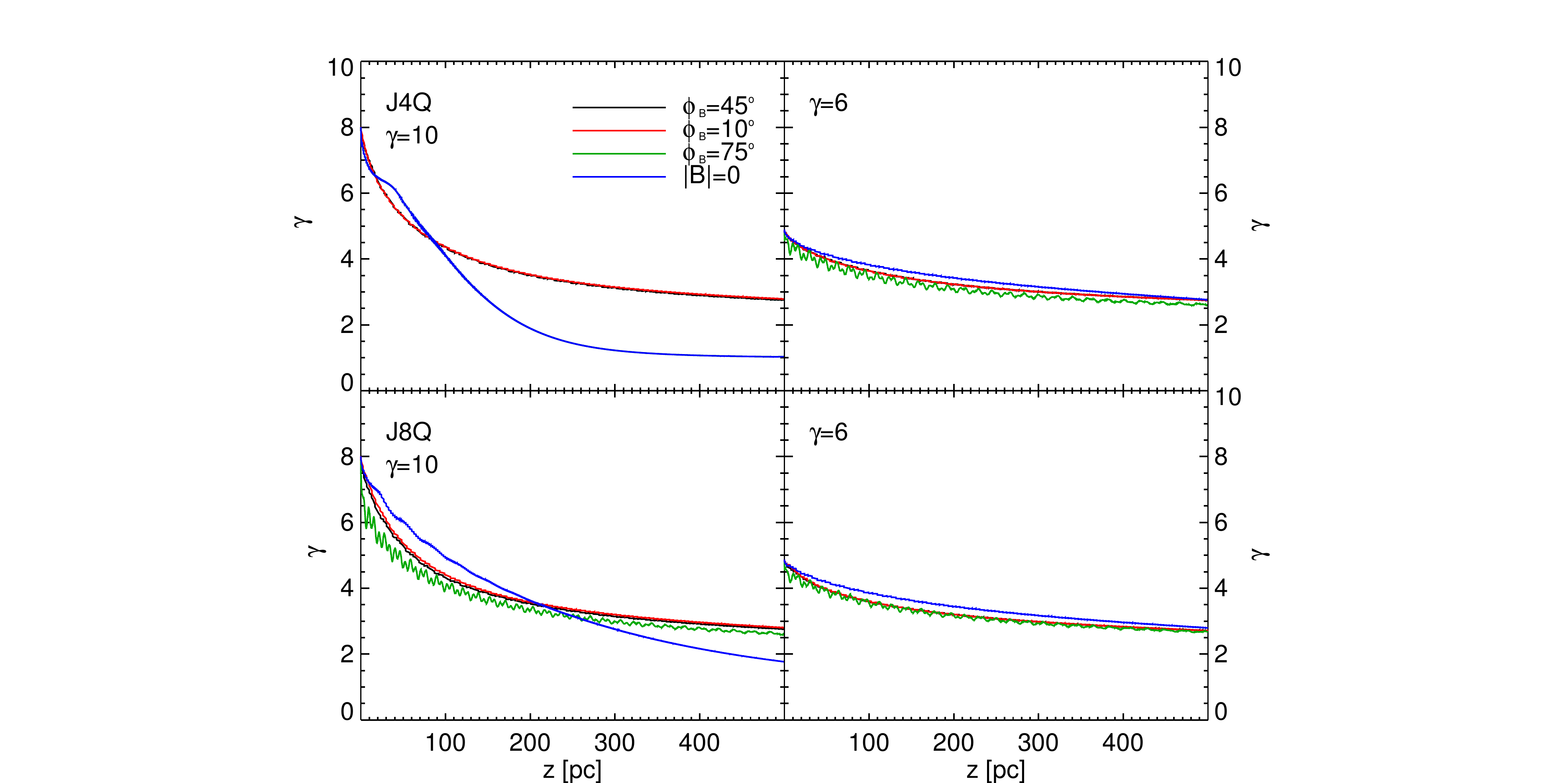} 
    \caption{Mean jet Lorentz factor versus distance for models J4Q\_ (top) and J8Q\_ (bottom) with Lorentz factor 10 (left column panels), and Lorentz factor 6 (right column panels). In each plot, the black, red, green and blue lines represent the jets with mean pitch angles $\phi_B=45^\circ$, $10^\circ$, $75^\circ$, and purely hydro models, respectively.}
    \label{fig:lorQ}
 \end{figure*}

  \begin{figure*} 
    \includegraphics[width=\textwidth]{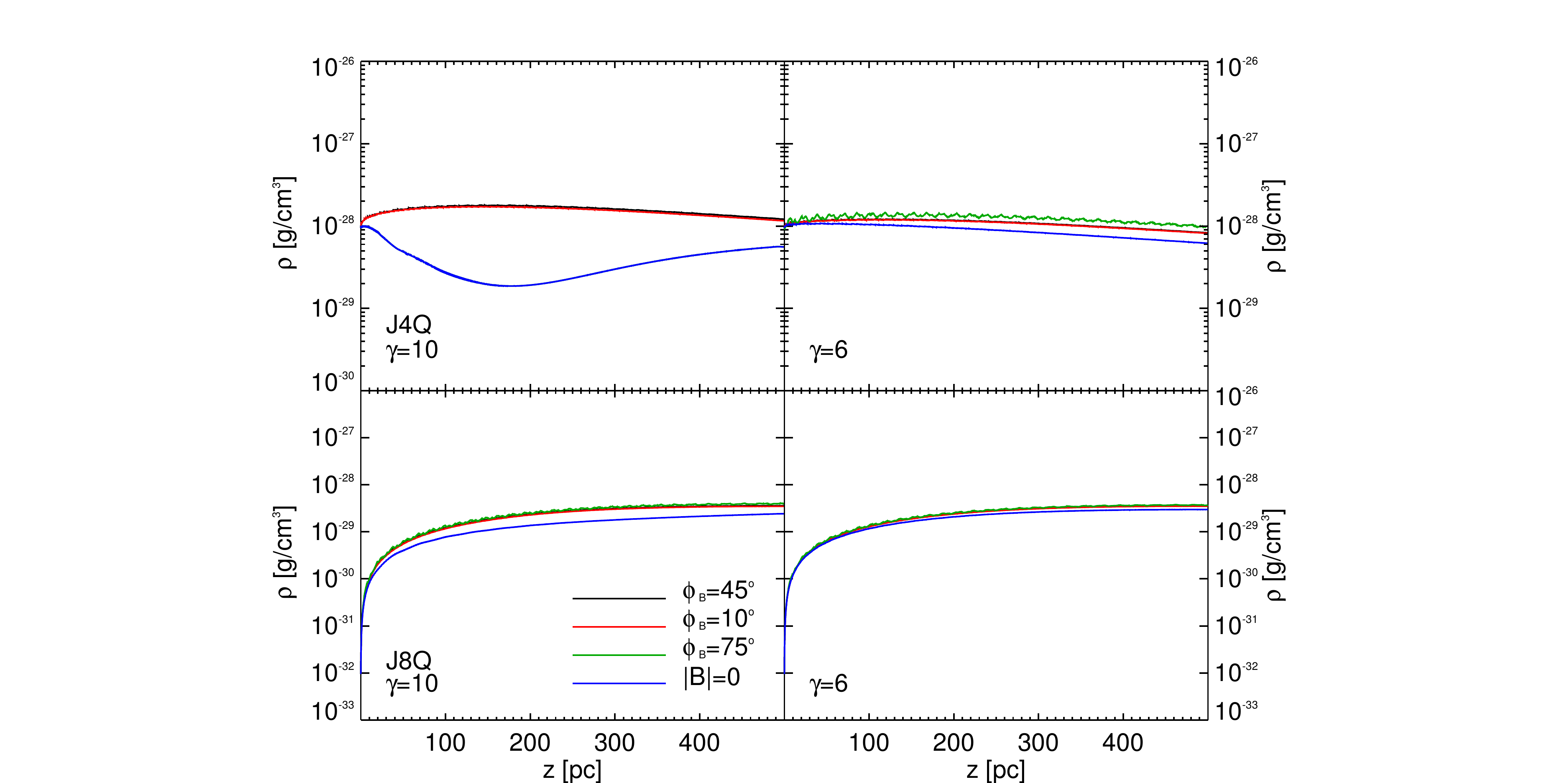} 
    \caption{Rest-mass density versus distance for models J4Q\_ (top) and J8Q\_ (bottom) with Lorentz factor 10 (left column panels), and Lorentz factor 6 (right column panels). In each plot, the black, red, green and blue lines represent the jets with mean pitch angles $\phi_B=45^\circ$, $10^\circ$, $75^\circ$, and purely hydro models, respectively.}
    \label{fig:rhoQ}
 \end{figure*}
 
 \begin{figure*} 
    \includegraphics[width=\textwidth]{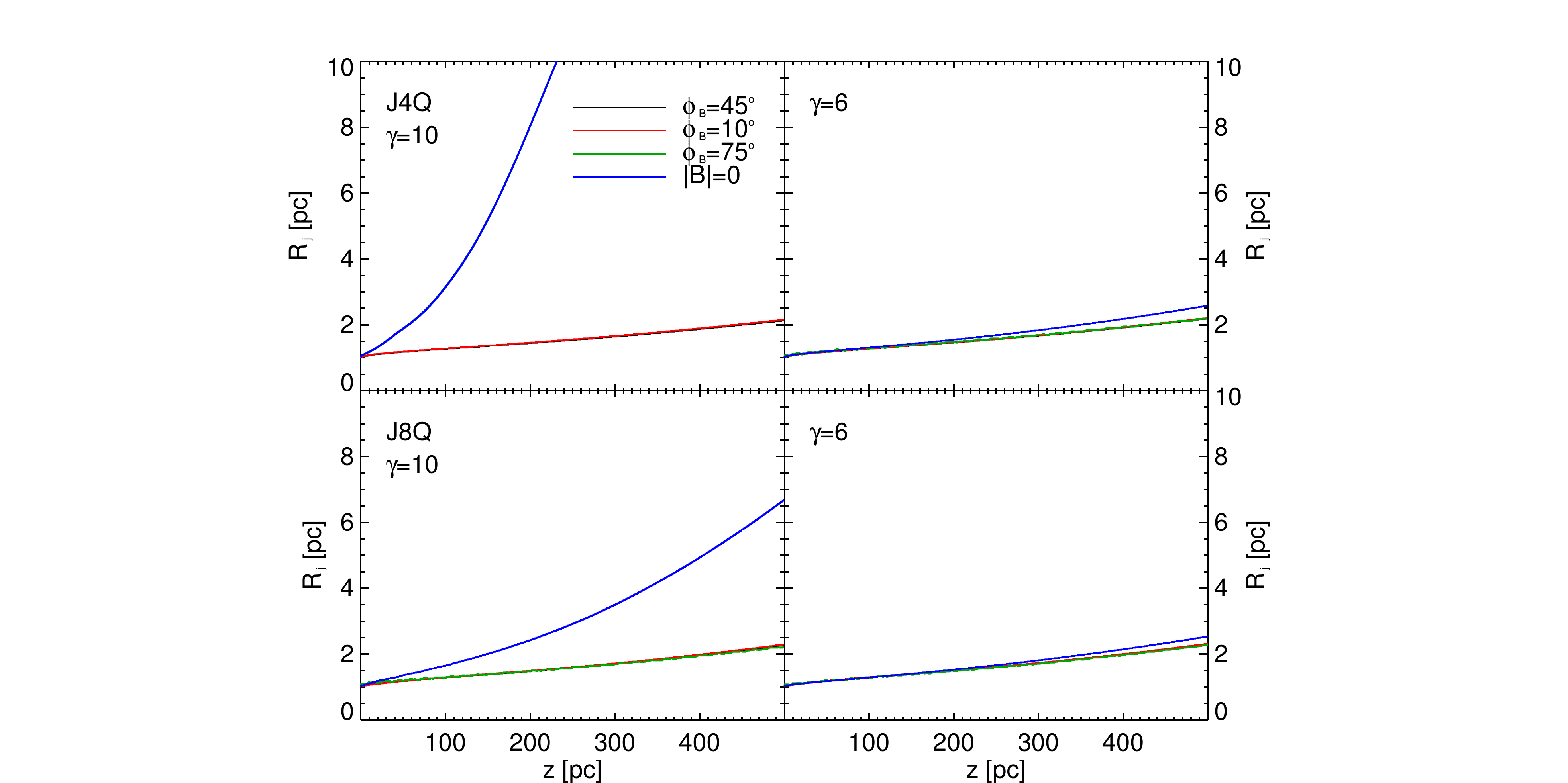} 
    \caption{Jet radius for models for models J4Q\_ (top) and J8Q\_ (bottom) with Lorentz factor 10 (left column panels), and Lorentz factor 6 (right column panels). In each plot, the black, red, green and blue lines represent the jets with mean pitch angles $\phi_B=45^\circ$, $10^\circ$, $75^\circ$, and purely hydro models, respectively.}
    \label{fig:radQ}
 \end{figure*}

 \begin{figure*} 
    \includegraphics[width=\textwidth]{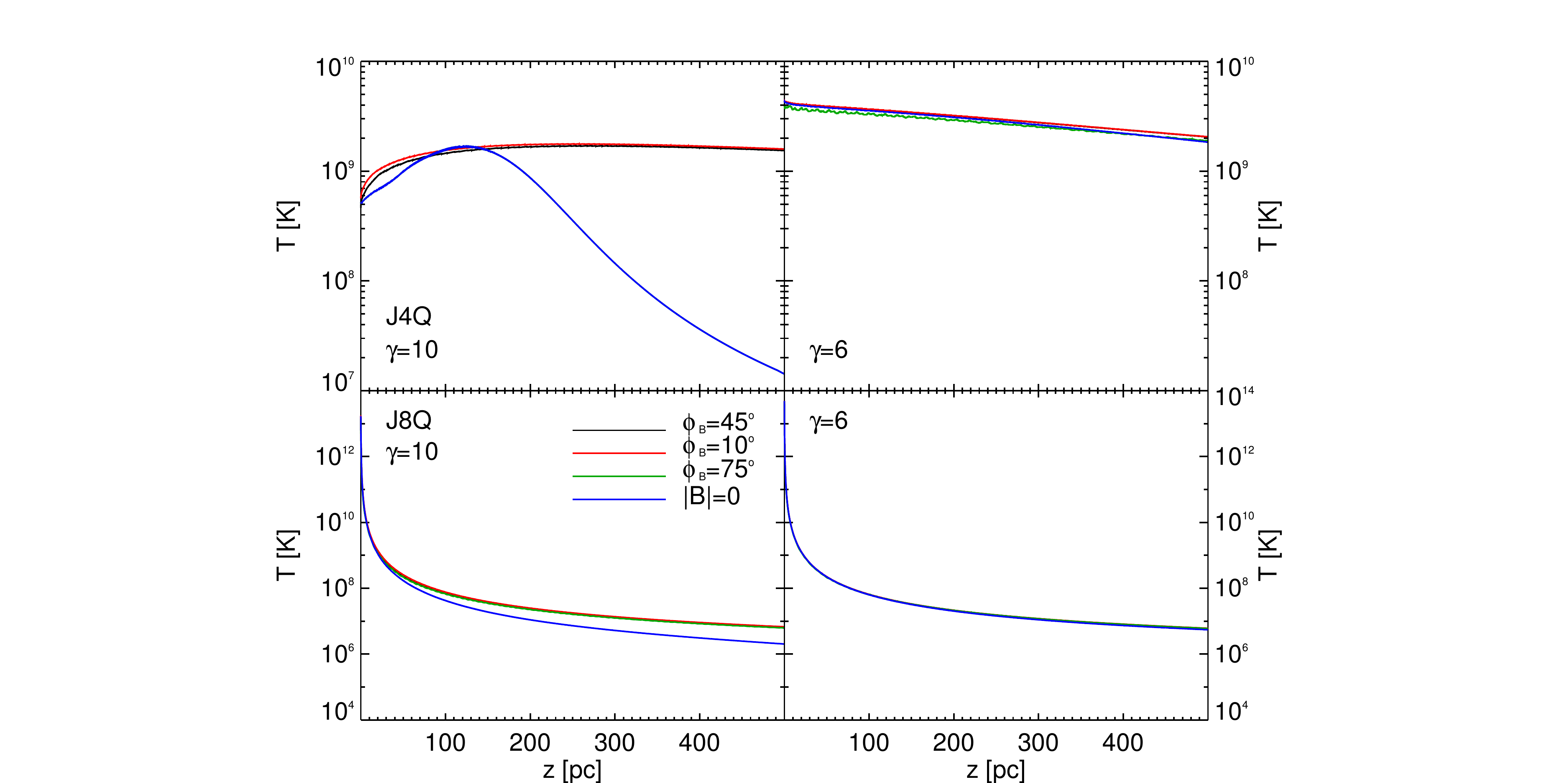} 
    \caption{Jet temperature for models for models J4Q\_ (top) and J8Q\_ (bottom) with Lorentz factor 10 (left column panels), and Lorentz factor 6 (right column panels).In each plot, the black, red, green and blue lines represent the jets with mean pitch angles $\phi_B=45^\circ$, $10^\circ$, $75^\circ$, and purely hydro models, respectively.}
    \label{fig:temQ}
 \end{figure*}

%\section{Strong mass-load} \label{app:str}
%In this Appendix we show the Figures of mean Lorentz factor, rest-mass density, radius and temperature for models with Lorentz factors 10 and 6. These results are discussed in the paper (Section~\ref{sec:isload}).

 \begin{figure*} 
    \includegraphics[width=\textwidth]{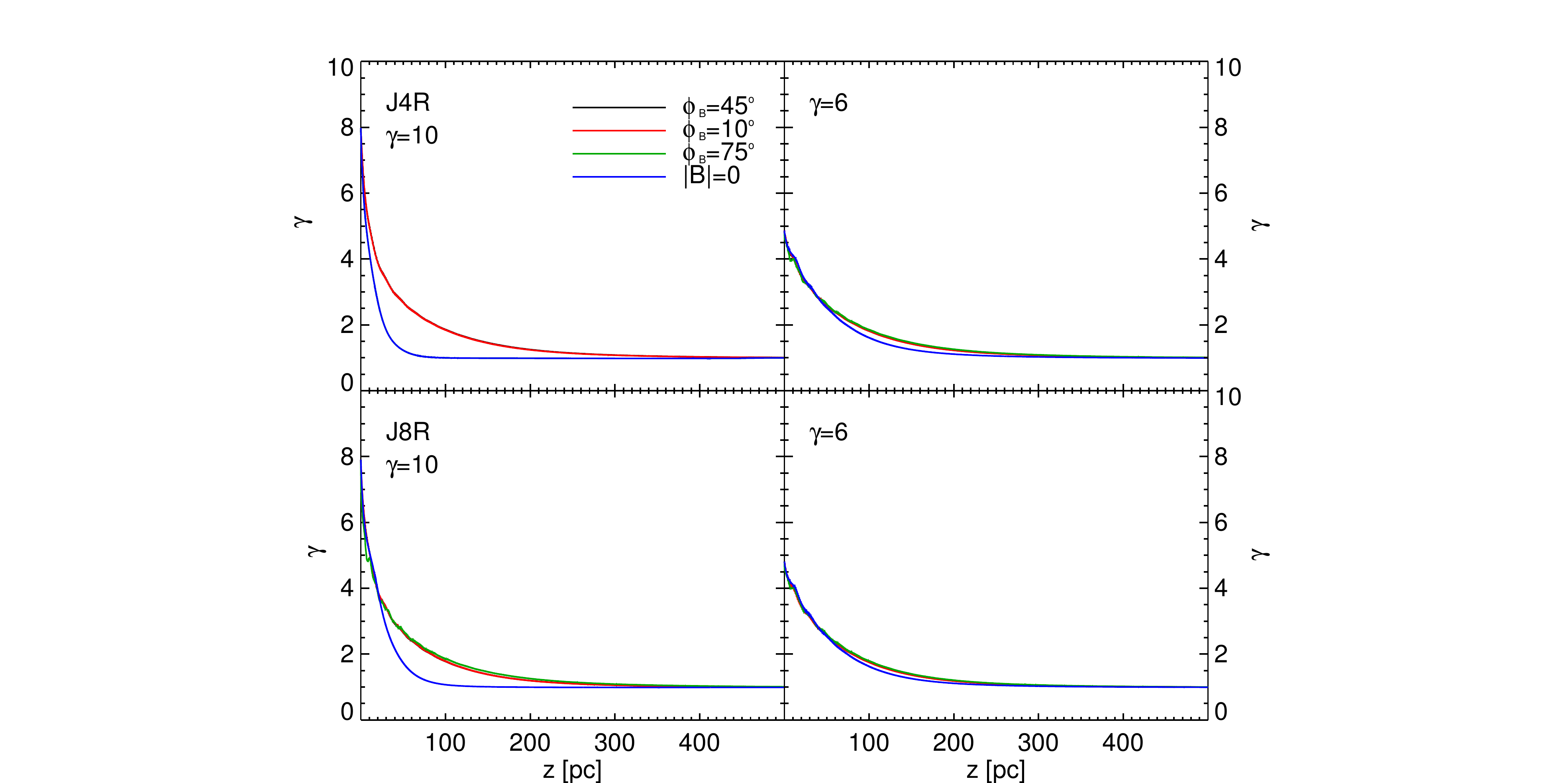} 
    \caption{Mean jet Lorentz factor versus distance for models J4R\_ (top) and J8R\_ (bottom) with Lorentz factor 10 (left column panels), and Lorentz factor 6 (right column panels). In each plot, the black, red, green and blue lines represent the jets with mean pitch angles $\phi_B=45^\circ$, $10^\circ$, $75^\circ$, and purely hydro models, respectively.}
    \label{fig:lorR}
 \end{figure*}

  \begin{figure*} 
    \includegraphics[width=\textwidth]{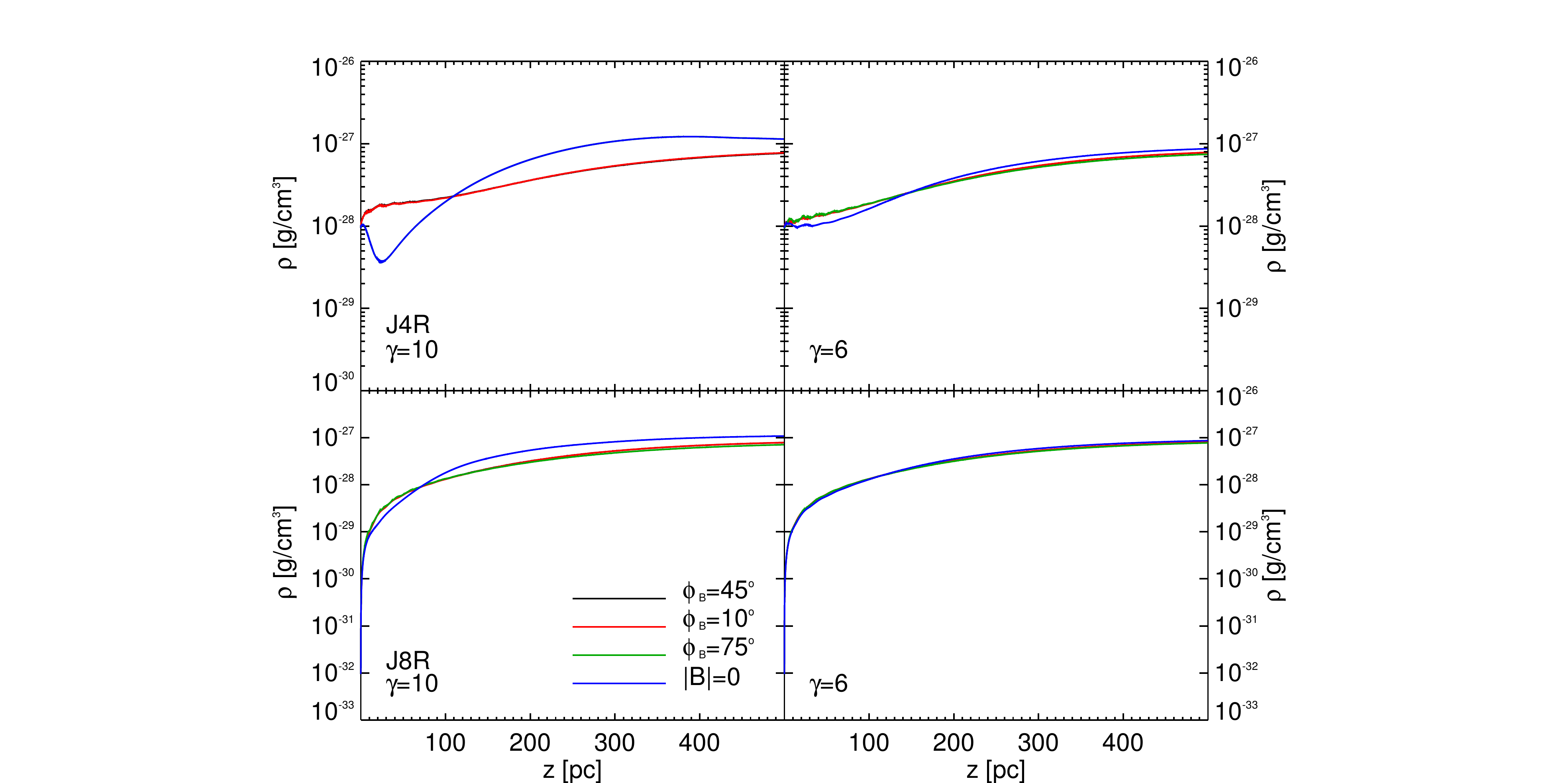} 
    \caption{Rest-mass density versus distance for models J4R (top) and J8R (bottom) with Lorentz factor 10 (left column panels), and Lorentz factor 6 (right column panels). In each plot, the black, red, green and blue lines represent the jets with mean pitch angles $\phi_B=45^\circ$, $10^\circ$, $75^\circ$, and purely hydro models, respectively.}
    \label{fig:rhoR}
 \end{figure*}
 
 \begin{figure*} 
    \includegraphics[width=\textwidth]{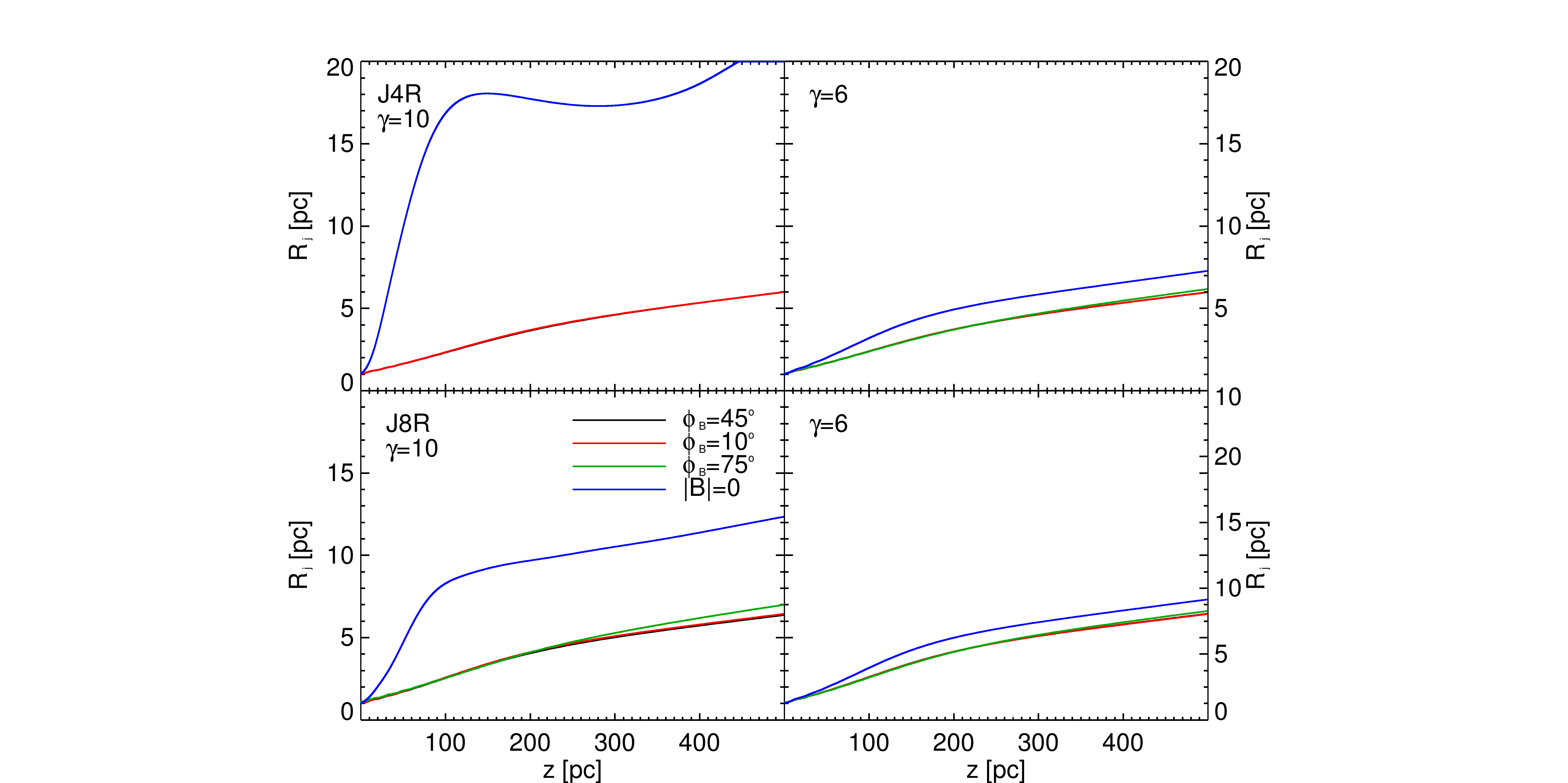} 
    \caption{Jet radius for models for models J4R (top) and J8R (bottom) with Lorentz factor 10 (left column panels), and Lorentz factor 6 (right column panels). In each plot, the black, red, green and blue lines represent the jets with mean pitch angles $\phi_B=45^\circ$, $10^\circ$, $75^\circ$, and purely hydro models, respectively.}
    \label{fig:radR}
 \end{figure*}

 \begin{figure*} 
    \includegraphics[width=\textwidth]{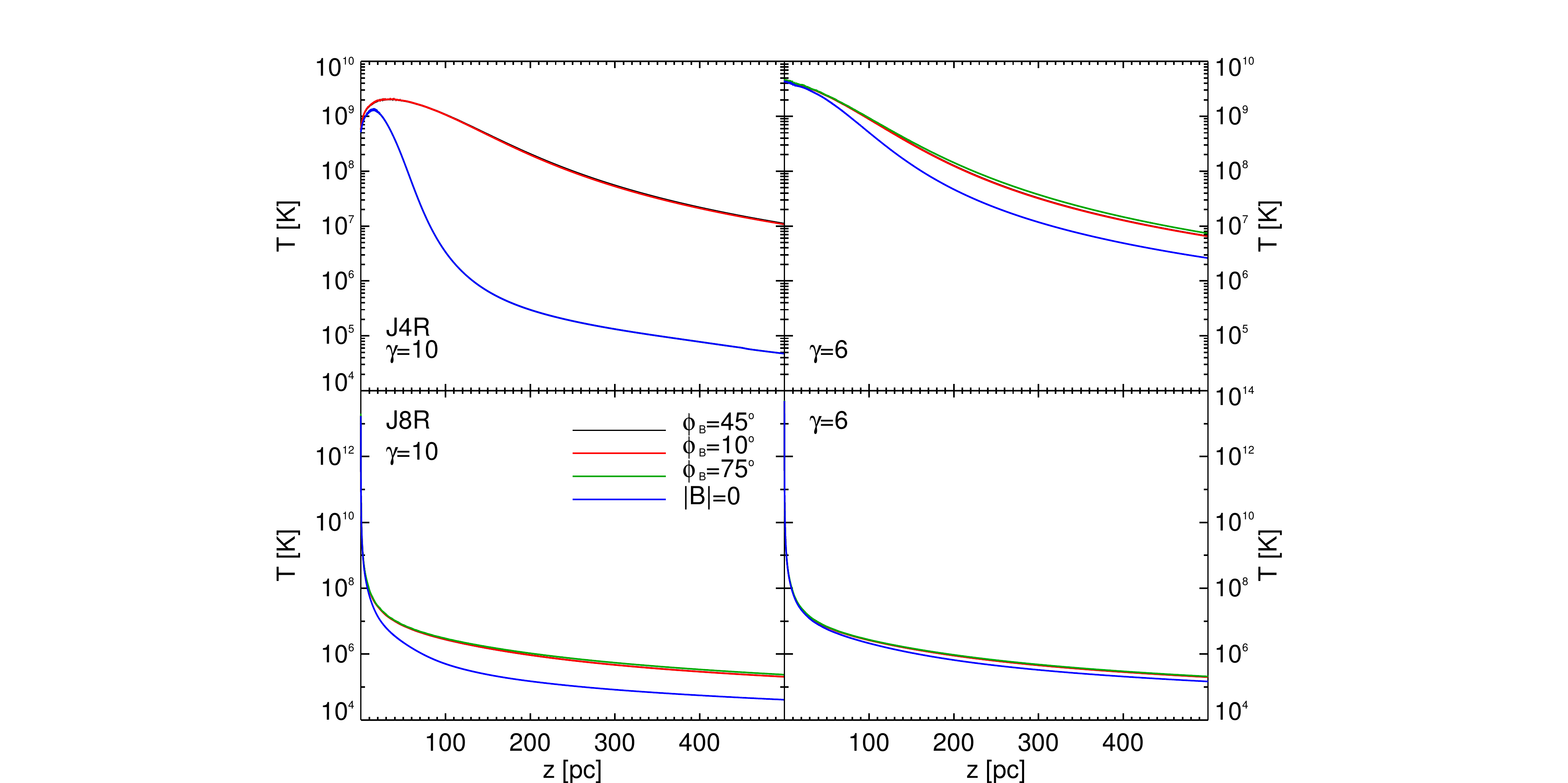} 
    \caption{Jet temperature for models for models J4R (top) and J8R (bottom) with Lorentz factor 10 (left column panels), and Lorentz factor 6 (right column panels).In each plot, the black, red, green and blue lines represent the jets with mean pitch angles $\phi_B=45^\circ$, $10^\circ$, $75^\circ$, and purely hydro models, respectively.}
    \label{fig:temR}
 \end{figure*}

% Don't change these lines
\bsp	% typesetting comment
\label{lastpage}
\end{document}